%% file: wme.tex
\documentclass[11pt,prd,aps,nofootinbib,floatfix]{revtex4-1} 

\usepackage{epsfig}
\usepackage{graphicx}
\usepackage{multirow} 
\usepackage{color}
\usepackage{hyperref} 
\usepackage{amsmath}
\usepackage{amssymb}
\usepackage{macros}
\usepackage{subfig} 
\usepackage[justification=justified,singlelinecheck=false]{caption}
\usepackage{rotating}

\usepackage{slashed}
\usepackage[normalem]{ulem}
\graphicspath{{./figs//}}

\newcommand\york{Department of Physics and Astronomy, York University, Toronto, Ontario, M3J 1P3, Canada}

\newcommand\glasgow{SUPA, School of Physics and Astronomy, University of Glasgow, Glasgow, G12 8QQ, UK}

\newcommand\liverpool{Theoretical Physics Division, Department of Mathematical Sciences, University of Liverpool, Liverpool L69 3BX, UK}

\begin{document}


\title{Neutral Kaon Mixing Beyond the Standard Model 
with \texorpdfstring{$n_f=2+1$}{nf=2+1} Chiral Fermions
Part 1: Bare Matrix Elements and Physical Results}
  
\author{N.~Garron$^a$, R.J.~Hudspith$^b$, A.T.~Lytle$^c$}
\affiliation{$^a$\liverpool,}
\affiliation{$^b$\york,}
\affiliation{$^c$\glasgow.}

\collaboration{The RBC-UKQCD Collaboration}

\date{\today}

\begin{abstract}  
We compute the hadronic matrix elements of the four-quark operators 
relevant for  $K^0-{\bar K^0}$ mixing beyond the Standard Model. 
Our results are from lattice QCD simulations with $n_f=2+1$ 
flavours of domain-wall fermion, which exhibit
continuum-like chiral-flavour symmetry. 
The simulations are performed at two different values of the lattice spacing 
($a\sim0.08$ and $a\sim 0.11 \, \fm $) 
and with lightest unitary pion mass $\sim 300\, \MeV$.      
For the first time, the full set of relevant four-quark operators
is renormalised non-perturbatively through RI-SMOM schemes;
a detailed description of the renormalisation procedure is presented in a companion paper.
We argue that the intermediate renormalisation scheme is
responsible for the discrepancies found by different collaborations.
We also study different normalisations and determine the matrix elements of the
relevant four-quark operators with a precision of $\sim 5\%$ or better.

\end{abstract}


\maketitle

\input ./Sections/s1_intro.tex
\input ./Sections/s2_methodology.tex

\input ./Sections/s3_bare.tex

\input ./Sections/s4_results.tex

\input ./Sections/s5_concl.tex

\input Aknow.tex

\clearpage
\newpage

\section{Appendices}
\input ./Appendices/app_chisqr.tex
\input ./Appendices/app_npr.tex
\input ./Appendices/app_chipt.tex

\input ./Appendices/app_bare_results.tex
\input ./Appendices/app_MethodAB.tex
\input ./Appendices/app_MethodC.tex
\input ./Appendices/app_correlations.tex

\bibliography{biblio}{}
\bibliographystyle{h-elsevier}

\end{document}

%% file: Sections/s1_intro.tex
\section{Introduction}\label{sec:intro}

The investigation of neutral Kaon mixing has 
been an important area for our understanding of the Standard Model (SM) of particle physics.
CP-violation was first observed in $K_S$ regeneration 
experiments~\cite{PhysRevLett.13.138} 
and the small value of the $K_L$-$K_S$ mass difference
led to the prediction of the charm quark at the GeV scale~\cite{Glashow:1970gm,Bjorken:1964gz}. 
\begin{figure}[t]
\begin{center}
\includegraphics[width=6cm]{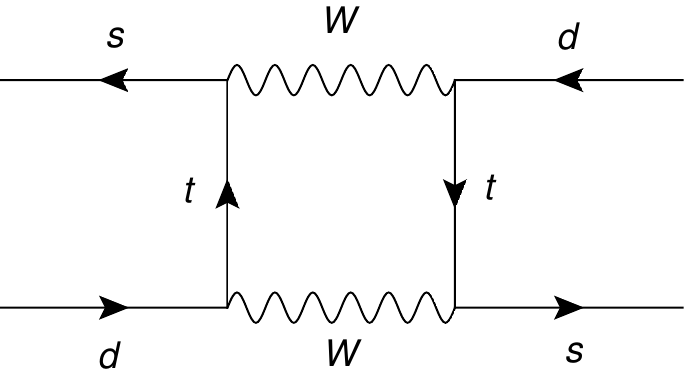}
\caption[]{Example of leading order box diagrams that contributes to $K^{0}-\bar{K}^{0}$
mixing in the SM.} 
\label{fig:box}
\end{center}
\end{figure}
Neutral Kaon mixing within the SM 
is dominated by $W$-exchange box diagrams as illustrated in Fig.~\ref{fig:box}.
By performing an operator product expansion, 
one can factorise the long-distance effects into  
the matrix element $\la \Kb | O_1 | \K \ra$
of the four quark operator.
\begin{equation}\label{eq:O1}
O_1 =  
(\overline s_a \gamma_\mu (1-\gamma_5) d_a)\,
(\overline s_b  \gamma_\mu (1-\gamma_5) d_b) \, 
\;,
\end{equation}
where $a$ and $b$ are colour indices and the summation over Dirac indices is implicit.
In the SM, the only Dirac structure which contributes is 
``$(\textrm{Vector-Axial})\times (\textrm{Vector-Axial})$ ''
arising from the W-vertices. 
The four-quark operator given in Eq.~\eqref{eq:O1} is invariant
under Fierz re-arrangement, therefore gluonic exchanges 
do not introduce new four-quark operators. 

In a massless renormalisation scheme which preserves chiral symmetry 
the four-quark operator $ O_1$ does not mix with other four-quark operators, nor with lower dimensional operators.
The importance of the matrix element given in Eq.~\eqref{eq:O1} has motivated
many lattice studies of the SM kaon bag parameter
(defined in some renormalisation scheme at some scale $\mu$) 
\begin{equation}\label{eq:BK}
B_K(\mu) \equiv \frac{\la \Kb | O_1 (\mu) | \K \ra}{\frac{8}{3}f_K^2 m_K^2}\;,
\end{equation}
which have now achieved accuracies at the few-percent level~\cite{Aoki:2010pe,Blum:2014tka,Durr:2011ap,Constantinou:2010qv}.
(Our convention for the decay constant is such that $f_{K^−} = 156.1 $ MeV.)
Combined with the value of the Wilson coefficient $C(\mu)$, computed in perturbation theory, and experimental observables,
such as the mass difference $\Delta_{M_K}= m_{K_L}-m_{K_S}$ and $\varepsilon_K$, the determination of
$B_K(\mu)$ provides important constraints on the Cabibbo-Kobayashi-Maskawa (CKM) matrix.
Schematically, one obtains
\begin{equation}\label{eq:epsilonSM}
\varepsilon_K = C(\mu) \times B_K(\mu) 
\times {\cal F}(V_{ij}^{CKM},m_K,f_K,\Delta M_K,\ldots)\;,
\end{equation}
where ${\cal F}$ is a known function of the CKM factors and of 
well-measured quantities.
In the framework of the SM, 
the experimental value of $\varepsilon_K$ (which parametrizes indirect CP violation)
together with the theoretical determination of $B_K$ 
provides an important constraint on the apex of one the CKM unitary triangles
- in the $(\bar\eta, \bar \rho)$ plane - and 
on the overall consistency of the CKM picture. 
$\varepsilon_K$ is also a powerful probe of potential new physics, with sensitivity
to energies well beyond those being explored directly at the 
LHC~(see for example \cite{Ligeti:2016qpi,Bertolini:2014sua,Buras:2013ooa,Mescia:2012fg}).

Beyond the SM, both left-handed and right-handed 
currents may contribute in the $K^0$-$\bar{K}^0$ mixing process and the CP-violation parameter 
$\varepsilon_K$ is sensitive to new CP violating phases generically predicted by these models.
Here we assume that the new-physics effects occur at energy scales much higher than the interaction scale of QCD
and that QCD remains a valid description of the strong interaction in the non-perturbative regime. 
In addition to the SM operator $O_1$ given in Eq.~\eqref{eq:O1},
seven four-quark operators appear in a generic effective $\Delta S=2$ Hamiltonian~\cite{Gabbiani:1996hi}
\footnote{
Several basis conventions exist in the literature, here we choose the 
``SUSY'' basis~\cite{Gabbiani:1996hi,Allton:1998sm,Ciuchini:1998ix}
}

\begin{equation}\label{eq:H}
H = \sum_{i=1}^{5} \, C_i(\mu) \, O_i(\mu) + \sum_{i=1}^{3} \, \tilde C_i(\mu) \, \tilde{O}_i(\mu) \,,
\end{equation}
where
\begin{equation}\label{eq:susybasis}
 \begin{aligned}
O_2&=\;(\overline s_a (1-\gamma_5) d_a)\,(\overline s_b (1-\gamma_5) d_b)\\
O_3&=\;(\overline s_a  (1-\gamma_5) d_b)\,(\overline s_b   (1-\gamma_5) d_a)\\
O_4&=\;(\overline s_a  (1-\gamma_5) d_a)\,(\overline s_b   (1+\gamma_5) d_b)\\
O_5&=\;(\overline s_a  (1-\gamma_5) d_b)\,(\overline s_b   (1+\gamma_5) d_a)\,,
 \end{aligned}
\end{equation}
and $\tilde{O}_{i=1,2,3}$ are obtained from the $O_{i=1,2,3}$ 
by swapping chirality $(1-\gamma_5)\to(1+\gamma_5)$.  
The Wilson coefficients $C_i(\mu)$ and $\tilde{C}_i(\mu)$  depend on the details of the new-physics model under consideration
but the matrix elements $\la \Kb | O_i | \K \ra$ are model independent. (In our framework parity is conserved,
therefore the operators  $\tilde{O}_{i=1,2,3}$ are redundant).
In terms of representation of $SU_L(3)\times SU_R(3)$,
it is straightforward to show that in the chiral limit  $O_2$ and $O_3$ transform like $(6,\bar 6)$
while $O_4$ and $O_5$ belong to $(8,8)$.
Therefore these new operators mix pair-wise under renormalisation: $O_2$ with $O_3$; 
and $O_4$ with $O_5$. 

In contrast to $B_K(\mu)$, studies of the extended set of matrix elements are relatively few.
The first computation performed with dynamical fermions was reported by our collaboration in~\cite{Boyle:2012qb}
and was done with $n_f=2+1$ DW fermions at a single lattice spacing. It was followed by a $n_f=2$ computation
by the European Twisted Mass (ETM) collaboration  using twisted-mass Wilson fermions with several
lattice spacings~\cite{Bertone:2012cu}. These two computations reported results in decent agreement
(the matrix elements of $O_{2,3,4}$ agree within errors, $O_5$ only within $\sim 2 \,\sigma$), suggesting
that these quantities are not very sensitive to the number of flavours.
However, another study by the Staggered Weak Matrix Element (SWME) collaboration using $n_f=2+1$ flavours of improved staggered 
fermions~\cite{Bae:2013tca} found a noticeable disagreement 
for two of these  matrix elements ($O_{4}$ and $O_5$). 
The ETM collaboration has since repeated their computation with $n_f=2+1+1$ flavours and
found bag parameters compatible with their $n_f=2$ results 
(only within $\sim 2 \,\sigma$ for $O_5$)\cite{Carrasco:2015pra}.
The SWME collaboration has extended their previous study by adding more ensembles
and improving extrapolations to the physical point~\cite{Jang:2015sla},
they confirmed their disagreement with the other studies.
Since the results have been extrapolated to the continuum limit, one does not expect
the fermion discretisation used (Domain-Wall, Twisted-Mass, or Staggered)
to be responsible for the discrepancy.

Central to this work is an explanation for this disagreement, 
our arguments and preliminary results have been presented 
in~\cite{Hudspith:2015wev} and discussed with the authors of~\cite{Jang:2015sla}.
We improve upon our earlier DWF result~\cite{Boyle:2012qb} in two important ways:
by adding a second lattice spacing, allowing us to take the continuum limit 
(with a resonable handle on the lattice artefacts)
and by renormalizing the four-quark operators through non-exceptional momentum schemes.

As we will show, the second point is of great importance and is often
overlooked. Some systematic errors in the original RI-MOM schemes
which are very hard to control are absent in the RI-SMOM schemes we present here.

In the next section, we give an overview of our strategy and make explicit our choice of conventions 
(choice of basis, normalisation). Sections~\ref{sec:methodology} and \ref{sec:bare}
contain our global fit procedure and the method for determining the bare hadronic matrix elements
$\la \Kb | O_i(\mu) \K \ra $. In section~\ref{sec:results} we present our 
final results and compare with previous works.

%% file: Sections/s2_methodology.tex
\section{Extrapolations to the {\em physical point}}\label{sec:methodology}

In this work we have considered data with pion masses in the range of 
$m_P \sim 300-430\, \MeV$ and performed a chiral extrapolation to the physical value of
$m_\pi = 140\, \MeV$ (we take the mass of the charged pions). The spatial extent our the simulated lattice is
$L\sim 2.66$ fm, so within this range of pion masses $L m_P> 4 $, therefore 
the finite volume effects are expected to be negligible compared to our systematic errors.
We work in the isospin limit, $m_u=m_d\equiv m_{ud}$ and for the same reason
we do not consider isospin corrections.
Furthermore, we also require a continuum extrapolation to reach the {\em physical point}
($a=0, m_\pi =140\, \MeV$).
Since we work with Domain-Wall fermions, we expect
the dominant lattice artefacts to be linear in $a^2$
(we remind the reader that $a^3$ corrections of the fermionic action are forbidden by chiral symmetry~\footnote{
  See the footnote in Section~\ref{sec:results} about the effects of the residual mass.}
).
Before the continuum extrapolation can be performed, a renormalisation step is also necessary:
we employ the non-perturbative Rome-Southampton method~\cite{Martinelli:1994ty},
as explained in detail in a companion paper~\cite{KKNPR}.
Below we list our strategy to extract the physical quantities of interest from our lattice simulations:
\begin{enumerate}
\item Compute the bare matrix elements, at two values of the lattice spacing and several values of
  the quark masses (on already existing RBC-UKQCD ensembles).
\item Renormalise these bare quantities.
\item Interpolate/extrapolate to the physical value of the strange quark mass.
\item Extrapolate to the physical point (Continuum/Chiral extrapolation in the light quark sector).
\end{enumerate}

Central to this work is an investigation of the extrapolations to the physical point
(details can be found in section~\ref{sec:results}).
In particular we have studied several parametrisations  of the four-quark operator matrix elements.
Ideally, one would like to find a dimensionless quantity 
which can smoothly be extrapolated to the physical point and be free of large systematic errors.
For the SM matrix element one usually defines the bag parameter $B_K$
as in Eq.~\eqref{eq:BK}: The matrix element of the four-quark operator
is normalised by its Vacuum Saturation Approximation (VSA).
This normalisation is widely accepted for the SM contribution, however
this is not the case for the BSM matrix elements,
for which different possibilities have been proposed
(see for example~\cite{Donini:1999nn, Becirevic:2004qd, Babich:2006bh, Bae:2013tca}).
We investigate several strategies which differ by the choice of normalisation and global fit procedure,
allowing us to estimate the systematic uncertainties of our work.

\subsection{The ratios $R_i$}

A possible parameterisation of the matrix elements has been proposed in~\cite{Babich:2006bh}.
Denoting by $\P$ the simulated strange-light pseudo-scalar particle (kaon) of mass $m_P$ and decay constant $f_P$,
the ratios $R_i$ are defined by
\begin{equation}\label{eq:R}
{\cal R}_i\left (\frac{m_P^2}{f_P^2}, \mu, a^2 \right)= 
\left[ {\frac{f_K^2}{m_K^2}} \right]_{\text{Exp.}}
\left[ {\frac{m_P^2}{f_P^2} } { \frac{\langle \Pb| O_i(\mu) | \P \rangle }{\langle \Pb|O_1(\mu) | \P \rangle } }\right]_{\text{Lat.}}\;,
\end{equation}
such that at the physical point $\left(m_P=m_K,a^2=0\right)$
\begin{equation}
\label{eq:Rphys}
R_i (\mu) = 
{\cal R}_i 
\left( \frac{m_K^2}{f_K^2}, \mu, 0 \right) = 
{
\frac{\langle \Kb| O_i(\mu) | \K \rangle}
{\langle \Kb|O_1(\mu) | \K \rangle} }\;,
\end{equation}
is 
the ratio of the BSM matrix element to the SM one. Previous studies have shown that these ratios are large ( $\sim O(10)$ )
as the BSM matrix elements are enhanced compared to the SM one \cite{Babich:2006bh,Bertone:2012cu,Garron:2012ex}
(this is expected from Chiral Perturbation Theory: the SM matrix element vanishes in the chiral limit
whereas the BSM matrix elements remain finite).
An advantage of this method compared to the bag parameters is that the denominators do not depend on the quark masses.
The BSM matrix elements can be reconstructed from the ratios $R_i$, the SM bag parameter $B_K$, 
the kaon mass and decay constant only.
Moreover, since the numerator and the denominator are very similar,
one expects some cancellations of the statistical and systematic errors to occur in the ratio.

\subsection{The Bag parameters $B_i$}

The renormalised bag parameters are defined as the ratio of the weak matrix elements normalised by their VSA values:
\begin{equation}
B_i(\mu) =\frac{ \la \Kb |O_i(\mu)| \K \ra } { \la \Kb |O_i(\mu)| \K \ra}_{\text{VSA}}\;.
\end{equation} 
For the SM bag parameter $B_1(\mu)=B_K(\mu)$ with our conventions,
\begin{equation}
\la \Kb |O_1(\mu)| \K \ra = \frac{8}{3} m_K^2 f_K^2 B_1(\mu)\;,
\end{equation}\
and for the BSM ones~\footnote{More precisely, the BSM matrix elements 
are normalised by a large $N$ approximation of the VSA, 
see for example the discussion in \cite{Allton:1998sm}.},
\begin{equation}
{\la \Kb |O_i(\mu)| \K \ra } = 
N_i \frac{m_K^4 f_K^2}{(m_s(\mu)+m_d(\mu))^2} B_i(\mu)\;,  \qquad  i>1\;.
\end{equation}
The factors $N_{i>1}$ depend on the basis, as we work in the SUSY basis we have $
N_{i>1} = \left\{ -\frac{5}{3},\frac{1}{3}, 2, \frac{2}{3} \right\}$.

For SM bag parameter $B_K$, the denominator consists of the precisely known quantities $f_K$ and $m_K$.
This contrasts with the BSM $B_i$, for which the denominator is not uniquely defined, it depends
on the scheme and the renormalisation scale.

\subsection{The Combinations $G_{ij}$}

Another possibility, advocated for example in~\cite{Becirevic:2004qd,Bae:2013tca} is to define products
and ratios of bag parameters such that the leading  chiral logarithms cancel out.
For some of these quantities (called ``golden combinations''), this cancellation actually occurs at every order
of the chiral expansion. For the other ones (``silver combinations''), only the leading logarithms cancel. 
Such quantities were introduced in~\cite{Becirevic:2004qd} for $SU(3)$ chiral perturbation theory and later in the context of $SU(2)$
staggered chiral perturbation theory in~\cite{Bailey:2012wb}.
The relevant NLO continuum $SU(2)$ chiral expansions can be found in Appendix~\ref{app:chipt}.
We follow~\cite{Bae:2013tca} and define~\footnote{
  Within our conventions, these definitions match the ones of ~\cite{Bae:2013tca},
  except for $G_{23}$. This is discussed in section~\ref{sec:results}.}
\begin{equation}\label{eq:G}
\begin{gathered}
  G_{21}(\mu)=\frac{B_2(\mu)}{B_{K}(\mu)}, \quad G_{23}(\mu)=\frac{B_2(\mu)}{B_3(\mu)} \,\\
  G_{24}(\mu)=B_2(\mu)B_4(\mu),\quad G_{45}(\mu)=\frac{B_4(\mu)}{B_5(\mu)}\;.
\end{gathered}
\end{equation}
  As can be seen in the Appendix~\ref{app:chipt}, the quantities $G_{23}$ and $G_{45}$ have no chiral logarithms,
  whereas in $G_{21}$, $G_{24}$ the cancellation only occurs for the leading logarithms.
  
\subsection{Continuum and chiral fitting strategies}

We start by adjusting our (renormalised) results to the physical strange mass.
On the coarse lattice we perform a linear interpolation whereas a tiny extrapolation is necessary on the fine one
(the numerical values are given in the next section).
Then we perform a combined chiral-continuum  extrapolation to the physical point.
In order obtain a reliable estimate of our systematic error we follow three different strategies:
\begin{itemize}
\item {\bf Method A.}
  We perform a global fit according to NLO $SU(2)$ chiral perturbation theory (see Appendix~\ref{app:chipt}).
  The general form of the fit function we use is (we drop the renormalisation scale dependence $\mu$ for clarity),
  \begin{equation}
    {\cal Y}_i(m_P^2,a^2) = Y_i(m_\pi^2,0)\left[ 1 + \alpha_i a^2 +
      \frac{m_P^2}{ f^2}\left( \beta_i + \frac{C_i}{16\pi^2} \log\left(\frac{m_P^2}{\Lambda^2}\right)\right)\right]\;.
  \end{equation}
  Where in this expression $m_P$ is the mass of the pseudoscalar meson made of two light quarks.  
  The values $Y_i$, $\alpha_i$ and $\beta_i$ are free parameters and fit simultaneously between ensembles of different lattice spacings.
  The values for $C_i$ are listed in Table~\ref{methodology:tab:chilogs} below.
  We have checked that for $f$, using the chiral value, the physical value or the simulated value $f_P$
  give compatible results.
  We apply the procedure to the ratios $R_i$ and to the bag parameters $B_i$.
  \\
  
\begin{table}[h]
\begin{tabular}{c | c c | c c }
\toprule
& $R_{2,3}$ & $R_{4,5}$ & $B_{1,2,3}$ & $B_{4,5}$ \\
\cline{2-5}
$C_i$ & $\frac{3}{2}$ & $\frac{5}{2}$ & $-\frac{1}{2}$ & $\frac{1}{2}$ \\ 
\botrule
\end{tabular}
\caption{
  {Chiral logarithm factors $C_i$ for $R_i$ and the $B_i$.}
}\label{methodology:tab:chilogs}
\end{table}

\item {\bf Method B.} We perform a continuum/chiral extrapolation of   $R_i$ and $B_i$ 
  using a global fit procedure according to the following ans\"atz ($\kappa_i$ and $\delta_i$
  are free parameters simultaneously fit  between ensembles)
\begin{equation}
{\cal Y}_i \left( m_P^2, a^2 \right) = 
Y_i ( m_\pi^2 , 0 ) 
+ \kappa_i  a^2 + \delta_i m_P^2\;.
\end{equation}

\item {\bf Method C.} We first extrapolate the combinations $G_{ij}$ according Method B (linearly in the pion mass squared),
  and then reconstruct the bag parameters. 
\end{itemize}

Methods A and B are equivalent up to the chiral logarithm terms, the difference allows us to estimate how strong the chiral effects
from being at non-physical pion mass are. The corresponding analysis is presented in great detail
in section~\ref{sec:results}.
Method C allows us to determine the bag parameters with no leading chiral logarithm, except from the standard model one,
whose effect is benign (as explained in section \ref{sec:results}). 
Furthermore, the quantities $G_{ij}$ have different statistical and systematic errors. 
Performing the analysis using different quantities and extrapolation methods allows us to check the consistency
of our final results and ensure our systematics are understood.
The results for Method C are presented in Appendix~\ref{app:methodC}.

%% file: Sections/s3_bare.tex
\section{Lattice implementation}
\label{sec:bare}

Our measurements are performed on $n_f=2+1$ gauge ensembles generated by RBC-UKQCD using the Iwasaki
gauge action~\cite{Iwasaki:1985we,Okamoto:1999hi} and the Shamir DWF formulation~\cite{Shamir:1993zy}.
These ensembles have been described extensively in~\cite{Arthur:2012opa} and references therein.

The finer of the two lattices used in this study has a lattice volume of $32^3 \times 64 \times 16$ with inverse lattice
spacing $a^{-1} = 2.383(9) \,\GeV$. 
There are three values of light sea quark masses $am^{\text{sea}}_{ud} = 0.004, 0.006,$ and $0.008$,
corresponding to unquenched pion masses of approximately $300$, $360$, and $410$ MeV
respectively. 
For the light valence quarks we use only unquenched data, $am^{\text{val}}_{ud} = am^{\text{sea}}_{ud}$.
The simulated strange quark mass for this ensemble is $am^{\text{sea}}_{s} = 0.03$. 
To reach the physical kaon mass we extrapolate using unitary ($am^{\text{val}}_{s} =am^{\text{sea}}_{s}= 0.03$)
and partially quenched ($am^{\text{val}}_{s}= 0.025$) data, which is close to its physical value
of $0.02477(18)$ \cite{Blum:2014tka}.

The coarser lattice has an extent of $24^3 \times 64 \times 16$, and inverse lattice spacing $a^{-1} = 1.785(5) \text{ GeV}$.
There are three values of light sea quark mass used in the simulations, $am^{\text{sea}}_{ud} = 0.005,  0.01$ and $0.02$
(we drop the heaviest of these in the chiral extrapolations). We again use only unquenched light valence quarks,
corresponding to pion masses of approximately $340$ and $430$ MeV. The simulated strange quark mass for the
ensemble is $am^{\text{sea}}_{s} = 0.04$, while the physical value has been determined to be $am^{\text{phys}}_{s} = 0.03224(18)$.
As with the fine ensemble, we interpolate between unitary ($am^{\text{val}}_{s} =am^{\text{sea}}_{s}=0.04$) and
partially-quenched ($am^{\text{val}}_{s}=0.035, 0.03$) data to the physical kaon mass.
The parameters for these ensembles are summarised in Table~\ref{tab:lattparam}.

\begin{table}[t]
\centering
\begin{tabular}{  c | c | c | c| c | c | c  }
\toprule
    Volume & $a^{-1}$ [GeV] & $am^{\text{sea}}_{ud} \, (= am^{\text{val}}_{ud})$ &
    $m_\pi$ [MeV] & $am^{\text{sea}}_{s}$ & $am^{\text{val}}_{s}$ &
    $am^{\text{phys}}_{s}$\\
    \hline
    $24^3 \times 64 \times 16$ & 1.785(5)  
    & 0.005, 0.01, 0.02 &340, 430, (560)& 0.04
    &0.04,0.035, 0.03 & 0.03224(18) \\ 
    $32^3 \times 64 \times 16$ & 2.383(9) 
    & 0.004, 0.006, 0.008 &300, 360, 410 &0.03
    & 0.03, 0.025 & 0.02477(18) \\ 
    \botrule
\end{tabular} 
\caption{\raggedright{Summary of our lattice ensembles. The heaviest mass of the coarse ensemble is not used 
    in the chiral extrapolations. For the coarse lattice, we use 155, 152 and 146 configurations for the 
    $am=0.005,0.01$ and $0.02$ ensembles respectively.
    For the fine lattice, we use $129, 186$ and $208$ configurations 
    for the $am=0.004, 0.006$ and $0.008$ ensembles respectively.
    The $560$ MeV pion-mass ensemble on the $24^4$ is deemed too heavy for use in the chiral extrapolations
    and is only shown in the plots for illustration purposes.}}
  \label{tab:lattparam}
\end{table}

\subsection{Correlation functions}

We have used Coulomb gauge fixed wall-source propagators, which allow for much greater statistical resolution
at similar cost to a point-source propagator inversion and should have better overlap of the ground state.
The fine ensemble results were generated as part of the calculation of $B_K$ in \cite{Aoki:2010pe}.
The coarse ensemble configurations were first Coulomb gauge fixed using the time-slice by time-slice
FASD algorithm of~\cite{Hudspith:2014oja} (to an accuracy of $\Theta<10^{-14}$).

Working in Euclidean space, we define the two-point functions,
\begin{equation}
c^{s_1 s_2}_{\mathcal{O}_1 \mathcal{O}_2} (t,t_i) 
= \sum _x \la \mathcal{O}_1^{s_1}(x,t) \mathcal{O}_2^{s_2}(0,t_i)^\dagger \ra \;,
\end{equation}
where $\mathcal{O}_i$ represents a bilinear operator. 
For the present analysis we only consider non-flavour singlet operators
with two different Dirac structures: either $\mathbb{P}$ the pseudo-scalar density,
or $\mathbb{A}_0$ the temporal component of the local axial current. 
The superscripts ($s_i$) denote the source type, either (L)ocal or (W)all source.
The two-point functions are fit to their asymptotic form
($T$ is the temporal extent of the lattice):
\begin{equation}
c^{s_1 s_2}_{\mathcal{O}_1 \mathcal{O}_2} (t,t_i) 
\underset{t_i \ll t \ll T } {\longrightarrow}
{a^3N^{s_1 s_2}_{\mathcal{O}_1 \mathcal{O}_2} \left( e^{-m_P(t-t_i)} \pm e^{-m_P (T-(t-t_i))}\right) }\;,
\end{equation}
Our conventions are such that 
\begin{equation}
a^3N^{s_1 s_2}_{\mathcal{O}_1 \mathcal{O}_2} =   
\frac{1}{2am_P} 
a^4 \la 0 | \mathcal{O}^{s_1}_1| \P \ra \la \P| \mathcal{O}^{s_2}_2|0\ra,
\end{equation}
and $\P = \bar\psi_1 \gamma_5 \psi_2$  (and therefore $\Pb = \bar\psi_2 \gamma_5 \psi_1$)
denotes a (flavour non-singlet) pseudo-scalar sate of mass $m_P$. 

The corresponding decay constant $f_P$ is defined (at finite lattice spacing and zero momentum) by 
\begin{equation}
\la 0 |\mathbb{A}^{\text R}_0 | \P \ra = m_P f_P,
\end{equation}
and can be extracted from an appropriate ratio of two-point functions.
The superscript $R$ denotes the fact that a finite (re)-normalisation
factor is required to connect the local axial current 
$\mathbb{A}_\mu^{\text{Local}}=\bar\psi_1 \gamma_\mu\gamma_5 \psi_2$
to the conserved current 
$\mathbb{A}^{\text R}_\mu$
\begin{equation}
\mathbb{A}^{\text R}_\mu = Z_V \mathbb{A}_\mu^{\text{Local}} \;.
\end{equation}
We prefer to renormalise the axial current with $Z_V$ rather than $Z_A$ for numerical reasons,
($Z_A$ and $Z_V$ should be identical if chiral symetry is exact, however $Z_V$ is numerically easier
to extract).
In a similar way, the bare matrix elements $\la \Pb | O_i | \P \ra $ are determined from three-point
correlation functions where the operator is inserted between two well separated wall sources,
\begin{equation}
c_k^{WLW}(t_f, t, t_i) =  \langle (\P^W(t_f))^\dagger \, \mathcal{O}_k^L(t) \, (\P^W(t_i))^\dagger \rangle  \;,
\end{equation}

\begin{figure}
\subfloat[${\cal R}_2^{\text{Lat}}(T/2, t, 0)$]
{
\includegraphics[type=pdf,ext=.pdf,read=.pdf,width=8.75cm]{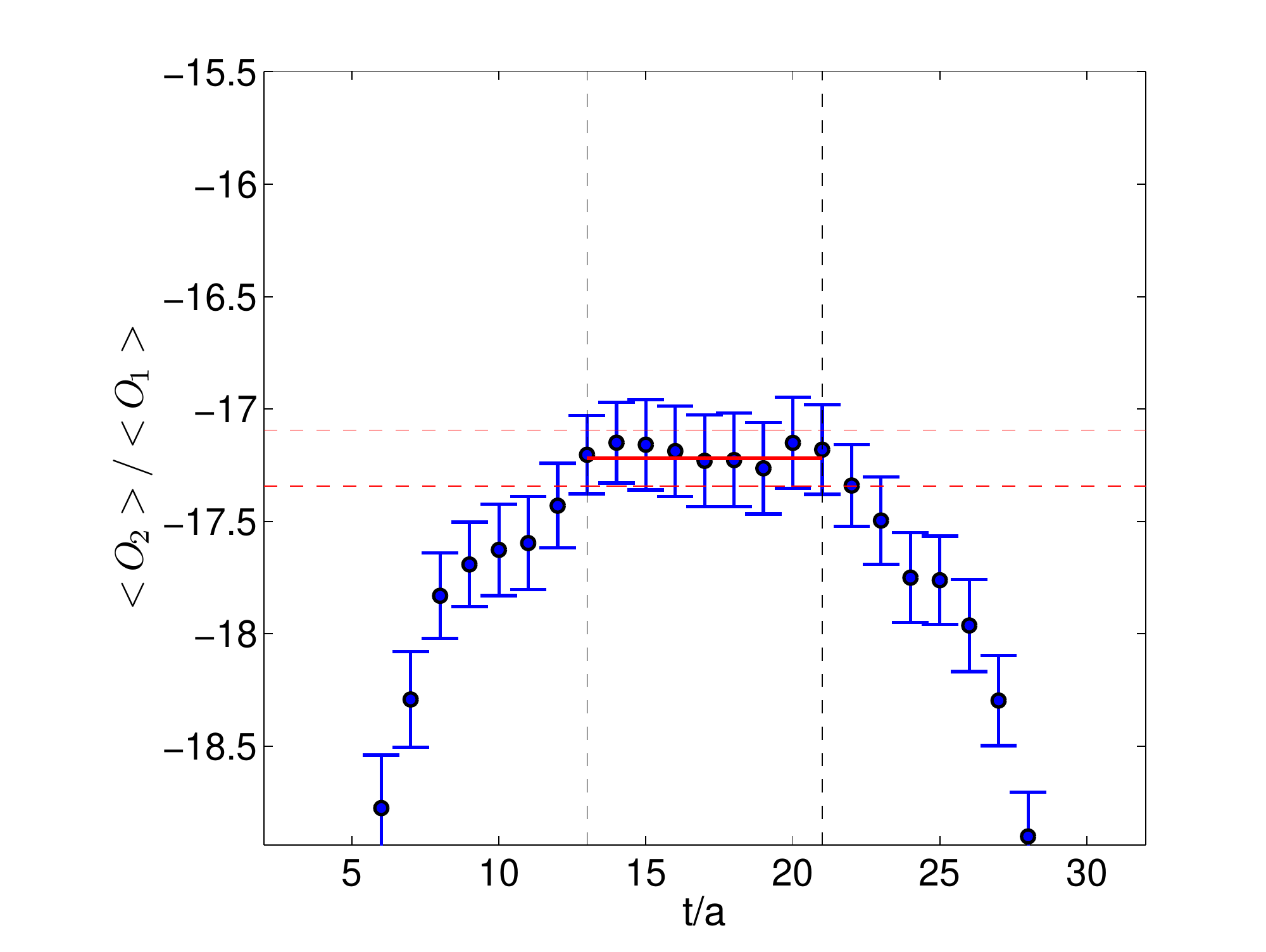}
}
\subfloat[${\cal R}_3^{\text{Lat}}(T/2, t, 0)$]
{
\includegraphics[type=pdf,ext=.pdf,read=.pdf,width=8.75cm]{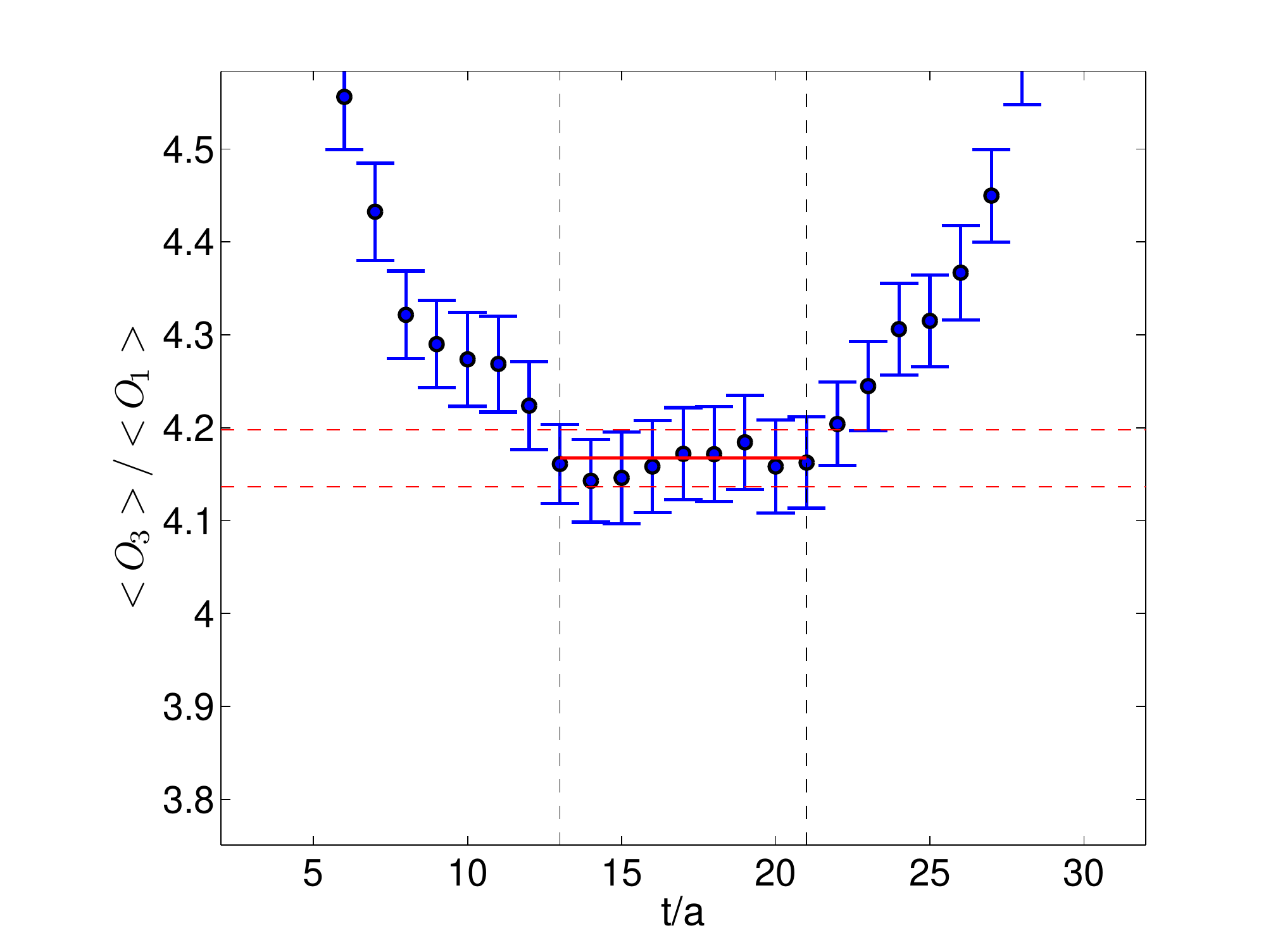}
}
\vspace{2pt}
\subfloat[${\cal R}_4^{\text{Lat}}(T/2, t, 0)$]
{
\includegraphics[type=pdf,ext=.pdf,read=.pdf,width=8.75cm]{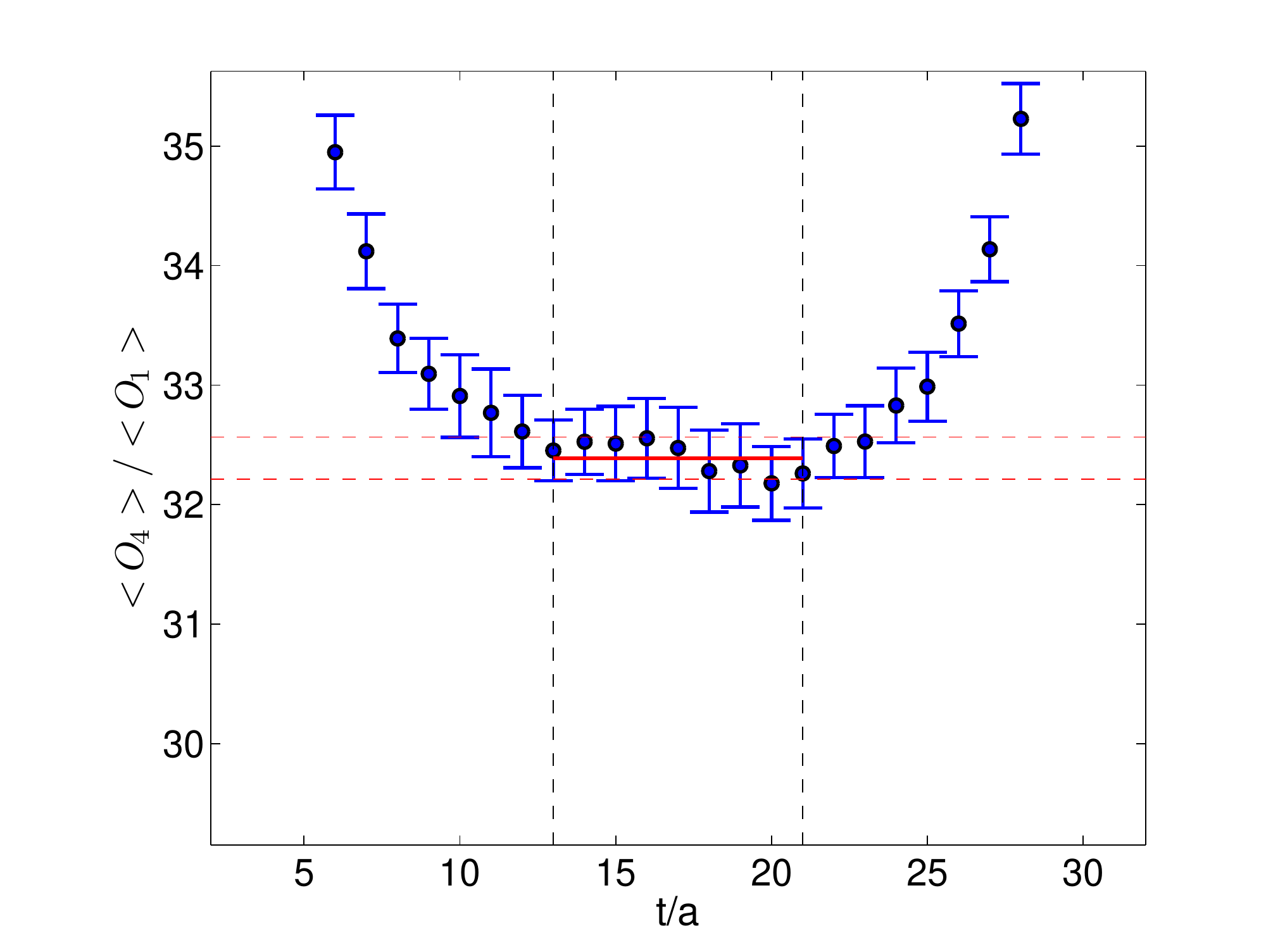}
}
\subfloat[${\cal R}_5^{\text{Lat}}(T/2, t, 0)$]
{
\includegraphics[type=pdf,ext=.pdf,read=.pdf,width=8.75cm]{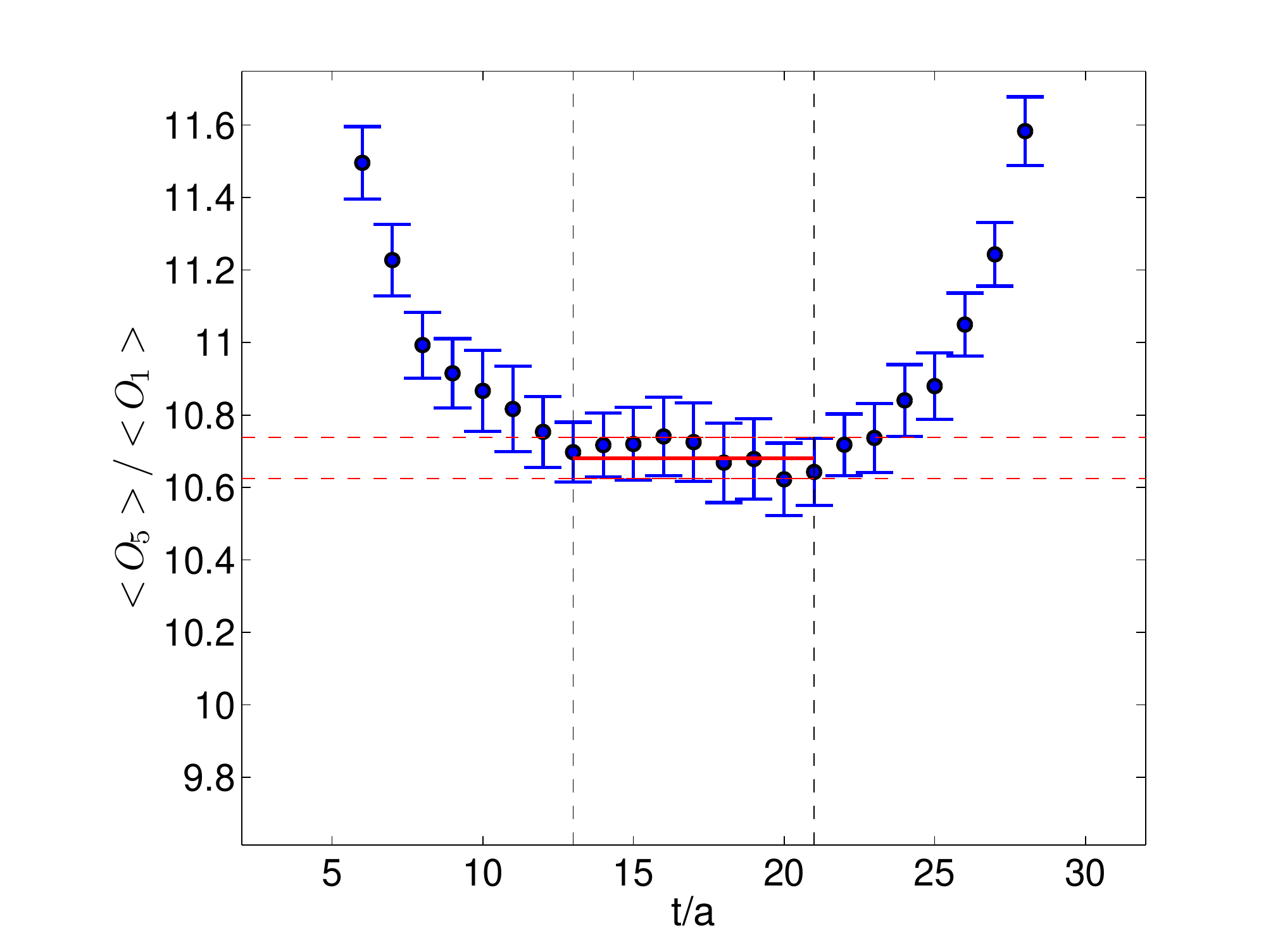}
}
\caption{\raggedright{Example of the plateau for ${\cal R}_i^{\text{Lat}}(T, t, 0)$ as a function of the operator insertion time $t/a$. We show our results for the lightest kaon mass on our coarse lattice.}}\label{fig:BSM_susy_rat_24IW}
\end{figure}

\begin{figure}
\subfloat[${\cal R}_2^{\text{Lat}}(T, t, 0)$]
{
\includegraphics[type=pdf,ext=.pdf,read=.pdf,width=8.75cm]{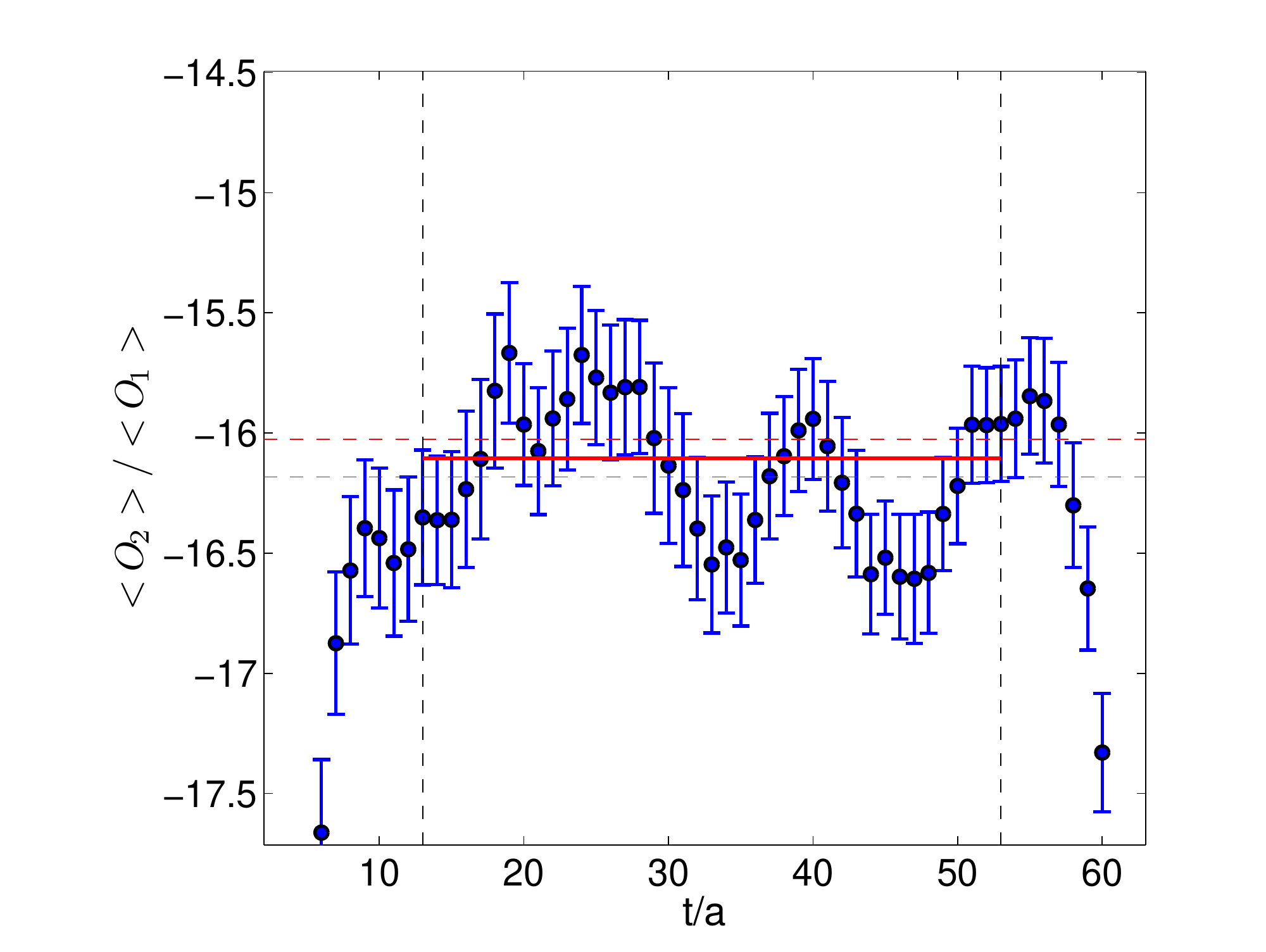}
}
\subfloat[${\cal R}_3^{\text{Lat}}(T, t, 0)$]
{
\includegraphics[type=pdf,ext=.pdf,read=.pdf,width=8.75cm]{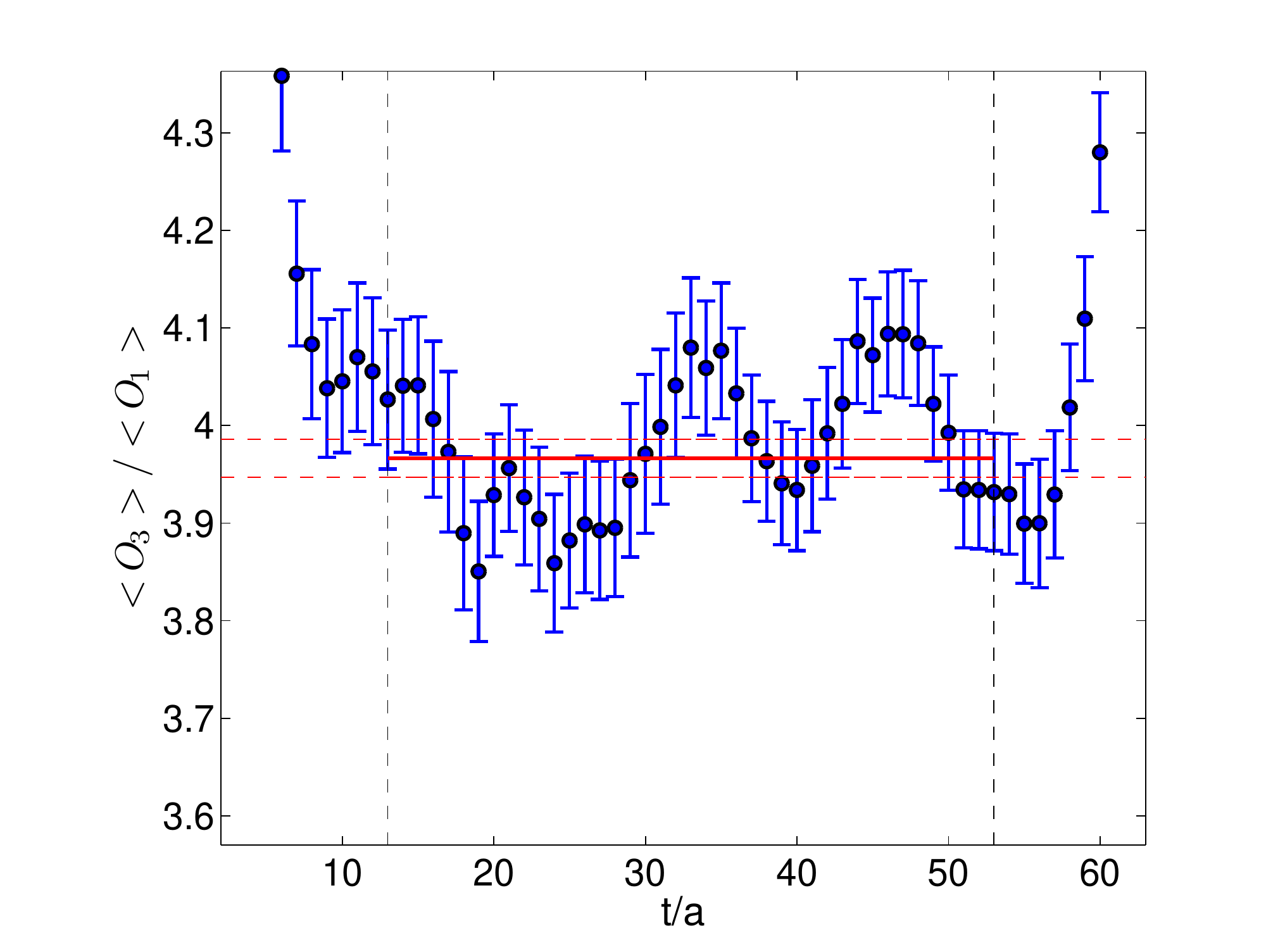}
}
\vspace{2pt}
\subfloat[${\cal R}_4^{\text{Lat}}(T, t, 0)$]
{
\includegraphics[type=pdf,ext=.pdf,read=.pdf,width=8.75cm]{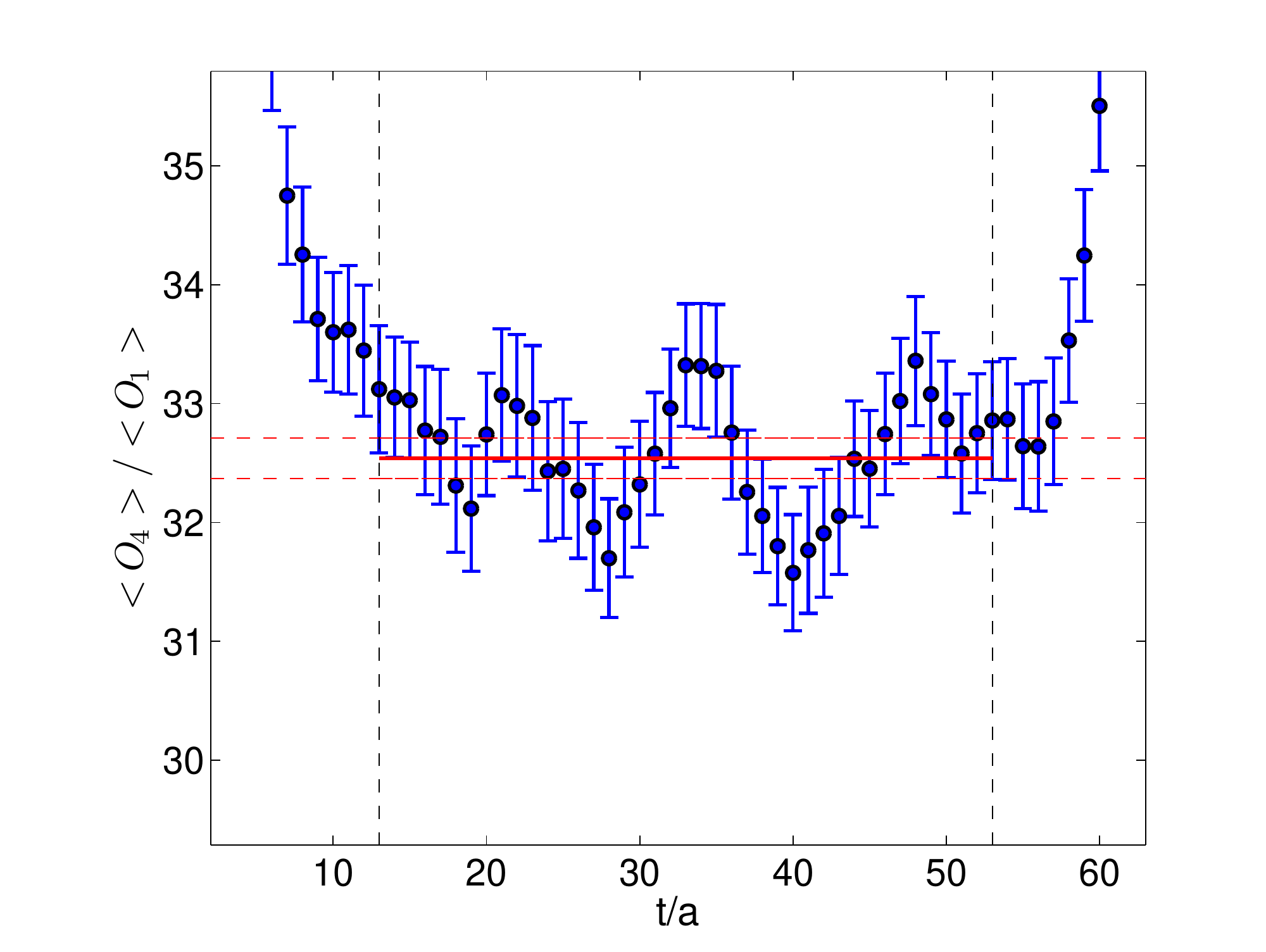}
}
\subfloat[${\cal R}_5^{\text{Lat}}(T, t, 0)$]
{
\includegraphics[type=pdf,ext=.pdf,read=.pdf,width=8.75cm]{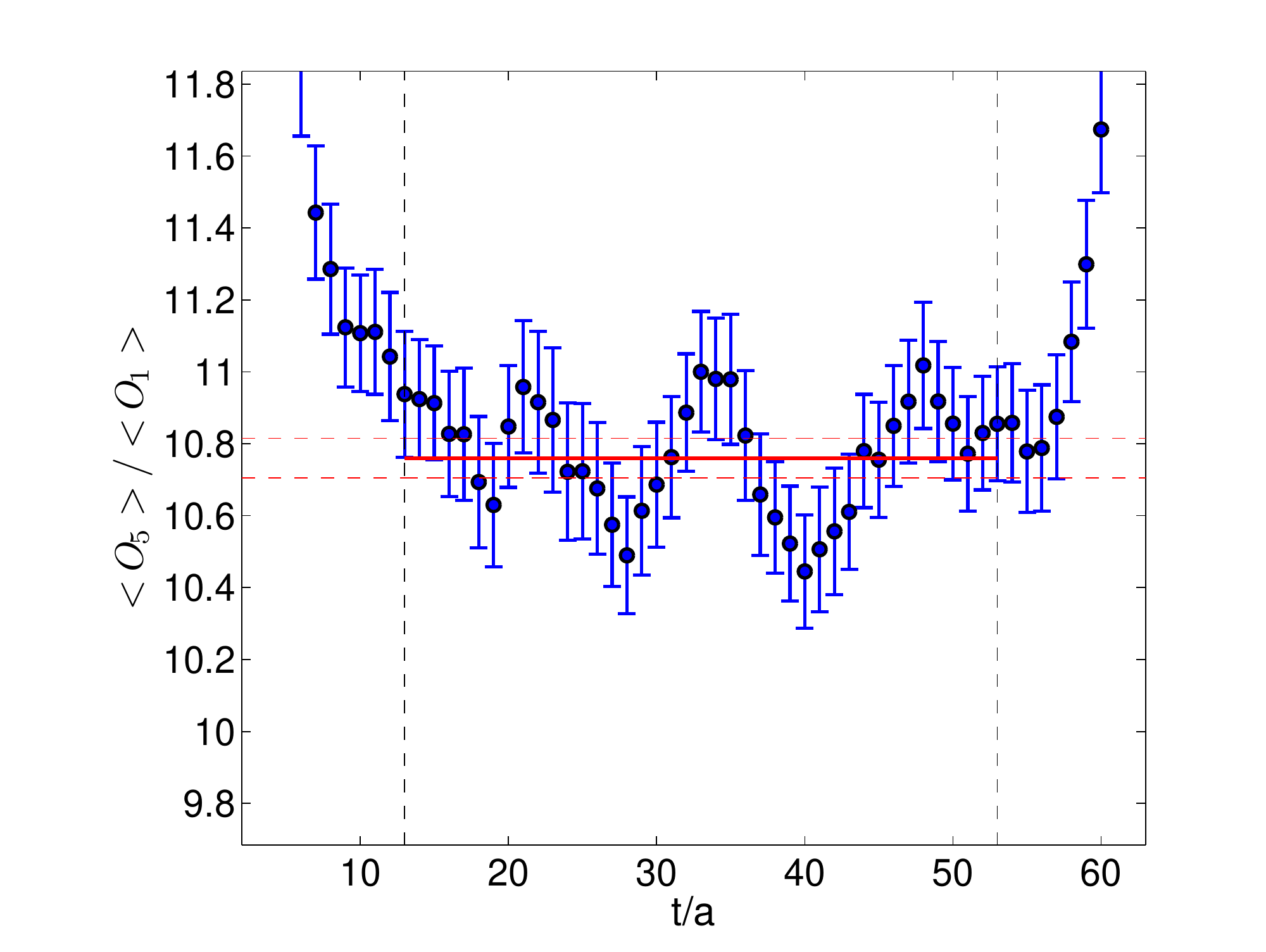}
}
\caption{Same as the previous figures but for our fine lattice.
}
\label{fig:BSM_susy_rat_32IW}
\end{figure}

In order to have a better handle on our systematics, 
we extract the quantities of interest in different ways
(which are in principle equivalent up to lattice artifacts). 
Our key results are obtained through the ratio of three-point functions ($k=2,\ldots,  5$)
which we fit to a constant in the asymptotic region: 
\begin{equation}\label{eqRbare}
{\cal R}_k^{\text{Lat}}(t_f, t, t_i) = \frac{c_k^{WLW}(t_f, t, t_i)}{ c_1^{WLW}(t_f, t, t_i)} \underset{t_i \ll t \ll t_f \ll T } {\longrightarrow} 
\frac 
{\la \Pb  | {O}^{\Delta S=2}_k | \P \ra}
{\la \Pb  | {O}^{\Delta S=2}_1 | \P \ra} =  R_k^{\text{Bare}} \;.
\end{equation}
We also define the ratios of three-point over two-point functions, which at large times allows us to obtain the bare BSM bag parameters:
\begin{equation}\label{eq:Bbare}
{\cal B}_{k}^{\text{Lat}}(t_f, t, t_i) =\frac{1}{N_k} \frac{c_k^{WLW}(t_f, t, t_i)} { c^{W L}_{\Pb P } (t_f,t)  c^{L W}_{P \Pb } (t,t_i)}
\underset{t_i \ll t \ll t_f \ll T } {\longrightarrow} 
\frac{1}{N_k}
\frac 
{\la \Pb  | O^{\Delta S=2}_k  | \P \ra}
{\la \Pb | \mathbb{P} | 0 \ra  \la 0 | \mathbb{P} | \P \ra  } = B_k^{\text{Bare}}\;, \qquad k>1\;.
\end{equation}
We show some examples of plateaux in Figs~\ref{fig:BSM_susy_rat_24IW} and~\ref{fig:BSM_susy_rat_32IW}.
The simulated time extent is $T/a=64$ on both lattices, but for the fine lattice we implement the 
Periodic $\pm$ Anti-periodic trick which is designed to reduce the round the world artifacts.
Effectively this trick doubles the number of accessible points~\cite{Sasaki:2003jh}
(see also the discussion in~\cite{Arthur:2012opa}).
Although the signal obtained from the coarse lattice time slice per time slice is different from the one of the fine lattice,
the precision obtained on the ratio $R_k^{\text{Lat}}$ (by a correlated fit) is of the same order.

\subsection{Non-Perturbative Renormalisation (NPR)}

Once the bare matrix elements have been obtained, they need to be renormalised
in order to have a well-defined continuum limit. 
We opt for the framework which is now standard within the RBC-UKQCD
collaboration: the non-perturbative Rome-Southampton renormalisation
method~\cite{Martinelli:1994ty}, with non-exceptional kinematics (we use
the symmetric RI-SMOM schemes)~\cite{Sturm:2009kb}, momentum sources~\cite{Gockeler:1998ye}
and twisted boundary conditions~\cite{Bedaque:2004kc,deDivitiis:2004kq,Sachrajda:2004mi}. 
Similarly to what was done for $B_K$ and $K\to\pi\pi$,  
we define two schemes: the RI-SMOM-$\gmugmu$ and RI-SMOM-$\qq$ schemes
\footnote{The running of the relevant operators in the RI-SMOM-$\gmugmu$ scheme
  has been discussed in~\cite{Arthur:2011cn}.}
(we drop the ``RI'' in the following).
We refer to these schemes as ``intermediate schemes''.
Our final results are the ones given in these SMOM schemes; however the matrix elements of
interest are conventionally given in a $\msbar$ scheme at a reference scale of 2 or 3 GeV.
Although the computation of the bare matrix elements and of the renormalisation factors
is done non-perturbatively, this matching step involves (continuum) perturbation theory. 
$\msbar$ results obtained via different intermediate schemes should be consistent,
up to higher-order PT matching corrections (and lattice artifacts if the resutls
are given at finite lattice spacing).
The use of multiple intermediate schemes allows one to gain a better handle on these uncertainties
\footnote{We thank Christoph Lehner for computing the matching factor of the $(6,\bar 6)$ operators,
  the details will be given in~\cite{KKNPR}.}.
  .

We also implement the original RI-MOM scheme~\cite{Martinelli:1994ty},
however we find that the results are not consistent with the SMOM ones.
As shown in detail in the companion paper~\cite{KKNPR}, we find that the RI-MOM $Z$-matrices
exhibit large violations of the block diagonal structure expected from the
chiral-flavour properties of the four-quark operators.
This seems to be due to important infrared artefact which go as
inverse powers of the quark mass. These pole ``contamination'' require a hard
subtraction, and render results significantly more unreliable than
their SMOM counterparts.  
This is indicated in Table~\ref{results:tab:collabcomparison} by the discrepancies 
of the RI-MOM scheme results with the SMOM ones and with the ones obtained by
the SWME collaboration, whose (1-loop) perturbative matching is free from IR contamination
(see~\cite{Hudspith:2015wev,Jang:2015sla}).
We do not advocate the use of these RI-MOM results, indeed we show that
this choice of intermediate scheme is probably
the cause of the disagreement observed between different collaborations.

Another advantage of the SMOM schemes is that the perturbative matching factors
connecting them to the $\msbar$ scheme are much closer to the identity matrix.
This suggests a better behaved perturbative series with less 
matching uncertainty than for the MOM case, which would demand a higher
matching scale.  Referring again to Table~\ref{results:tab:collabcomparison},
the close compatibility of the SMOM-$(\gamma_\mu, \gamma_\mu)$
and $(\slashed{q}, \slashed{q})$ results provides strong evidence that
the matching uncertainty for the SMOM schemes is negligible within our error budget.

%% file: Sections/s4_results.tex
\section{Results at the physical point and discussions}\label{sec:results}

We report here our main results for the ratios $R_i$, the bag parameters $B_i$
and the combinations $G_{ij}$. We consider the main results of this work to be the ratios $R_i$,
because at the physical point they directly provide the ratio of the BSM matrix element to the SM one.
They do not depend on the quark masses, nor on our ability to renormalise the pseudo-scalar
density as the bag parameters and some of the combinations $G_{ij}$ are. 
The results for the bag parameters extracted from the combinations $G_{ij}$ (method C)
are reported in Appendix~\ref{app:methodC}. 
We also compute the matrix elements $\la \Kb |O_i |\K\ra$ using the different strategies.
The quality of the fits can be judged from the $\chi^2$ reported in Appendix~\ref{appendix:app_table},
Table~\ref{app_table:tab:chisq}

\subsection{The ratios \texorpdfstring{$R_i$}{Ri}}

\begin{figure}[t]
\begin{center}
\begin{tabular}{cc}
\includegraphics[type=pdf,ext=.pdf,read=.pdf,width=7cm]{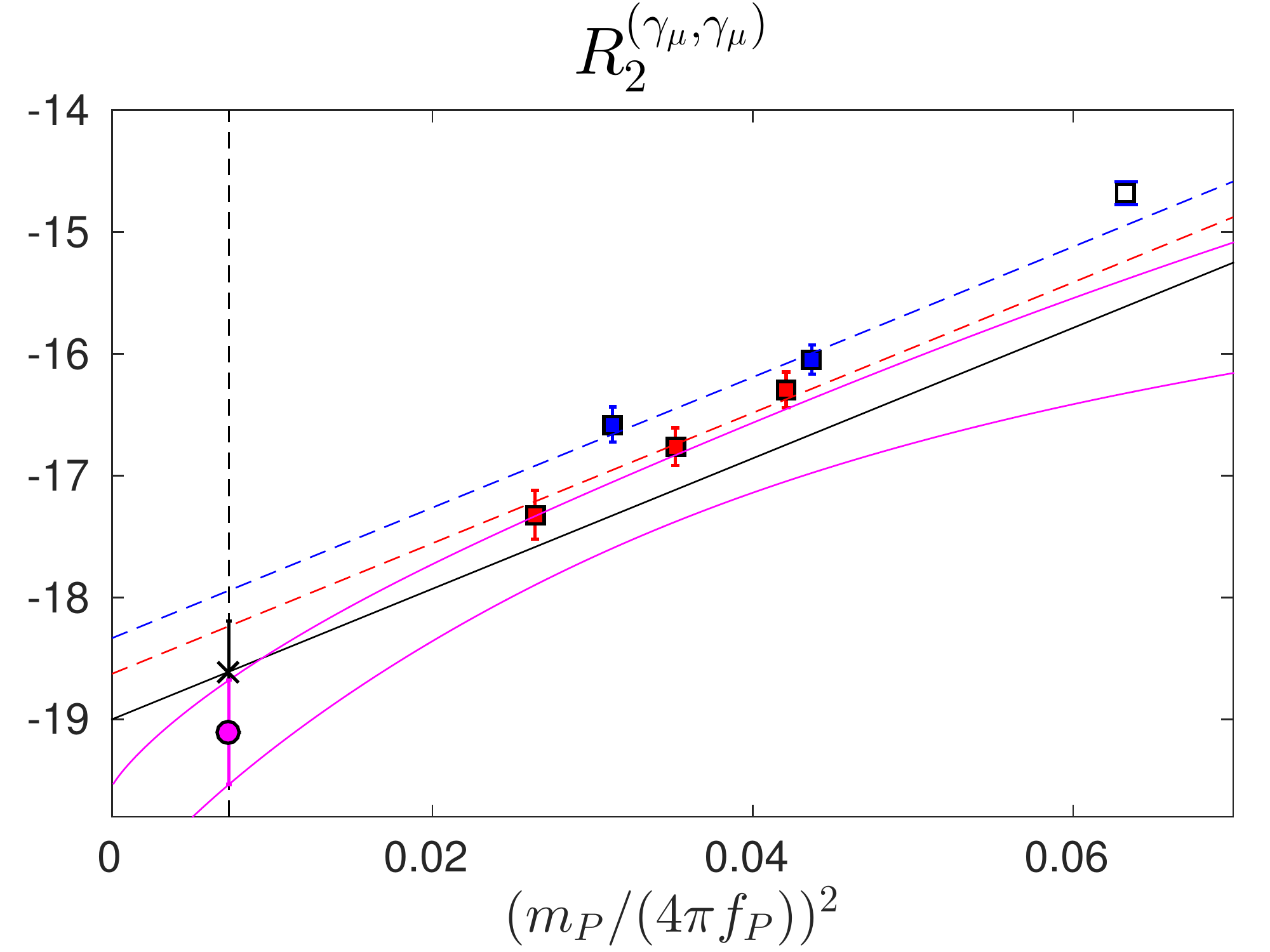} &
\includegraphics[type=pdf,ext=.pdf,read=.pdf,width=7cm]{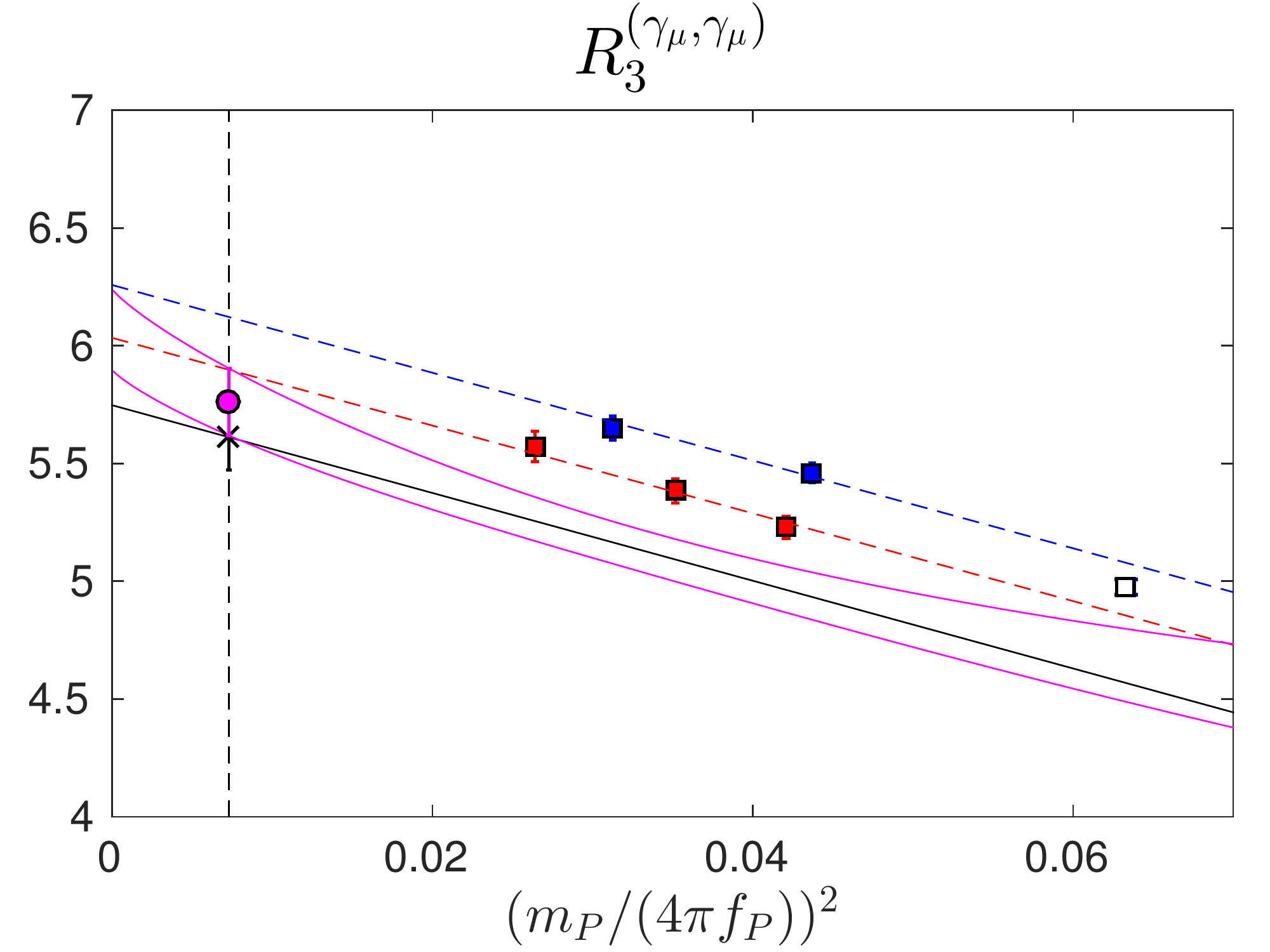}
\vspace{0.2cm}\\
\includegraphics[type=pdf,ext=.pdf,read=.pdf,width=7cm]{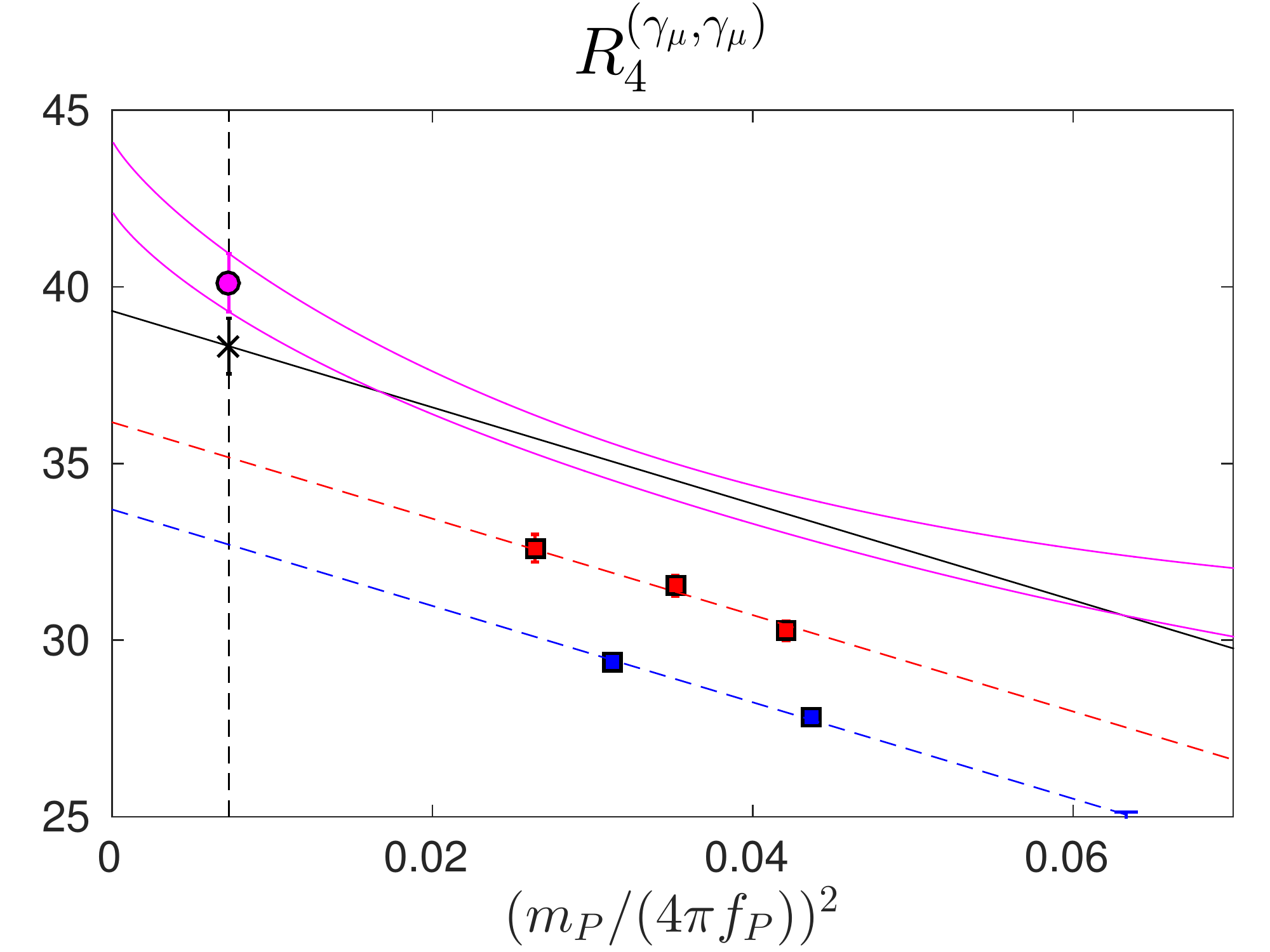} & 
\includegraphics[type=pdf,ext=.pdf,read=.pdf,width=7cm]{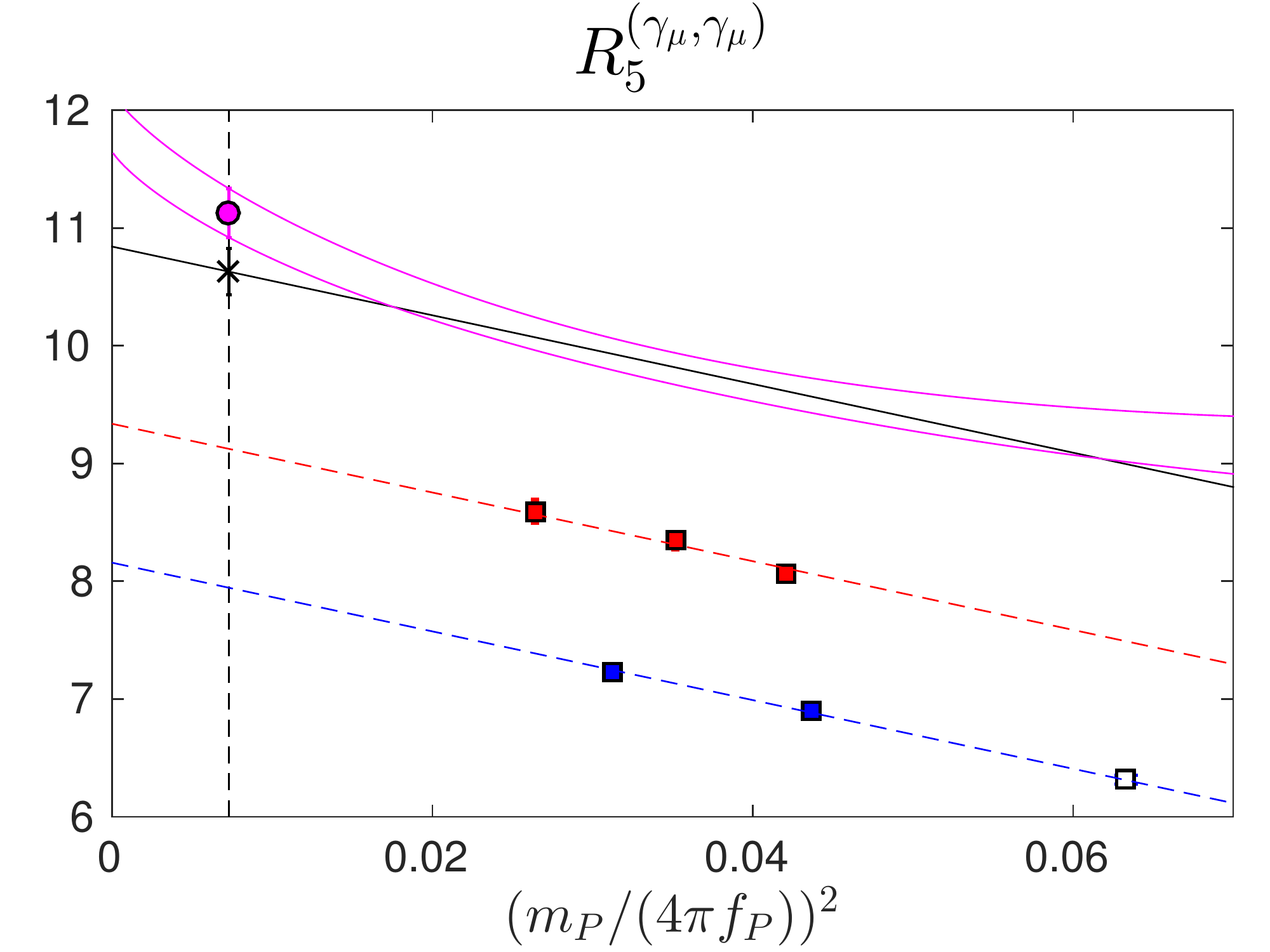}  
\end{tabular} 
\caption{\raggedright{Continuum/chiral extrapolation {\bf Method A and B} of the ratio $R_i$
    in the SUSY basis and renormalised in the $(\gamma_\mu,\gamma_\mu)$-scheme.
The conventions here and in the following plots are: 
red squares are the fine lattice data points, the blue squares the coarse ones. 
Open symbols represent a point which was omitted in the fit procedure.
All the points have been interpolated/extrapolated to the corresponding physical strange
quark mass. The magenta curve is the chiral fit and the solid point is its chiral-continuum value.
The black line is the linear fit at $a^2=0$.
We keep the relative scale constant for the vertical axis (around fifty percent of the extrapolated
value).
}}\label{results:fig:globalfitRgg_susybasis}
\end{center}
\end{figure}

In Fig.~\ref{results:fig:globalfitRgg_susybasis}, we show the results using the combined continuum-chiral
fits discussed in section~\ref{sec:methodology}, both Method A and Method B
in the non-exceptional
SMOM-$\gmugmu$ scheme. We show all of our results in this scheme, however we note that the SMOM-$\qq$ scheme
gives very similar results. The RI-MOM results have already been presented in~\cite{Boyle:2012qb,Lytle:2013oqa},
they are just reported for comparison with previous work.
In the figures, the dashed line represents the chiral extrapolation performed linearly in $m_P^2$
(the pion mass squared) at fixed lattice spacing and the $a^2\rightarrow 0$ extrapolation is shown
as a solid black line. The magenta line represents the Method A fit, in which we take
the leading chiral logarithms into account.
Our physical results obtained by this chiral-continuum extrapolation
is the filled circle.

We note that the fit quality is very good with chi-square per degree-of-freedom ($\chi^2/d.o.f)$
of order one or less as shown in Table~\ref{app_table:tab:chisq} of Appendix~\ref{appendix:app_table}.
We also note that although the ratios $R_i$ have the largest coefficients for the chiral logarithms,
the effect of these terms is mild and the difference between the linear fit in $m_P^2$ and the chiral
one is at most of the order of a few per cent. 
The fits for Method A and Method B show similar quality as indicated by by their $\chi^2/d.o.f$,
hence we do not see significant curvature.
We take the fact that the fit quality for Method A is good as an indication that NLO Chiral Perturbation Theory
is a decent description of the mass dependence of our data,
this is our choice for our central values.
We use the difference of the results obtained from Methods A and B to estimate the effects of the chiral logarithms.
As shown in the plots, the two methods give very close results.
This might be because the ensembles we have used are at relatively heavy pion mass.
However we give another argument below based on Method C, to justify that
the chiral extrapolations to the physical quark masses are well under control,
and that the chiral extrapolation effect is one of the most benign 
compared to the other systematics in this calculation.

For some of these quantities we see significant cut-off effects, especially $R_5$ which requires an extrapolation
of the order of $15\%$ from the fine ensemble's data to reach $a^2=0$. We observe that this is largely due to
the $3$-GeV renormalisation factors (for this quantity our estimate for the discretisation
error is almost a factor two smaller at $2$ GeV).
From Fig.~\ref{results:fig:globalfitRgg_susybasis} it is interesting
to note that as we approach the continuum limit $R_2$, $R_4$ and $R_5$ of our BSM matrix elements are larger
(in magnitude) than we previously determined just from our fine ensemble's data in~\cite{Boyle:2012qb}.
As other previous studies have noted, the BSM matrix elements are an order of magnitude larger than the SM one.

\subsection{The bag parameters \texorpdfstring{$B_i$}{Bi}}

\begin{figure}[t]
  \begin{tabular}{cc}
    \includegraphics[type=pdf,ext=.pdf,read=.pdf,width=8.75cm]{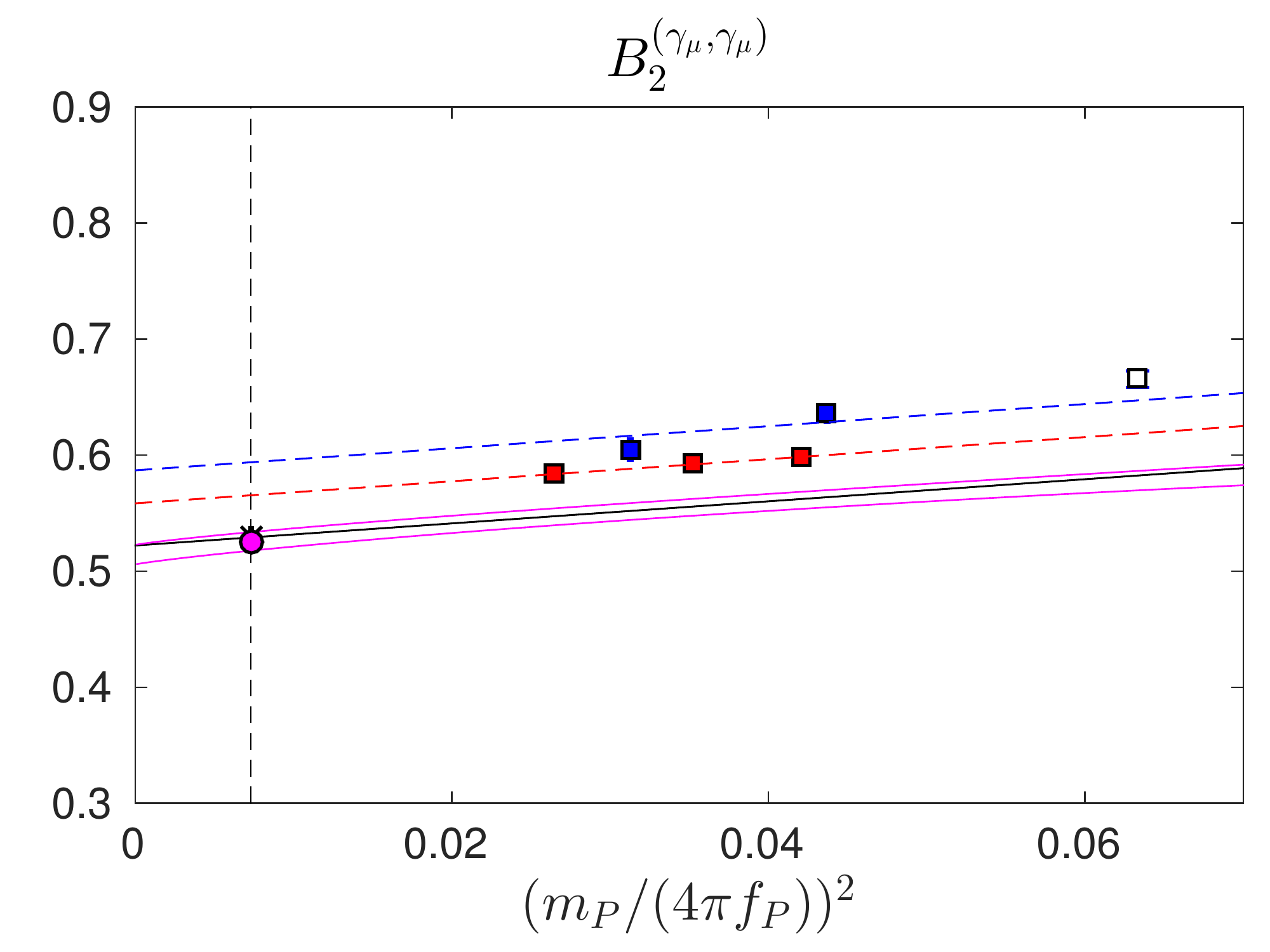} &
    \includegraphics[type=pdf,ext=.pdf,read=.pdf,width=8.75cm]{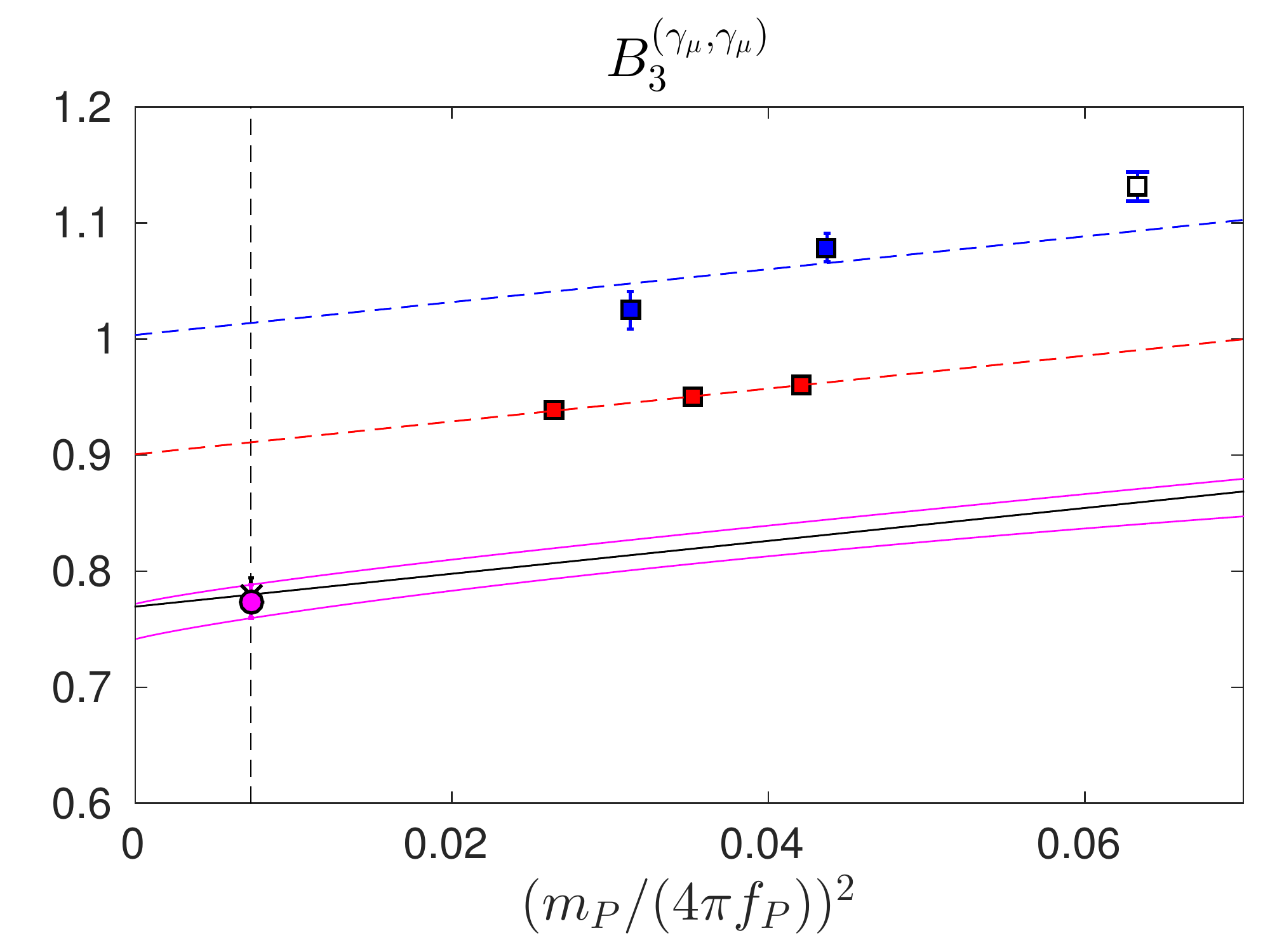}
    \vspace{0.2cm}\\
    \includegraphics[type=pdf,ext=.pdf,read=.pdf,width=8.75cm]{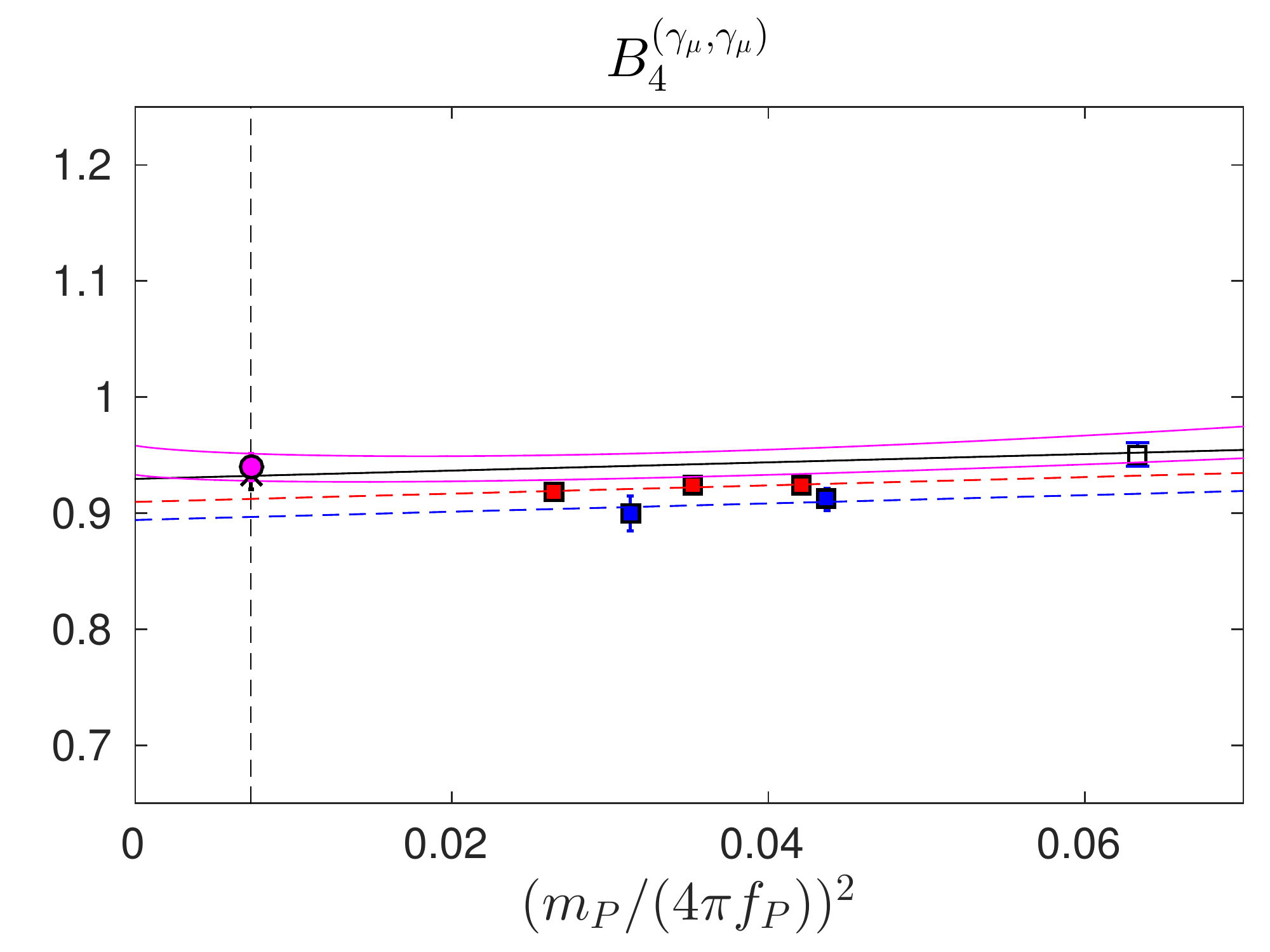} &
    \includegraphics[type=pdf,ext=.pdf,read=.pdf,width=8.75cm]{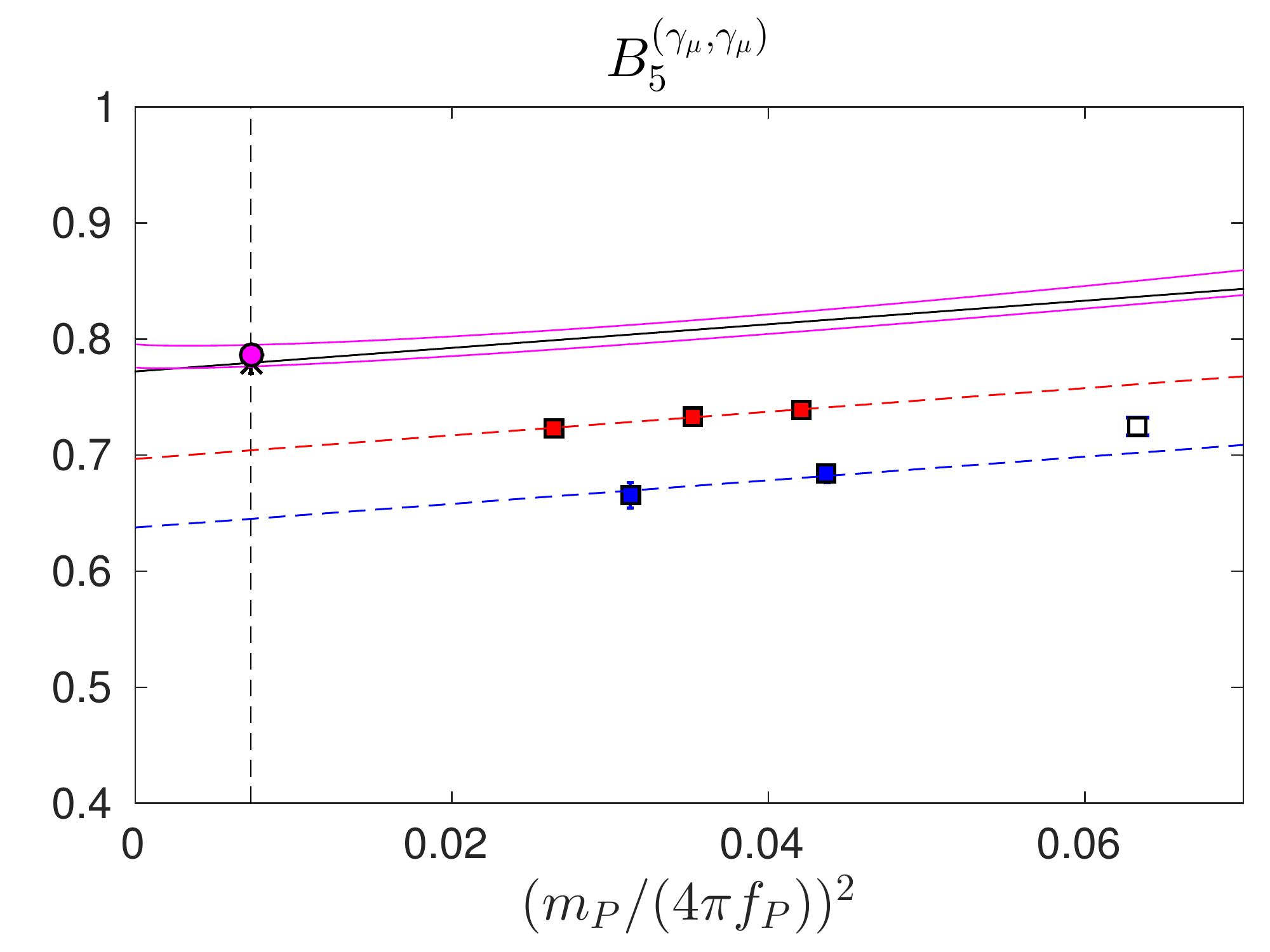}
  \end{tabular}
  \caption{\raggedright{Continuum/chiral extrapolation of the bag parameters renormalised at $\mu=3\text{ GeV}$.
      Results are renormalised in the $(\gamma_\mu,\gamma_\mu)$-scheme.
      The absolute scale on the vertical axis is kept constant.
  }}\label{results:fig:globalfitBgg_susybasis}
\end{figure}

The combined chiral-continuum plots for the $B_i$ are shown in Fig.~\ref{results:fig:globalfitBgg_susybasis}
using the same conventions as in the previous section. We show our results again for the $\gmugmu$ scheme.
We observe that the fit quality is a bit worse for the $B_i$ compared to the $R_i$ with $\chi^2/d.o.f$
ranging between $0.4$ to $1.9$ (Table~\ref{app_table:tab:chisq}).
We also note that while the effect of chiral logarithms is almost invisible, the discretisation effects
are larger than anticipated for two of these quantities $B_3$ and $B_5$:
we observe a deviation of more than $10\%$ between the fine ensemble and
the $a^2\rightarrow0$ extrapolation.

\subsection{The combinations $G_{ij}$}

\begin{figure}[t]
  \begin{tabular}{cc}
    \includegraphics[type=pdf,ext=.pdf,read=.pdf,width=8.5cm]{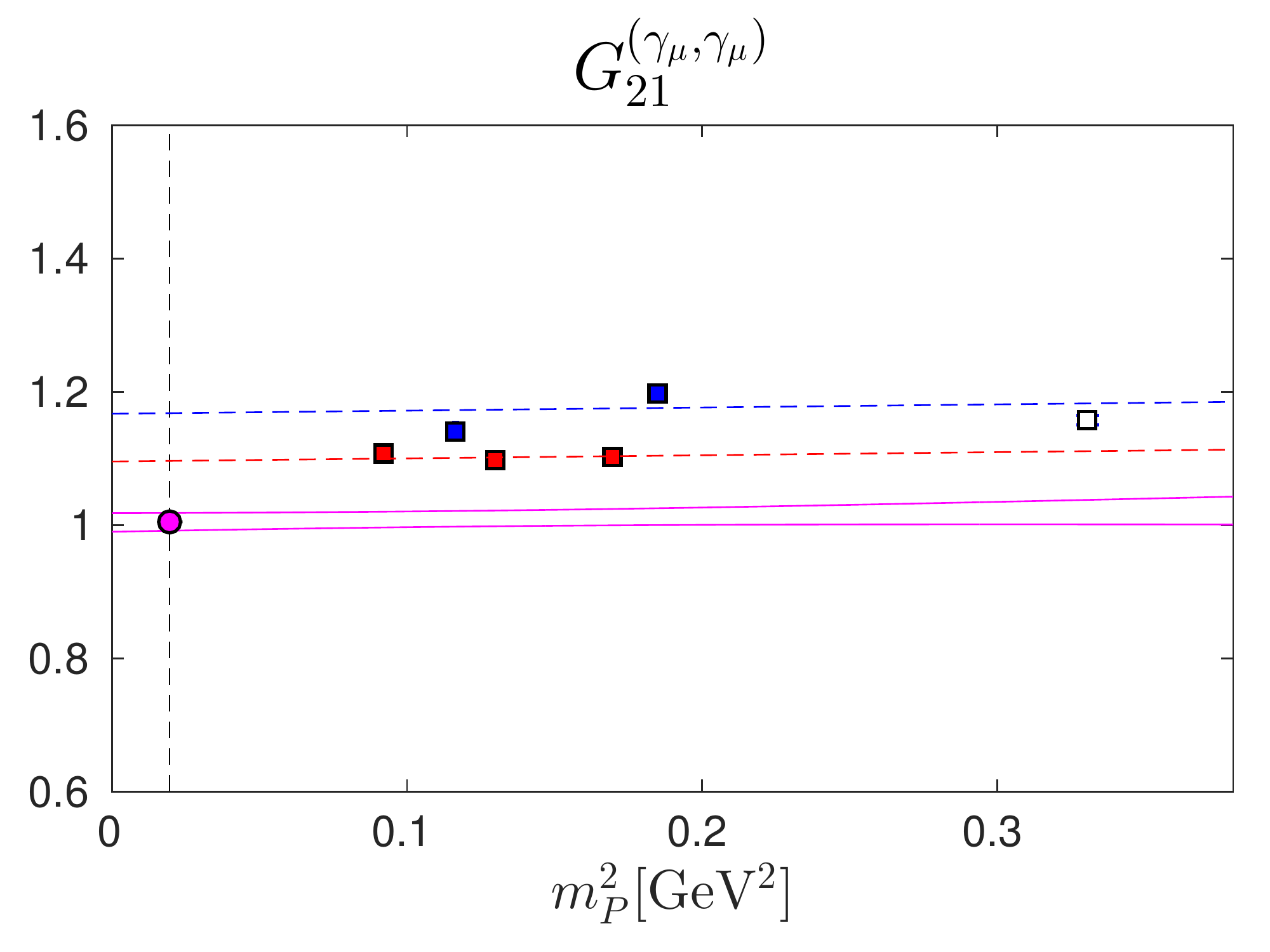} & 
    \includegraphics[type=pdf,ext=.pdf,read=.pdf,width=8.5cm]{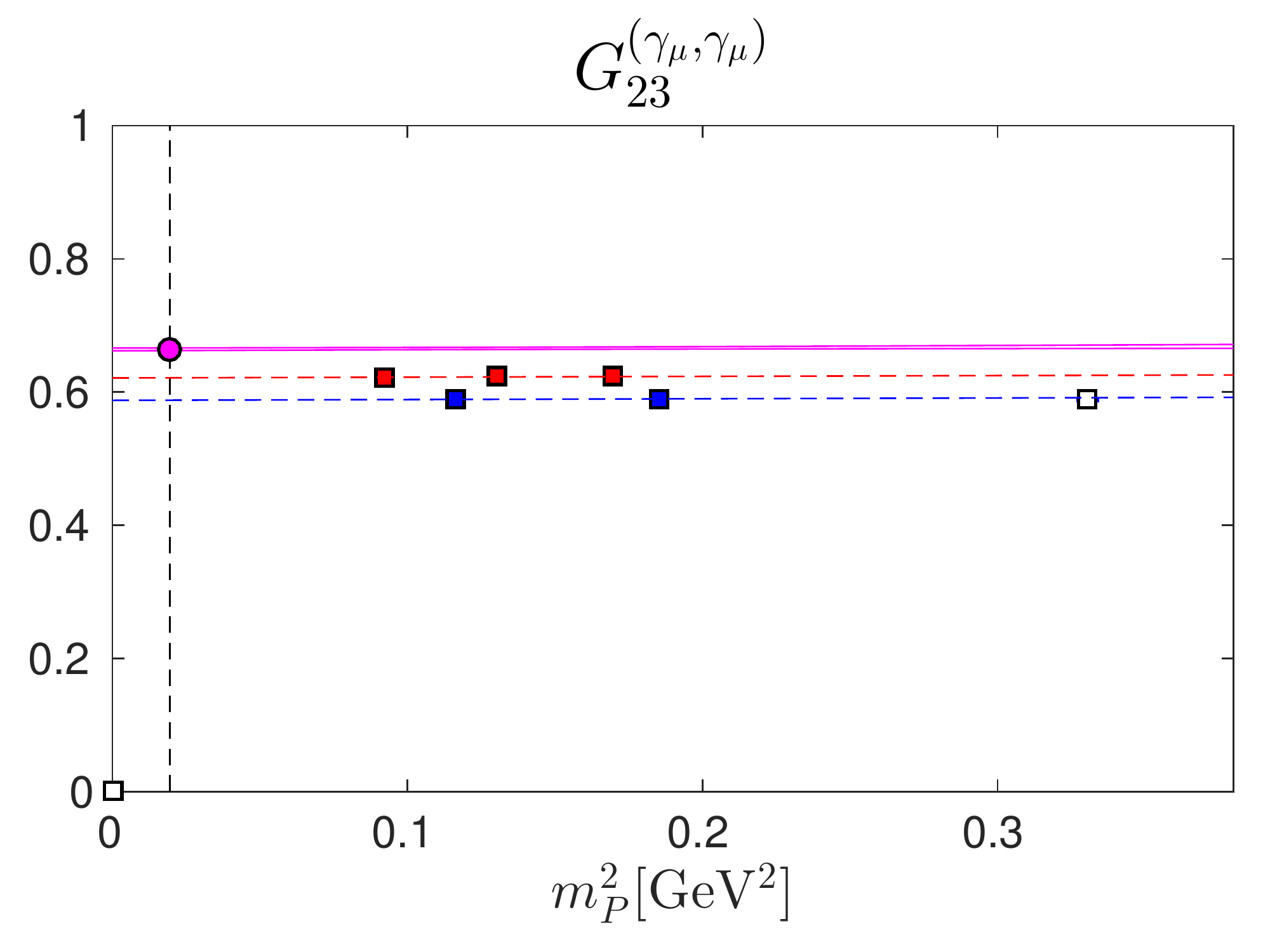}
    \vspace{0.2cm}\\
    \includegraphics[type=pdf,ext=.pdf,read=.pdf,width=8.5cm]{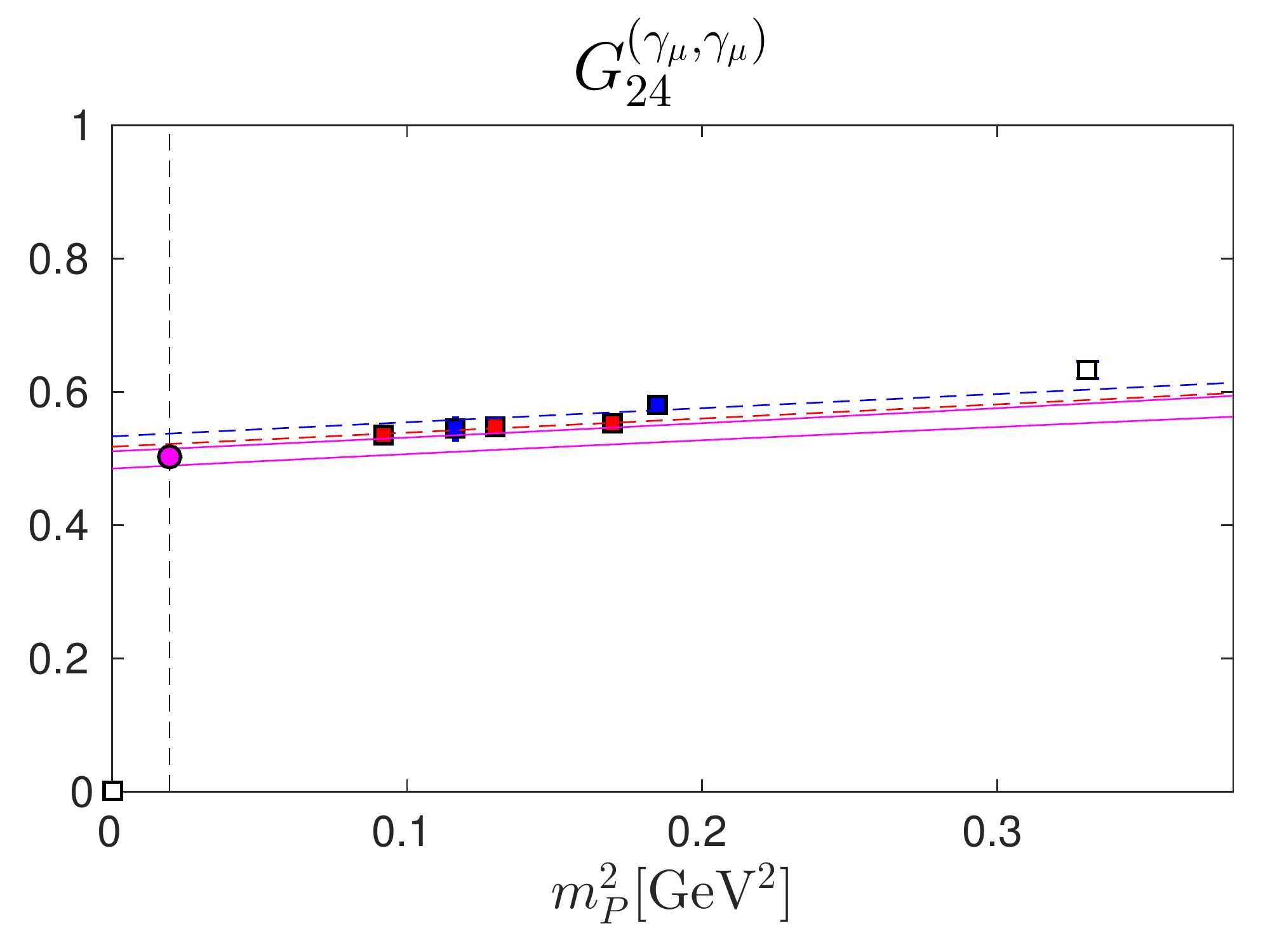} &
    \includegraphics[type=pdf,ext=.pdf,read=.pdf,width=8.5cm]{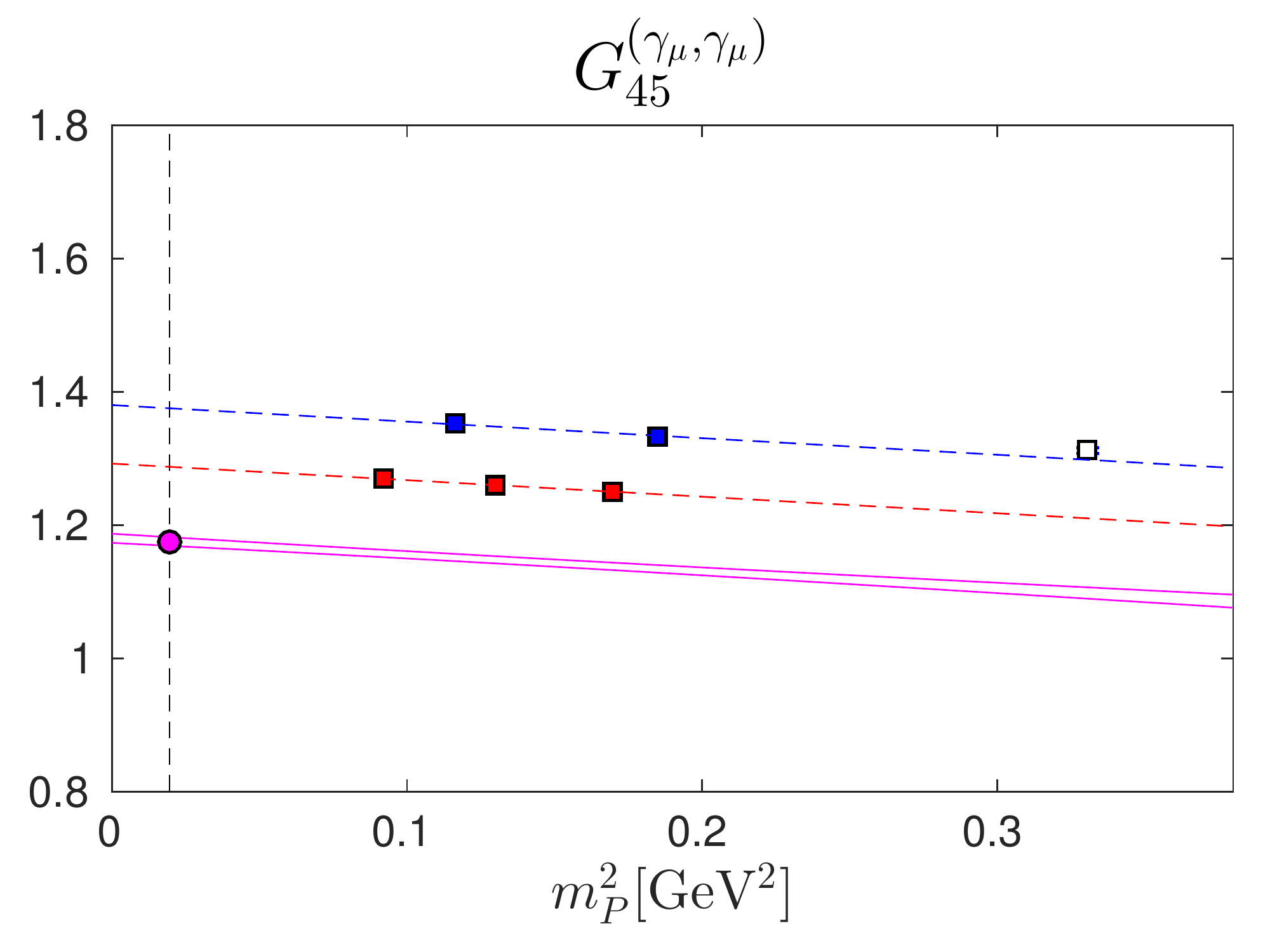} 
  \end{tabular}
  \caption{\raggedright{Continuum/Chiral extrapolation of the combinations $G_{ij}$
      renormalised at $\mu=3\,\text{GeV}$. in the  $(\gamma_\mu,\gamma_\mu)$ scheme.
  }}
  \label{results:fig:G_MOM_gg_susy_basis}
\end{figure}

\begin{figure}[t]
\begin{center}
  \includegraphics[type=pdf,ext=.pdf,read=.pdf,width=8.5cm]{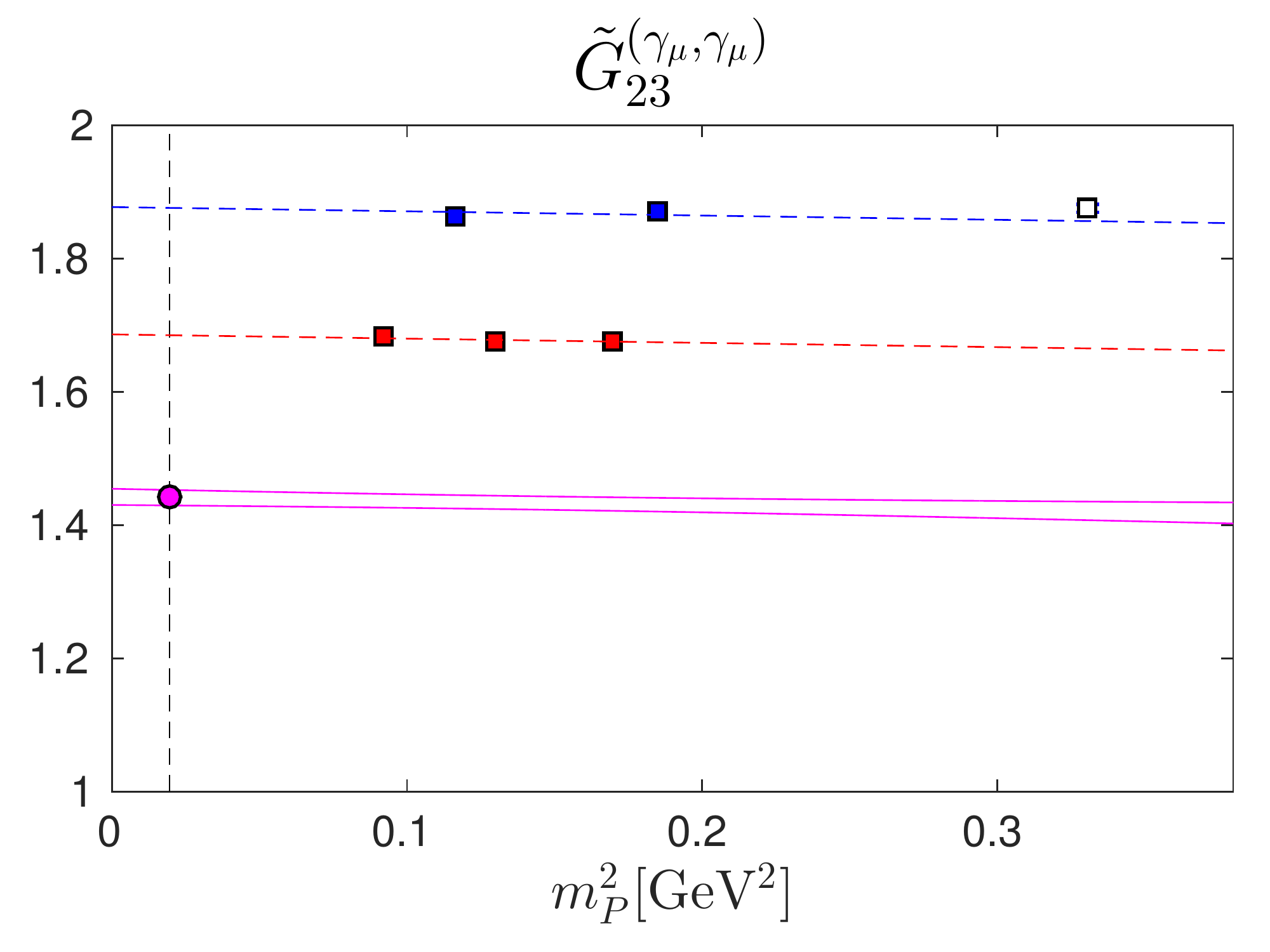}
\end{center}
\caption{\raggedright{Alternative definition for one for the combinations,
    $\tilde G_{23} = B_2/B_3$ where $B_2$ and $B_3$ are
    computed in the BMU basis. 
    The discretisation effects are enhanced in the ratio, illustrating the fact the size of the cutoff
    effects depends on the choice of basis, see the discussion in the text. }}
      \label{fig:tildeG23}
\end{figure}

Fig.~\ref{results:fig:G_MOM_gg_susy_basis} shows the results obtained
in the $\gmugmu$ scheme, using the same conventions as in the previous figures.
Firstly, we see that there is no noticeable chiral curvature which is unsurprising as these quantities
were designed for this purpose.
We observe that the combinations $G_{ij}$ can be numerically very different.
For $G_{23}$ and $G_{45}$, we find a rather good $\chi^2/d.o.f.$,
a linear behaviour in $m_P^2$ (with a very small slope), however the lattice
artefacts for $G_{45}$ are clearly visible (with again a difference of order $10\%$
between the fine lattice and the extrapolated value).
We have also computed an alternative combination, $\tilde G_{23}$,
in order to compare our results with the SWME collaboration.
Similarly to $G_{23}$, it is defined as the ratio of the two bag parameters
$B_2$ and $B_3$, but computed in a different basis,
the one introduced by Buras, Misiak, and Urban in \cite{Buras:2000if}.
We call this basis the ``BMU basis'' in the following.
This is also the choice of the SWME collaboration,
therefore what we call $\tilde G_{23}$ here is called
$G_{23}$ in~\cite{Bae:2013tca} and~\cite{Jang:2015sla}. 
Only $B_3$ differs between the two sets of operators.
Within our convention
the operator $O_3$ is defined as the colour partner of $O_2$,
whereas in the BMU basis, it is purely a ``tensor-tensor'' operator.
Although in principle the two definitions are equivalent  (thanks to Fierz theorem),
the  cutoff effects can be very different.
Indeed we observe that the sign of the $a^2$ coefficient of two-colour
partner operators are identical : positive for $B_2, B_3$
and negative for $B_4 ,B_5$. This results in some cancellation
of these artefact in the ratio $G_{23}$
(whereas in $G_{45}$, the cutoff effects are completely dominated by
$B_5$ and taking the ratio does improve very much from that point of view).
We now turn to $\tilde G_{23}$, which reads in terms
of bag parameters 
\be
\tilde G_{23} = G_{23}^{BMU} = \frac{B_2^{BMU}}{B_3^{BMU}}
=\frac{3 B_2 }{ 5 B_2 - 2 B_3}
\ee
where $B_2$ and $B_3$ refer to the SUSY basis.
In this peculiar combination, the cutoff-effects do not cancel,
but on the contrary they add up, as illustrated in Fig.~\ref{fig:tildeG23}.
We note that the authors of~\cite{Jang:2015sla} also found
this combination difficult to fit.

The fit of the product $G_{24}$ is very reasonable with a $\chi^2 / d.o.f.$ of
around $1.4$, the pion mass dependence is very mild,
and there is clearly an important cancellation of the lattice artefacts in the product
as the $a^2$ coefficients have a different sign. However, we believe that this cancellation
is purely accidental.

We find that the ratio $G_{21}$ is much more difficult to fit, with a $\chi^2/d.o.f$
of order seven. The difficulty comes mainly from the coarse ensemble, where the results seem
to fluctuate around a constant value of the mass. This effect could be due to some unfortunate
statistical fluctuation or lattice artefact and need to investigated further in the future.
This is rather unfortunate because the quantity $G_{21}$ is needed to reconstruct the
bag parameters from the $G_{ij}$. Therefore in Appendix~\ref{app:methodC}, we propose alternative
combinations of bag parameters, which improve the determination of $B_4$ and $B_5$
(with respect to the combinations $G_{ij}$ used in this section).
In the same Appendix~\ref{app:methodC}, we compare the results for the bag parameters
extrapolated directly (Methods A and B), to the ones extracted from
the combinations $G_{ij}$. We find that the combinations $G$ do not provide more precise
results (within our sytematic error budget)
except for one quantity, $B_3$ (if $G_{23}$ is computed in the SUSY basis).

Finally, we point out that one could also first perform a continuum extrapolation of the
bag parameters in the range of simulated pion mass,
then compute the combinations $G_{ij}$ and finally perform the chiral extrapolation.
We leave this for future investigations.

\subsection{Error budget}

Our central results are the BSM quantities non-perturbatively renormalised
through the SMOM-$(\gamma_\mu,\gamma_\mu)$ and $\qq$ schemes,
given in Tables~\ref{tab:syst3GeV} and \ref{tab:syst2GeV}.
For these quantities, we have identified two main sources of systematic error:
discretisation effects and chiral extrapolation to the physical pion mass.
We have illustrated that some of our results have larger than expected $O(a^2)$ lattice artefacts;
since we have only two lattice spacings,
we take half the difference between the fine ensemble's result (extrapolated to the physical pion mass)
and the continuum extrapolation's result as an estimate of a potential curvature due to
$O(a^4)$ artefacts~\footnote{
  Exact chiral symmetry would guarantee the absence of $O(a)$ and $O(a^3)$ artefacts.
  Strictly speaking with Domain-Wall fermions there could be $O(am_{\rm res})$ and $O((am_{\rm res})^3)$ terms,
  however all our numerical studies show that these terms are numerically irrelevant, if not absent,
  as expected from naive power counting.}
In the future, it will be crucial to include a third lattice spacing
to reduce (or eliminate) this error and check that
these quantities approach their continuum values linearly in $a^2$.

\input Tables/table_results_3GeV
\input Tables/table_results_2GeV

Our chiral extrapolations are well under control,
as illustrated in Figs.~\ref{results:fig:globalfitRgg_susybasis} and \ref{results:fig:globalfitBgg_susybasis}.
We find that both a chiral perturbation theory prediction (Method A)
and a linear Ansatz (Method B) in $m_P^2$ give
very good  $\chi^2$ per degree-of-freedom.
We take half the difference between these to estimate our chiral extrapolation error.
We also observe that the results of the bag parameters extrapolated with a chiral fit
give are very similar to those obtained  from the combinations $G_{ij}$ (Method C), see Appendix~\ref{app:methodC}.
Since the combinations $G_{ij}$ are free from leading chiral logarithms we
conclude that the chiral extrapolation to the physical quark masses 
are well under control.
In the future, we plan to perform the computation at physical values of the quark mass~\cite{Blum:2014tka}
and therefore eliminate this error.

In Tables~\ref{tab:syst3GeV} and~\ref{tab:syst2GeV}, we give the breakdown of our error budget.
For our main results, the ratios $R_i$ renormalised in SMOM-$\gmugmu$ and $\qq$ 
schemes at $\mu=3$ GeV, we give the statistical errors together with our estimate of the discretisation
and chiral errors. We emphasise that these quantities are completely non-perturbative.
We determine $R_2$ and $R_3$ with a precision better than $5\%$, whereas $R_4$
and $R_5$ have an error of $5 \%$  and $8\%$ respectively. The latter are largely
dominated by the discretisation errors, therefore we expect an important
improvement with the future inclusion of a third lattice spacing in our analysis.

We have also converted our results to $\msbar$;
since this matching is done in perturbation theory, there is an uncertainty 
due to the truncation of the perturbative series, in this case of order $O(\alpha_s^2)$.
We estimate this error by taking the difference:
\be
\delta_i^{\rm PT }(\mu) = \frac{|R_i^{\msbar\leftarrow \gmugmu}(\mu) - R_i^{\msbar\leftarrow \qq}(\mu)|}
      {\frac 12 (R_i^{\msbar\leftarrow \gmugmu}(\mu) + R_i^{\msbar\leftarrow \qq}(\mu))}
\ee
In Tables~\ref{tab:syst3GeV} and~\ref{tab:syst2GeV}, this error refers to as ``PT'' (Perturbation Theory).
Although the conversion can be done in the continuum limit,
we checked that the applying the conversion to $\msbar$ on the data before continuum/chiral extrapolation
give the same results as if we apply it in the continuum (the difference is smaller than
our statistical errors). In Table~\ref{tab:syst3GeV}, these results are denoted by
($\msbar \leftarrow \text{SMOM}$). We observe that the matching has very little effect
on the central values and on the error budget
(except of course that there is a perturbative error in addition). 
For the central value and the errors given in Tables~\ref{tab:syst3GeV} and~\ref{tab:syst2GeV},
we quote the results obtained using SMOM-$\gmugmu$ as an intermediate
scheme~\footnote{If we use the SMOM-$\qq$ as an intermediate scheme,
  the results are very close and the error budget
  almost identical, therefore we do not repeat it here.
  The interested reader can find the corresponding central values in Table~\ref{results:tab:final_results} }.
The effect of the intermediate SMOM scheme is less than $3\%$ for $\mu=3$ GeV
and $4-5\%$ for $\mu=2$ GeV.
Regarding the total error, we find that all together, after conversion to $\msbar$,
the $\mu=2$ GeV results are of the same size as the $3$ GeV ones.
(Although we also note that in general if we lower the scale,
the perturbative errors increase and the discretisation errors decrease,
as expected).

We also give the error budget for the bag parameters $B$ and their combinations
$G$. Not surprisingly, we also find that the discretisation effects
are larger than anticipated. In particular for the quantities $B_3$
and $B_5$ we quote an error of $\sim 8\%$ and $\sim 5\%$ at $\mu=3$ GeV.
Clearly these errors come mainly from the NPR procedure as
we observe a reduction of a factor two when we lower the scale
to $\mu=2$ GeV. However, as for the ratios $R_i$, the perturbative errors increase
if we lower the scale and - apart from $B_3$ - we observe that the $\mu=2$ and $\mu=3$ results have
similar total uncertainty, after conversion to $\msbar$.
We expect the systematic uncertainty associated with the discretisation effects
to drop drastically in the future with the inclusion of a third (finer) lattice spacing.
The $\mu=3$ GeV results should then have
have significantly reduced systematic errors in comparison to results renormalised at $\mu=2\text{ GeV}$

\subsection{Final results and comparison with previous works}

We report our final results for the ratios $R$, the bag parameters $B$ and the combinations $G$
in Table~\ref{results:tab:final_results}.  The first error is statistical and the second combines
the various systematic errors. Our main results are those given in the intermediate SMOM-$\gmugmu$
and $\qq$ schemes. The RI-MOM results are only given for comparison with previous work. All these
results are purely non-perturbative.
The corresponding correlation matrices are given in Appendix~\ref{app:correlations}.

For completeness, we also give our results after conversion to $\msbar$; in order to
keep track of the intermediate scheme dependence, we denote them by $\msbar \leftarrow scheme$,
where $scheme$ can be one of the three intermediate schemes. We remind the reader that this conversion
is done in perturbation theory, therefore the systematic errors also include
an estimate of the perturbative error (except for the RI-MOM scheme as we do not find these results to be reliable).
After conversion to $\msbar$ at $\mu=3\,\GeV$, one expects
the results to be independent from the intermediate scheme, up to small perturbative corrections. 
Table~\ref{results:tab:final_results} shows that upon matching to $\msbar$ the conversion has very little effect
on the ratios for the non-exceptional schemes. Furthermore we the $\msbar \leftarrow \gmugmu$
and $\msbar \leftarrow \qq$ are compatible within statistical fluctuations
(in the worst case within $\sim1.5$ standard deviations).
This is highly suggestive that the perturbative series for these schemes are well-behaved at this matching scale.

However, as shown in Tables~\ref{tab:syst3GeV}, \ref{tab:syst2GeV} and \ref{results:tab:final_results},
we observe that our new results using the non-exceptional schemes differ significantly from the ones
renormalised though the RI-MOM scheme.
This could be due to large higher order terms in the perturbative series for the matching of RI-MOM
to $\overline{\text{MS}}$ that we neglect, although at the high matching scale we use this seems unlikely,
which leaves this discrepancy to being some systematic inherent to the exceptional scheme renormalisation technique itself,
such as the subtraction of the Goldstone pole (absent in the SMOM schemes).
We argue below that the non-perturbative renormalisation procedure is the cause of the disagreement
between the different collaborations and that it is due to systematic errors inherent in the RI-MOM scheme.


We finalise this section with a comparison of our results with previous measurements shown in Table~\ref{results:tab:collabcomparison}.
\input Tables/table_schemes.tex
We report the two most recent results of the ETM collaboration, who renormalised their results non-perturbatively
using the intermediate, exceptional, RI-MOM scheme.
We also compare our results to those of the SWME collaboration, who used 
1-loop continuum perturbation theory.
We choose to compare the bag parameters because the ratios $R_i$ are in general not reported by these collaborations.
First, looking at the first three columns, ETM 12, ETM 15, and RBC-UKQCD 12, 
we see that the $n_f=2$ results are compatible with the $n_f=2+1$ and $n_f=2+1+1$ ones
(only within $\sim 2.8 \sigma$ for  $B_5$),
suggesting that these quantities do not depend strongly on the number of flavours.
However the values of $B_4$ and $B_5$ quoted by the SWME collaboration differ significantly 
from the other determinations.
In this work we show that 
our values of $B_4$ and $B_5$ are compatible with those of the ETM collaboration if we use the RI-MOM
intermediate scheme.  However, if we use an SMOM scheme (as we strongly advocate in this work)
our results are then compatible with the SWME collaboration .
The fact that we are compatible with ETM whilst using the same renormalisation scheme suggests that the scheme dependence
we see is legitimate.

\input Tables/table_compareB.tex

\subsection{Matrix elements of the BSM four-quark operators}

\input Tables/table_me.tex
\input Tables/table_me_cov.tex

We end this section with the matrix elements of interest $\ME{i}$. 
They can obtained from the ratios $R_i$, the bag $B_i$
or the combinations $G_{ij}$ with different source of systematic errors.
We find that the methods give consistent results, but the error can be very
different. We find that the $(6,\bar 6)$ operators are more precise when computed
from the $R_i$ whereas the bag $B_i$ give smaller systematic errors for the the $(8,8)$
operators. Our best estimates are given in Table~\ref{table:MEFinal},
where we also convert to $\msbar$ for the reader's convenience.
The corresponding correlation matrix is given in Table~\ref{table:CovME}.
As expected, there are important correlations between operators of same chirality
which have to be taken into account in phenomenological application.
The non-perturbative results are obtained with a precision of $5\%$ or better,
this is the most precise computation of these matrix elements. 
The details of this computation are given in Appendix~\ref{appendix:MethodAB}.

%% file: Tables/table_results_3GeV.tex
\begin{table}[t]
  \begin{tabular}{ c | c | c c c c | c c c c c | c c c c }
 \toprule 
 Scheme &  & $ R_2 $  & $ R_3 $  & $ R_4 $  & $ R_5 $  & $ B_1 $  & $ B_2 $  & $ B_3 $  & $ B_4 $  & $ B_5 $  & $ G_{21} $  & $G_{23} $  & $ G_{24} $  & $ G_{45} $ \\
 \hline 
 \multirow{6}{*}{ $(\gamma_\mu, \gamma_\mu)$ } 
 & central & $ -19.11 $ & $ 5.76 $ & $ 40.12 $ & $ 11.13 $ & $ 0.523 $ & $ 0.526 $ & $ 0.774 $ & $ 0.940 $ & $ 0.786 $ & $ 1.005 $  & $ 0.664 $ & $ 0.502 $  & $ 1.175 $ \\
 \cline{2-15} 
 & Stat. &  2.2\% &  2.5\% &  2.1\% &  1.9\% &  1.7\% &  1.5\% &  1.9\% &  1.2\% &  1.2\% &  1.3\% &  0.3\% &  2.5\% &  0.6\%\\
 & Discr. &  1.0\% &  2.5\% &  4.1\% &  7.1\% &  1.3\% &  3.4\% &  8.4\% &  1.1\% &  4.8\% &  4.6\% &  3.2\% &  2.0\% &  4.8\%\\
 & Chiral &  1.3\% &  1.3\% &  2.2\% &  2.2\% &  0.4\% &  0.3\% &  0.4\% &  0.4\% &  0.4\% &  -  &  -  &  -  &  - \\
 &  Total$^*$  &  2.8\% &  3.8\% &  5.1\% &  7.7\% &  2.2\% &  3.8\% &  8.6\% &  1.7\% &  5.0\% &  4.7\% &  3.2\% &  3.2\% &  4.8\%\\
 \hline 
 \multirow{6}{*}{ $(\slash \!\!\!q, \slash \!\!\!q)$ } 
 & central & $ -20.31 $ & $ 6.12 $ & $ 42.74 $ & $ 10.68 $ & $ 0.541 $ & $ 0.523 $ & $ 0.770 $ & $ 0.937 $ & $ 0.708 $ & $ 0.967 $  & $ 0.664 $  & $ 0.498 $  & $ 1.296 $ \\
 \cline{2-15} 
 & Stat. &  2.3\% &  2.5\% &  2.1\% &  1.9\% &  1.8\% &  1.5\% &  1.9\% &  1.2\% &  1.2\% &  1.3\% &  0.3\% &  2.6\% &  0.6\%\\
 & Discr. &  0.8\% &  2.9\% &  4.0\% &  7.3\% &  1.0\% &  3.5\% &  8.5\% &  1.1\% &  5.3\% &  4.3\% &  3.2\% &  2.1\% &  5.6\%\\
 & Chiral &  1.3\% &  1.3\% &  2.2\% &  2.2\% &  0.4\% &  0.4\% &  0.4\% &  0.4\% &  0.4\% &  -  &  -  &  -  &  - \\
 &  Total$^*$  &  2.7\% &  4.0\% &  5.0\% &  7.9\% &  2.1\% &  3.9\% &  8.8\% &  1.7\% &  5.5\% &  4.5\% &  3.2\% &  3.3\% &  5.6\%\\
 \hline 
 \botrule 
 \multirow{6}{*}{ $\msbar\leftarrow \text{SMOM}$ } 
 & central & $ -19.48 $ & $ 6.08 $ & $ 43.11 $ & $ 10.99 $ & $ 0.525 $ & $ 0.488 $ & $ 0.743 $ & $ 0.920 $ & $ 0.707 $ & $ 0.930 $  & $ 0.642 $  & $ 0.456 $  & $ 1.278 $ \\
 \cline{2-15} 
 & Stat. &  2.3\% &  2.5\% &  2.1\% &  1.9\% &  1.7\% &  1.5\% &  1.9\% &  1.3\% &  1.2\% &  1.3\% &  0.3\% &  2.5\% &  0.6\%\\
 & Discr. &  1.0\% &  2.7\% &  4.1\% &  7.1\% &  1.3\% &  3.4\% &  8.6\% &  1.0\% &  4.9\% &  4.5\% & 3.4\% &  2.0\% &  4.8\%\\
 & Chiral &  1.3\% &  1.3\% &  2.2\% &  2.2\% &  0.4\% &  0.4\% &  0.4\% &  0.4\% &  0.4\% &  -  &  -  &  -  &  - \\
 & PT &  2.2\%&  2.3\%&  2.6\%&  2.9\%&  2.1\%&  0.9\%&  1.0\%&  1.4\%&  3.9\%&  2.3\%&  1.4\%&  1.7\%&  4.1\%\\
 & Total &  3.5\% &  4.5\% &  5.7\% &  8.2\% &  3.0\% &  3.9\% &  8.9\% &  2.2\% &  6.3\% &  5.2\% & 3.6\% &  3.6\% &  6.4\% \\
 \botrule 
 \multirow{5}{*}{ RI-MOM } 
 & central & $ -15.77 $ & $ 5.39 $ & $ 30.75 $ & $ 7.24 $ & $ 0.517 $ & $ 0.571 $ & $ 0.950 $ & $ 0.947 $ & $ 0.677 $ & $ 1.105 $  & $ 0.590 $  & $ 0.549 $  & $ 1.266 $ \\
 \cline{2-15} 
 & Stat. &  2.1\% &  2.4\% &  1.9\% &  1.6\% &  1.7\% &  1.3\% &  1.7\% &  1.1\% &  1.2\% &  1.2\% &  0.4\% &  1.9\% &  1.4\%\\
 & Discr. &  3.6\% &  1.2\% &  6.7\% & 12\% &  1.7\% &  1.0\% &  5.1\% &  5.2\% & 12\% &  0.5\% & 4.6\% &  6.3\% & 14\%\\
 & Chiral &  1.3\% &  1.3\% &  2.2\% &  2.2\% &  0.4\% &  0.4\% &  0.4\% &  0.4\% &  0.4\% &  -  &  -  &  -  &  - \\
 &Total$^*$  &  4.3\% &  3.0\% &  7.3\% & 13\% &  2.5\% &  1.7\% &  5.4\% &  5.3\% & 12\% &  1.3\% & 4.6\% &  6.5\% & 14\%\\
 \hline 
 \multirow{5}{*}{ $\msbar\leftarrow \text{RI-MOM}$ } 
 & central & $ -16.44 $ & $ 5.31 $ & $ 34.56 $ & $ 8.50 $ & $ 0.526 $ & $ 0.417 $ & $ 0.655 $ & $ 0.745 $ & $ 0.555 $ & $ 0.793 $  & $ 0.621 $  & $ 0.316 $  & $ 1.267 $ \\
 \cline{2-15} 
 & Stat. &  2.2\% &  2.5\% &  2.0\% &  1.7\% &  1.7\% &  1.4\% &  1.8\% &  1.1\% &  1.1\% &  1.2\% &  0.4\% &  2.1\% &  1.0\%\\
 & Discr. &  2.4\% &  2.5\% &  5.5\% & 10.2\% &  1.7\% &  0.4\% &  6.8\% &  3.7\% &  9.5\% &  2.0\% & 4.5\% &  3.7\% &  9.9\%\\
 & Chiral &  1.3\% &  1.3\% &  2.2\% &  2.2\% &  0.4\% &  0.4\% &  0.4\% &  0.4\% &  0.4\% &  -  &  -  &  -  &  - \\
 \botrule 

  \end{tabular}
 \caption{\raggedright{
    Central values and error budget for our final results renormalised at $\mu=3$ GeV. Note that for our non-perturbatively
    renormalised results in the SMOM-$(\gamma_\mu,\gamma_\mu)$ and $\qq$ scheme, the error Total$^*$
    does not include any perturbative uncertainty (PT). We also show the error budget for our $\msbar$ results
    where only SMOM-schemes have been considered. The central value is obtained using  SMOM-$(\gamma_\mu,\gamma_\mu)$
    as intermediate scheme.  For illustration, in the second part of the table, we  give the error budget
    if we only use the RI-MOM scheme. See text for details.  
}}
\label{tab:syst3GeV}
\end{table}

%% file: Tables/table_results_2GeV.tex
\begin{table}[t]
  \begin{tabular}{ c | c | c c c c | c c c c c | c c c c }
\toprule 
 Scheme &  & $ R_2 $  & $ R_3 $  & $ R_4 $  & $ R_5 $  & $ B_1 $  & $ B_2 $  & $ B_3 $  & $ B_4 $  & $ B_5 $  & $ G_{21} $  & $ G_{23} $  & $ G_{24} $  & $ G_{45} $ \\
 \hline 
 \multirow{6}{*}{ $(\gamma_\mu, \gamma_\mu)$ } 
 & central & $ -15.77 $ & $ 4.88 $ & $ 30.68 $ & $ 8.27 $ & $ 0.533 $ & $ 0.563 $ & $ 0.866 $ & $ 0.922 $ & $ 0.736 $ & $ 1.057 $  & $ 0.647 $  & $ 0.527 $  & $ 1.240 $ \\
 \cline{2-15} 
 & Stat. &  2.3\% &  2.4\% &  2.1\% &  2.0\% &  1.7\% &  1.5\% &  1.7\% &  1.3\% &  1.3\% &  1.3\% &  0.4\% &  2.5\% &  0.6\%\\
 & Discr. &  0.6\% &  1.1\% &  2.9\% &  4.4\% &  1.3\% &  2.2\% &  4.1\% &  0.8\% &  2.5\% &  3.3\% &  1.5\% &  1.1\% &  1.8\%\\
 & Chiral &  1.3\% &  1.3\% &  2.2\% &  2.2\% &  0.4\% &  0.4\% &  0.4\% &  0.4\% &  0.4\% &  -  &  -  &  -  &  - \\
 &  Total$^*$  &  2.7\% &  2.9\% &  4.2\% &  5.4\% &  2.2\% &  2.6\% &  4.4\% &  1.6\% &  2.9\% &  3.5\% &  1.5\% &  2.7\% &  1.9\%\\
 \hline 
 \multirow{6}{*}{ $(\slash \!\!\!q, \slash \!\!\!q)$ } 
 & central & $ -17.19 $ & $ 5.30 $ & $ 33.43 $ & $ 7.79 $ & $ 0.565 $ & $ 0.561 $ & $ 0.862 $ & $ 0.920 $ & $ 0.635 $ & $ 0.994 $  & $ 0.648 $  & $ 0.524 $  & $ 1.434 $ \\
 \cline{2-15} 
 & Stat. &  2.3\% &  2.4\% &  2.1\% &  2.0\% &  1.7\% &  1.5\% &  1.7\% &  1.3\% &  1.3\% &  1.3\% &  0.4\% &  2.5\% &  0.6\%\\
 & Discr. &  0.6\% &  1.1\% &  2.9\% &  4.5\% &  1.3\% &  2.3\% &  4.2\% &  0.8\% &  2.6\% &  3.4\% &  1.5\% &  1.2\% &  2.0\%\\
 & Chiral &  1.3\% &  1.3\% &  2.2\% &  2.2\% &  0.4\% &  0.4\% &  0.4\% &  0.4\% &  0.4\% &  -  &  -  &  -  &  - \\
 &  Total$^*$  &  2.7\% &  2.9\% &  4.2\% &  5.4\% &  2.2\% &  2.7\% &  4.6\% &  1.5\% &  2.9\% &  3.6\% &  1.6\% &  2.8\% &  2.1\%\\
 \hline 
 \botrule 
 \multirow{6}{*}{ $\msbar\leftarrow \text{SMOM}$ } 
 & central & $ -16.14 $ & $ 5.20 $ & $ 33.45 $ & $ 8.15 $ & $ 0.536 $ & $ 0.509 $ & $ 0.816 $ & $ 0.888 $ & $ 0.640 $ & $ 0.950 $  & $ 0.621 $  & $ 0.459 $  & $ 1.373 $ \\
 \cline{2-15} 
 & Stat. &  2.3\% &  2.4\% &  2.1\% &  2.0\% &  1.7\% &  1.5\% &  1.7\% &  1.3\% &  1.3\% &  1.3\% &  0.4\% &  2.5\% &  0.6\%\\
 & Discr. &  0.6\% &  1.1\% &  2.9\% &  4.5\% &  1.3\% &  2.1\% &  4.1\% &  0.8\% &  2.5\% &  3.2\% &  1.6\% &  1.0\% &  1.9\%\\
 & Chiral &  1.3\% &  1.3\% &  2.2\% &  2.2\% &  0.4\% &  0.4\% &  0.4\% &  0.4\% &  0.4\% &  -  &  -  &  -  &  - \\
 & PT &  3.9\%&  3.9\%&  4.2\%&  4.6\%&  4.3\%&  2.1\%&  2.2\%&  2.5\%&  6.2\%&  4.8\%&  3.1\%&  3.3\%&  6.7\%\\
 & Total &  4.7\% &  4.9\% &  5.9\% &  7.1\% &  4.8\% &  3.4\% &  5.0\% &  2.9\% &  6.8\% &  5.9\% &  3.5\% &  4.2\% &  7.0\% \\
 \botrule 
 \multirow{5}{*}{ RI-MOM } 
 & central & $ -14.16 $ & $ 5.00 $ & $ 26.24 $ & $ 5.62 $ & $ 0.530 $ & $ 0.536 $ & $ 0.940 $ & $ 0.841 $ & $ 0.529 $ & $ 1.010 $  & $ 0.572 $  & $ 0.448 $  & $ 1.555 $ \\
 \cline{2-15} 
 & Stat. &  1.9\% &  2.1\% &  1.8\% &  1.7\% &  1.7\% &  1.2\% &  1.5\% &  1.0\% &  1.3\% &  1.1\% &  0.5\% &  1.6\% &  1.1\%\\
 & Discr. &  6.1\% &  4.3\% &  8.7\% & 10.8\% &  1.7\% &  4.3\% &  2.3\% &  7.6\% & 10.0\% &  2.9\% &  2.2\% & 10.7\% &  3.3\%\\
 & Chiral &  1.3\% &  1.3\% &  2.3\% &  2.2\% &  0.4\% &  0.4\% &  0.4\% &  0.4\% &  0.4\% &  -  &  -  &  -  &  - \\
 &Total$^*$  &  6.5\% &  4.9\% &  9.1\% & 11.2\% &  2.4\% &  4.5\% &  2.8\% &  7.7\% & 10.1\% &  3.1\% &  2.2\% & 10.8\% &  3.5\%\\
 \hline 
 \multirow{5}{*}{ $\msbar\leftarrow \text{RI-MOM}$ } 
 & central & $ -15.80 $ & $ 5.20 $ & $ 32.21 $ & $ 7.41 $ & $ 0.541 $ & $ 0.423 $ & $ 0.693 $ & $ 0.731 $ & $ 0.497 $ & $ 0.782 $  & $ 0.613 $  & $ 0.308 $  & $ 1.448 $ \\
 \cline{2-15} 
 & Stat. &  1.9\% &  2.1\% &  1.8\% &  1.7\% &  1.7\% &  1.2\% &  1.5\% &  1.0\% &  1.2\% &  1.1\% &  0.5\% &  1.6\% &  0.8\%\\
 & Discr. &  6.1\% &  4.3\% &  8.7\% & 10.2\% &  1.7\% &  4.5\% &  2.5\% &  7.7\% &  9.4\% &  3.1\% &  2.2\% & 10.9\% &  2.4\%\\
 & Chiral &  1.3\% &  1.3\% &  2.3\% &  2.2\% &  0.4\% &  0.4\% &  0.4\% &  0.4\% &  0.4\% &  -  &  -  &  -  &  - \\
 \botrule
 \end{tabular}
  \caption{ 
    {Same as table~\ref{tab:syst3GeV} for our $\mu=2$ GeV results.}}
\label{tab:syst2GeV}
\end{table}

%% file: Tables/table_schemes.tex
\begin{table}[t]
\begin{tabular}{l | ccccc }
\toprule
& - & $R_2$ & $R_3$ & $R_4$ & $R_5$ \\
\cline{2-6}
$\gmugmu$ & - & -19.11(43)(31) & 5.76(14)(16) & 40.12(82)(188) & 11.13(21)(83) \\
$\qq$     & - & -20.31(46)(31) & 6.12(15)(19) & 42.74(88)(195) & 10.68(20)(82) \\
RI-MOM    & - & -15.77(33)(60) & 5.39(13)(9)  & 30.75(59)(217) &  7.24(11)(91) \\
\cline{2-6}
& $B_1$ & $B_2$ & $B_3$ & $B_4$ & $B_5$ \\
\cline{2-6}
$\gmugmu$ & 0.523(9)(7)  & 0.526(8)(18)  & 0.774(14)(65)  & 0.940(12)(11) & 0.786(9)(38) \\
$\qq$     & 0.541(9)(6)  & 0.523(8)(19)  & 0.770(14)(66)  & 0.937(12)(11) & 0.708(8)(38) \\
RI-MOM    & 0.517(9)(9)  & 0.571(8)(6)   & 0.950(17)(49)  & 0.947(10)(49) & 0.677(8)(81) \\
\cline{2-6}
& - & $G_{21}$ & $G_{23}$ & $G_{24}$ & $G_{45}$ \\
\cline{2-6}
$\gmugmu$ & - & 1.005(13)(46) & 0.664(2)(21) & 0.502(13)(10) & 1.175(6)(56)\\
$\qq$     & - & 0.967(13)(42) & 0.664(2)(21) & 0.498(13)(10) & 1.296(8)(72)\\
RI-MOM    & - & 1.105(13)(6) & 0.590(2)(27) & 0.549(11)(34) & 1.266(18)(181)\\
\botrule
& - & $R_2$ & $R_3$ & $R_4$ & $R_5$ \\
\cline{2-6}
$\msbar \leftarrow \gmugmu$ & - & -19.48(44)(32)(42)  & 6.08(15)(18)(14)  & 43.11(89)(201)(112)  & 10.99(20)(82)(32) \\
$\msbar \leftarrow \qq$     & - & -19.91(45)(30)(43)  & 6.22(16)(20)(14)  & 44.25(91)(202)(115)  & 10.68(20)(82)(31) \\
$\msbar \leftarrow$ RI-MOM  & - & -16.44(36)(44)\pp   & 5.31(13)(15)\pp   & 34.56(68)(204)\ppp   &  8.50(14)(89)\p   \\
\cline{2-6}
& $B_1$ & $B_2$ & $B_3$ & $B_4$ & $B_5$ \\
\cline{2-6}
$\msbar \leftarrow \gmugmu$ &\, 0.525(9)(7)(11)  & 0.488(7)(17)(4)  & 0.743(14)(64)(8)  & 0.920(12)(10)(13)  & 0.707(8)(35)(27) \\
$\msbar \leftarrow \qq$     &\, 0.536(9)(6)(11)  & 0.492(7)(17)(5)  & 0.751(14)(66)(8)  & 0.932(12)(17)(13)  & 0.680(8)(37)(26) \\
$\msbar \leftarrow$ RI-MOM  &\, 0.526(9)(9)\pp   & 0.417(6)(2)\pp   & 0.655(12)(44)\p  & 0.745(9)(28)\ppp    & 0.555(6)(53)\pp \\
\cline{2-6}
& - & $G_{21}$ & $G_{23}$ & $G_{24}$ & $G_{45}$ \\
\cline{2-6}
$\msbar \leftarrow \gmugmu$ & - & 0.930(12)(42)(41) & 0.642(2)(22)(26) & 0.456(12)(9)(18)\m  & 1.278(7)(62)(15)   \\
$\msbar \leftarrow \qq$     & - & 0.920(12)(40)(40) & 0.641(2)(22)(26) & 0.467(12)(10)(19)   & 1.342(8)(76)(16)   \\
$\msbar \leftarrow$  RI-MOM & - & 0.793(10)(16)\pp  & 0.621(2)(28)\pp  & 0.316(7)(12)\ppp     & 1.267(12)(125)\m\m \\
\botrule
\end{tabular}
\caption{\raggedright{Final results for $R_i$, $B_i$ and $G_{ij}$ renormalised at $\mu=3\,\text{GeV}$.
    The first error is statistical, the second one is an estimate of the systematic error (chiral and discretisation errors
    combined in quadrature). When present, the third one is the perturbative error coming from the matching to $\msbar$.
    Our best results are the ones obtained through the SMOM-$\gmugmu$ and $\qq$ schemes.
    The RI-MOM results are  presented here only for illustration and comparison purposes,
    we did not attempt to estimate the perturbative error for the $\msbar \leftarrow \rm RI-MOM$ case.
  }\label{results:tab:final_results}}
\end{table}

%% file: Tables/table_compareB.tex
\begin{table}[t]
  \begin{center}
$$
\small
\begin{array}{c| c | c | c | c  |c c}
  \toprule
    &   \rm{ETM\, 12} & \quad  \rm{ETM\, 15} \quad & \rm{RBC-UKQCD\, 12} & \quad \rm{SWME\, 15} \quad& \multicolumn{2}{c}{\qquad \text{This work}}  
\\
\hline
n_f & 2 & 2+1+1 & 2+1 & 2+1 & 2+1 & 2+1\\
interm. & \multirow{2}{*}{\rm RI-MOM} & \multirow{2}{*}{\rm RI-MOM} & \multirow{2}{*}{\rm RI-MOM} & \multirow{2}{*}{1-loop} & \multirow{2}{*}{\rm RI-SMOM} & \multirow{2}{*}{\rm RI-MOM} \\
scheme & & & & & & \\
      \hline 
      B_2 & 0.47(2) & 0.46(3)(1) & 0.43(5) &  0.525(1)(23) & 0.488(7)(17)  & 0.417(6)(2)   \\ 
      B_3 & 0.78(4) & 0.79(5)(1) & 0.75(9) &  0.772(5)(35) & 0.743(14)(65) & 0.655(12)(44) \\ 
      B_4 & 0.76(3) & 0.78(4)(3) & 0.69(7) &  0.981(3)(61) & 0.920(12)(16) & 0.745(9)(28)  \\ 
      B_5 & 0.58(3) & 0.49(4)(1) & 0.47(6) &  0.751(8)(68) & 0.707(8)(44)  & 0.555(6)(53)  \\
	  \hline
      \botrule
    \end{array}
$$ 
\caption{\raggedright{
    Comparison of the bag parameters $B_i$ at $3\,\GeV$ in the SUSY basis in the $\msbar$ scheme of \cite{Buras:2000if}.
    When only one error is quoted, it means that the errors have been already combined.
      If not, the first errors are statistical and the second systematic.
      We argue that the renormalisation procedure is the cause of the disagreement observed 
      for $B_4$ and $B_5$ between the different collaborations and that it is due to some underestimated 
      systematic errors present in the RI-MOM scheme.
      For the RI-SMOM results, we choose the $\gmugmu$ scheme.
      } \label{results:tab:collabcomparison}}
  \end{center}
\end{table}

%% file: Tables/table_me.tex
\begin{table}[t]
\centering
\begin{tabular}{c|l r|l r}
  \toprule
  & \multicolumn{2}{| c}{  SMOM-$\gmugmu$ } & \multicolumn{2}{|c}{ $\msbar$ } \\
  \hline
 $\la \Kb | O_2 | \K \ra$ & $-0.1597(42)_{\rm stat}(34)_{\rm syst}$  & $3.4\%$ & $ -0.1636(43)_{\rm stat}(49)_{\rm syst}(36)_{\rm PT}$  & 4.5\% \\
 $\la \Kb | O_3 | \K \ra$ & $\m0.0482(14)_{\rm stat}(15)_{\rm syst}$ & $4.2\%$ & $\m0.0510(14)_{\rm stat}(20)_{\rm syst}(12)_{\rm PT}$  & 5.3\% \\ 
 $\la \Kb | O_4 | \K \ra$ & $\m0.3377(42)_{\rm stat}(77)_{\rm syst}$ & $2.6\%$ & $\m0.3781(47)_{\rm stat}(113)_{\rm syst}(48)_{\rm PT}$ & 3.5\% \\
 $\la \Kb | O_5 | \K \ra$ & $\m0.0941(11)_{\rm stat}(49)_{\rm syst}$ & $5.4\%$ & $\m0.0969(12)_{\rm stat}(54)_{\rm syst}(27)_{\rm PT}$  & 6.9\% \\
  \botrule
\end{tabular}
\caption{\raggedright{Our best results for the matrix elements of the BSM four-quark operators.
    The numbers are given in units of GeV$^4$, in the SMOM-$\gmugmu$ scheme (left)   and in $\msbar$ (right) at $\mu=3$ GeV.
    Results are obtained from the ratios $R_i$ for $O_{2,3}$ and from the bag parameters $B_i$  for $O_{4,5}$
    as explained in the text.
    The systematic errors combine the chiral and the discretisation errors,
    the percentage error is obtained by adding all the different errors in quadrature.
}}
\label{table:MEFinal}
\end{table}
%

%% file: Tables/table_me_cov.tex
\begin{table}[t]
  \be
  \begin{array}{c | c c c | c c c }
    \toprule
     & \multicolumn{3}{c}{  \mbox{ SMOM-}\gmugmu } & \multicolumn{3}{|c}{ \msbar }\\
       & \ME{3}  & \ME{4}    & \ME{5}         & \ME{3}  & \ME{4}    & \ME{5}    \\
\hline
\ME{2} & -0.9950 &   -0.3400 &   -0.2762      & -0.9902 &  -0.3384  &   -0.2763  \\
\ME{3} &         &  \m0.3210 &  \m0.2480      &         &  \m0.3202 &  \m0.2466  \\
\ME{4} &         &           &  \m0.9016      &         &           &  \m0.8984  \\
    \botrule
  \end{array}
  \nn
  \ee
  \caption{
Correlation matrix for the matrix elements given in Table~\ref{table:MEFinal}.
}  
\label{table:CovME}
\end{table}

%% file: Sections/s5_concl.tex
\section*{Conclusions}\label{sec:concl}

We have computed the matrix elements necessary for the study of
neutral Kaon mixing beyond the Standard Model. 
We confirm that the ratio of BSM contribution to SM is of order $O(10)$,
as previous studies have shown and as expected from Chiral Perturbation Theory.
We also find that the colour mixed operators are significantly
smaller than their colour unmixed partners, as one would naively
expect from the VSA. However some bag parameters differs significantly from their
VSA (up to a factor $2$), showing the importance of using lattice QCD for such a computation.

This work improves on previous studies in various ways:
\begin{itemize}
\item We use a $n_f=2+1$ fermion discretisation that has good chiral-flavour properties at finite lattice spacing.
\item We have extended our previous work with the addition of a second lattice spacing,
  allowing us to extrapolate our results to the continuum.
\item The renormalisation is performed non-perturbatively and we have introduced two new SMOM schemes
  which use non-exceptional kinematics rather than the previously used $\rimom$ (exceptional) scheme.
\item We used different parametrisations of the matrix elements in order to control the
  extrapolation to the physical point (extrapolation to the continuum and to physical values
  of the quark masses). We show that the choice of parametrisation can affect the systematic
  errors in a drastic way (see for example the difference between $G_{23}$ and $\tilde G_{23}$).
\end{itemize}

We see that our systematics are dominated by the continuum extrapolation.
One could argue that our estimate of the discretisation effects is rather conservative
because - in principle -  $O(a^3)$ lattice artefacts are absent with chiral fermions.
However we believe that our choice is appropriate because we have only two lattice
spacings and we observe that the lattice artefacts are larger than anticipated.
We do not believe an increase in statistics or simulation at physical pion and strange masses
will be as beneficial as a third, finer lattice spacing.
\\

A very important point of this work comes from the renormalisation.
We argue that discrepancies observed between previous results are due to the choice of
intermediate momentum scheme. We show that if use the RI-MOM scheme
we can reproduce the ETMc results and that the RI-SMOM results are compatible with those
of the SWME collaboration.
We strongly advocate the use of the non-exceptional schemes defined in this work.
We show in a companion paper that the RI-MOM results rely strongly on a pole-subtraction procedure 
which is hard to control, whereas such an infra-red contamination is highly suppressed
in the RI-SMOM vertex functions. It is highly desirable that other collaborations
check this statement.

%% file: Aknow.tex
\section*{Acknowledgements}
The coarse ensemble results were computed on the STFC funded “DiRAC” BG/Q system
in the Advanced Computing Facility at the University of Edinburgh. 
We thank our colleagues of the RBC and UKQCD collaborations for stimulating discussions.
We are particularly indebted to Peter Boyle for his help with the UKQCD hadron software
(used to to generate the  coarse ensemble data) and to Christoph Lehner
for computing the matching factor between the SMOM schemes and $\msbar$. 
N.G. is supported by the Leverhulme Research grant RPG-2014-118.
R.J.H. is supported by the Natural Sciences and Engineering Research Council of Canada.

%% file: Appendices/app_chisqr.tex
\subsection{\texorpdfstring{$\chi^2/d.o.f$}{chi-squared per d.o.f} for our measurements}
\label{appendix:app_table}

In Table~\ref{app_table:tab:chisq} we give the $\chi^2$ per degree-of-freedom
of the global fit used in Method A, B and C (see Section~\ref{sec:methodology}).
Method A corresponds to fitting the ratios $R_i$ or the bag parameters $B_i$
using Chiral Perturbation theory ($\chi \rm PT$).
Method B uses a linear fit in $m_P^2$, where $m_P$ is the simulated pion mass.
We find that the fit of the ratios $R_i$ are of very good quality with
a $\chi^2$ per-degree-of freedom of order $0.5$.
The fits for $B_2$ and $B_3$ are a bit worse, although the $\chi^2$
are still reasonable (of order $1.5$).
It is important to stress that our data do not seem to prefer either of the method,
the effects of the chiral logs are not statistically significant. 
We also show the $\chi^2$ for the linear fits of the combinations $G_{ij}$, 
(Method C). There we find that $G_{21}$ is very hard to fit, with a $\chi^2$ per degree
of freedom of order $6$. We can see from Fig.~\ref{results:fig:G_MOM_gg_susy_basis}
that the problem seems to come from the coarse ensemble and could be
due to some lattice artefacts. The other $G_{ij}$ have a much more reasonable $\chi^2$.

\begin{table}[t]
\begin{tabular}{c | cc | cc | cc | cc }
\toprule
\multirow{2}{*}{Scheme} & $\chi\rm  PT$ \;& Linear & $\chi\rm PT$\;  & Linear & $\chi\rm PT$ \; & Linear & $\chi\rm PT$\; & Linear\\
\cline{2-9}
& \multicolumn{2}{|c|}{$R_2$} & \multicolumn{2}{|c|}{$R_3$} & \multicolumn{2}{|c|}{$R_4$} & \multicolumn{2}{|c}{$R_5$} \\
\hline
$\gmugmu$ & 0.45 & 0.55 & 0.35 & 0.43 & 0.51 & 0.45 & 0.45 & 0.45 \\
$\qq$     & 0.44 & 0.53 & 0.34 & 0.41 & 0.51 & 0.44 & 0.46 & 0.47 \\
RI-MOM    &  0.56 & 0.68 & 0.39 & 0.48 & 0.64 & 0.63 & 0.71 & 0.88 \\
\cline{2-9}
& \multicolumn{2}{|c|}{$B_2$} & \multicolumn{2}{|c|}{$B_3$} & \multicolumn{2}{|c|}{$B_4$} & \multicolumn{2}{|c}{$B_5$} \\
\cline{2-9}
$\gmugmu$ & 1.48 & 1.39 & 1.72 & 1.66 & 0.71 & 0.55 & 0.49 & 0.37 \\
$\qq$     & 1.42 & 1.40 & 1.72 & 1.66 & 0.71 & 0.55 & 0.48 & 0.36 \\
RI-MOM    & 1.32 & 1.29 & 1.72 & 1.64 & 0.71 & 0.54 & 0.25 & 0.18 \\
\cline{2-9}
& \multicolumn{2}{|c|}{$G_{21}$} & \multicolumn{2}{|c|}{$G_{23}$} & \multicolumn{2}{|c|}{$G_{24}$} & \multicolumn{2}{|c}{$G_{45}$} \\
\cline{2-9}
$\gmugmu$ & - & 6.86 & -& 0.97 & - & 1.37 & - & 0.09  \\
$\qq$     & - & 6.86 & - & 0.97 & - & 1.36 & - & 0.10 \\
RI-MOM    & - & 6.84 & - & 0.87 & - & 1.25 & - & 0.46 \\
\botrule
\end{tabular}
\caption{\raggedright{$\chi^2/d.o.f$ of the global fits  
    using a chiral fit ($\chi\rm PT$) or a linear fit in $m_P^2$. 
    Since the combinations $G_{ij}$  are designed to cancel (at least) the leading chiral logarithms,
    we did not perform a chiral fit on these quantities.
    The results presented here are for the fits performed on quantities renormalised in the RI-SMOM and RI-MOM schemes.
  }\label{app_table:tab:chisq}}
\end{table}

%% file: Appendices/app_npr.tex
\subsection{Renormalisation factors}
\label{app:npr}

We give the $Z$ matrices obtained though the SMOM-$\gmugmu$ and SMOM-$\qq$
schemes, together with their conversion to $\msbar$
in Tables~\ref{tab:ZBK_24}-\ref{tab:Z88_32}.
\input Tables/table_ZBK_24
\input Tables/table_Z66_24

\input Tables/table_Z88_24

\input Tables/table_ZBK_32

\input Tables/table_Z66_32

\input Tables/table_Z88_32

%% file: Tables/table_ZBK_24.tex
\begin{table}[h]
\begin{tabular}{c c c c}
\toprule
$Z^{\msbar}$     & $C^{\msbar \leftarrow {\rm RI-SMOM}}$ &   $Z^{\rm RI-SMOM}$  &   scheme  \\
\hline
$0.92022(26)$  & $1.00414$  & $0.91642(26)$        &   $(\gamma_\mu,\gamma_\mu)$\\
$0.94796(34)$  & $0.99112$  & $0.95645(34)$        &   $(\s{q},\s{q})$          \\
\botrule
\end{tabular}
\caption[]{$Z/Z_V^2$ factors for the $(27,1)$ operator $O_1$ at $3\, \GeV$
  on the coarse lattice, $a=a_{\bf 24}$.}
\label{tab:ZBK_24}
\end{table}

%% file: Tables/table_Z66_24.tex
\begin{table}[h]
\ars{1.1}
\begin{tabular}{c c c c}
\toprule
$Z^{\msbar}$     & $C^{\msbar \leftarrow {\rm RI-SMOM}}$ &   $Z^{\rm RI-SMOM}$  &   scheme  \\
\hline
$
\begin{pmatrix}
 0.9066(14)  & -0.05376(52) \\
-0.03801(99) &  1.18811(69) \\
\end{pmatrix}$ \;
& 
$\begin{pmatrix}
 1.02973 & 0.01937 \\
 0.01306 & 1.10237 \\ 
\end{pmatrix}$ \;
&
$\begin{pmatrix}
 0.8813(14) & -0.07249(49) \\            
-0.04493(91) & 1.07864(62) \\
\end{pmatrix}$ \;
&   $(\gamma_\mu,\gamma_\mu)$ 
\vspace{0.1cm} \\
$\begin{pmatrix}
 0.9635(13) & -0.05595(54)  \\
-0.0399(10) &  1.26728(103) \\
  \end{pmatrix}$ \;
&
$\begin{pmatrix}
 0.97764 & 0.01937 \\
 0.01306 & 1.05029 \\
  \end{pmatrix}$ \;
&
$\begin{pmatrix}
  0.9866(14)  & -0.08115(54)  \\
  -0.0502(10) &  1.20761(97)  \\
\end{pmatrix}$ \;
&
$(\s{q},\s{q})$ \\
\botrule
\end{tabular}
\caption[]{\raggedright{
  $Z/Z_V^2$ matrices for the $(6,\bar 6)$ operators $O_2$ and $O_3$ at $\mu=3\, \GeV$
    on the coarse lattice, $a=a_{\bf 24}$.}
  }
\label{tab:Z66_24}
\end{table}

%% file: Tables/table_Z88_24.tex
\begin{table}[h]
\ars{1.1}
\begin{tabular}{c c c c}
\toprule
$Z^{\msbar}$     & $C^{\msbar \leftarrow {\rm RI-SMOM}}$ &   $Z^{\rm RI-SMOM}$  &   scheme  \\
\hline
$
\begin{pmatrix}
   0.9535(19)  &  -0.11307(46)  \\
  -0.14099(19) &   1.050434(66) \\
\end{pmatrix}$ \;
& 
$\begin{pmatrix}
 1.08781 & -0.03152 \\
-0.00253 &  1.00084 \\
\end{pmatrix}$ \;
&
$\begin{pmatrix}
  0.8725(17) & -0.07354(42)  \\
 -0.13866(19)&  1.049363(66) \\
  \end{pmatrix}$ \;
&   $(\gamma_\mu,\gamma_\mu)$ 
\vspace{0.1cm} \\
$\begin{pmatrix}
  1.0195(18)  & -0.13876(41)  \\
 -0.14372(20) &  1.051161(65)  \\
\end{pmatrix}$ \;
&
$\begin{pmatrix}
 1.02921 & -0.01199 \\
-0.00253 &  1.00084 \\
\end{pmatrix}$ \;
&
$\begin{pmatrix}
 0.9889(17)  & -0.12259(40)  \\
-0.14110(20) &  1.049965(65) \\
\end{pmatrix}$ \;
&
$(\s{q},\s{q})$ 
\vspace{0.1cm} \\
\botrule
\end{tabular}
\caption[]{$Z/Z_V^2$ matrices for the $(8,8)$ operators $O_4$ and $O_5$ at $\mu=3\, \GeV$
  on the coarse lattice, $a=a_{\bf 24}$.}
\label{tab:Z88_24}
\end{table}

%% file: Tables/table_ZBK_32.tex
\begin{table}[t]
\begin{tabular}{c c c c}
\toprule
$Z^{\msbar}$     & $C^{\msbar \leftarrow {\rm RI-SMOM}}$ &   $Z^{\rm RI-SMOM}$  &   scheme  \\
\hline
$0.94526(26)$  & $1.00414$  & $0.94137(26)$        &   $(\gamma_\mu,\gamma_\mu)$\\
$0.96999(32)$  & $0.99112$  & $0.95645(34)$        &   $(\s{q},\s{q})$          \\
\botrule
\end{tabular}
\caption[]{$Z/Z_V^2$ factors for the $(27,1)$ operators at $3\, \GeV$
  on the fine lattice $a=a_{\bf 32}$.}
\label{tab:ZBK_32}
\end{table}

%% file: Tables/table_Z66_32.tex
\begin{table}[h]
\ars{1.1}
\begin{tabular}{c c c c}
\toprule
$Z^{\msbar}$     & $C^{\msbar \leftarrow {\rm RI-SMOM}}$ &   $Z^{\rm RI-SMOM}$  &   scheme  \\
\hline
$
\begin{pmatrix}
 0.8535(12)  & -0.02489(35) \\
 0.01553(70) &  1.22329(79) \\
\end{pmatrix}$ \;
& 
$\begin{pmatrix}
 1.02973 & 0.01937 \\ 
 0.01306 & 1.10237   \\
\end{pmatrix}$ \;
&
$\begin{pmatrix}
 0.8288(11)  & -0.04505(34)  \\
 0.00426(65) &  1.11022(72) \\
\end{pmatrix}$ \;
&   $(\gamma_\mu,\gamma_\mu)$ 
\vspace{0.1cm} \\
$\begin{pmatrix}
  0.8996(11)  & -0.02511(39) \\
  0.01719(73) & 1.2945(14)  \\
\end{pmatrix}$ \;
&
$\begin{pmatrix}
 0.97764 & 0.01937 \\
 0.01306 & 1.05029 \\
\end{pmatrix}$ \;
&
$\begin{pmatrix}
0.9201(12)  & -0.05011(39) \\
0.00492(71) &  1.2331(13)  \\
\end{pmatrix}$ \;
&
$(\s{q},\s{q})$ \\
\botrule
\end{tabular}
\caption[]{$Z/Z_V^2$ matrices for the $(6,\bar 6)$ operators at $\mu=3\, \GeV$
  on the fine lattice, $a=a_{\bf 32}$.}
\label{tab:Z66_32}
\end{table}

%% file: Tables/table_Z88_32.tex
\begin{table}[h!]
\ars{1.1}
\begin{tabular}{c c c c}
\toprule
$Z^{\msbar}$     & $C^{\msbar \leftarrow {\rm RI-SMOM}}$ &   $Z^{\rm RI-SMOM}$  &   scheme  \\
\hline
$
\begin{pmatrix}
 0.8739(16)  & -0.08782(29)  \\
-0.13909(38) &  1.04740(14)  \\  
\end{pmatrix}$ \;
& 
$\begin{pmatrix}
 1.08781 & -0.03152 \\
-0.00253 &  1.00084 \\
  \end{pmatrix}$ \;
&
$\begin{pmatrix}
 0.7994(15)  & -0.05041(27) \\
-0.13695(38) &  1.04639(14) \\
  \end{pmatrix}$ \;
&   $(\gamma_\mu,\gamma_\mu)$ 
\vspace{0.1cm} \\
$\begin{pmatrix}
 0.9282(16)  & -0.10992(50) \\
-0.14143(42) &  1.04908(16) \\
\end{pmatrix}$ \;
&
$\begin{pmatrix}
 1.02921 & -0.01199 \\
-0.00253 &  1.00084 \\
\end{pmatrix}$ \;
&
$\begin{pmatrix}
  0.9002(15)  & -0.09459(49) \\
 -0.13903(42) &  1.04795(16) \\
\end{pmatrix}$ \;
&
$(\s{q},\s{q})$ 
\vspace{0.1cm} \\
\botrule
\end{tabular}
\caption[]{$Z/Z_V^2$ matrices for the $(8,8)$ operators at $\mu=3\, \GeV$
  on the fine lattice, $a=a_{\bf 32}$.}
\label{tab:Z88_32}
\end{table}

%% file: Appendices/app_chipt.tex
\subsection{Chiral extrapolations}
\label{app:chipt}

We only consider physical (unitary) quarks, so $m^{val} = m^{sea}$. 
We use the following notation 
\begin{equation}
\begin{aligned}
m_l &= m_u=m_d,\\
\chi_l &= 2 \bar B_0^{\chi} m_l.
\end{aligned}
\end{equation}
such that at leading order (LO)
\begin{equation}
m_\pi^2 = 2 \bar B^0_{\chi} m_l = \chi_l.
\end{equation}
The parameter $\bar B$ related to the chiral condensate should not be confused with the bag parameter
(noted $\B$ in this appendix).
We consider kaon $SU(2)_L\times SU(2)_R$ $\chi$PT, ie $m_u=m_d \ll m_s, \Lambda_{QCD} $.
At next to leading order (NLO) we have

\begin{equation}\label{eq:MK} 
 \begin{aligned}
  m_K^2 &= \bar B^\chi m_s \left( 1 + \frac {a}{f^2}  \chi_l \right)  \;,\\
f_K  &= f^{\chi} \left(
1 + \frac{b}{f^2}\chi_l - \frac{3}{4} \frac{\chi_l}{(4\pi f)^2} \log \frac{\chi_l}{\Lambda^2} 
\right) \;, \\
\B_K = \B_1  &= \B_1^\chi \left( 
1 + \frac{c_1}{f^2}\chi_l - \frac{\chi_l}{2(4\pi f)^2} \log \frac{\chi_l}{\Lambda^2} 
\right).
 \end{aligned}
\end{equation}
Denoting the matrix element $\la \Kb | O_i | \K \ra $ by $\la O_i \ra$, we have
\begin{equation}
\la O_1 \ra = \frac{8}{3} m_K^2 f_K^2 \B_1 \;, 
\end{equation}
thus for the Standard Model matrix element, we find 
\begin{equation}
 \begin{aligned}
  \la O_1 \ra &=  \frac{8}{3} \B_1^\chi  \bar {B^{\chi}}^2 m_s {f^{\chi}}^2
\left( 1 + \frac{a+2b+c_1}{f^2} \chi_l - 2 \frac{\chi_l}{(4\pi f)^2} \log \frac{\chi_l}{\Lambda^2}  \right), \;
\\
&\equiv \la O_1 \ra^{\chi}  m_s 
\left( 1 + \frac{A_1}{f^2} \chi_l - 2 \frac{\chi_l}{(4\pi f)^2} \log \frac{\chi_l}{\Lambda^2}  \right).
 \end{aligned}
\end{equation}

We now turn to the BSM operators ($O_{i>1}$) in the SUSY basis. They read
\begin{equation}
\la O_i \ra = N_i \B_i \left( \frac{m_K^2 f_K}{m_s + m_l} \right)^2 \;,\quad
N_{2,\ldots,5} = \left\{ -\frac{5}{3},  \frac{1}{3}, 2, \frac{2}{3} \right\}.
\end{equation}
Rewriting Eq.~(\ref{eq:MK})
\begin{equation}
\frac{m_K^2}{m_s + m_l} = \bar B^\chi  \left( 1 + \frac {\tilde a}{f^2}  \chi_l \right).  \;
\end{equation}
The expansions for the Bag parameters read
\begin{equation}
\B_i = {\B_i}^{\chi} \left( 1 
+ \frac{c_i}{f^2}\chi_l  + s_i \frac {\chi_l}{2(4\pi f)^2} \log \frac{\chi_l}{\Lambda^2}  
\right).
\end{equation}
where $s_{2,3} = -1$ and $s_{4,5}=1$.

It is then clear that the combinations,
\begin{equation}
\frac{B_2}{B_3}, \quad \frac{B_4}{B_5}, \quad \frac{B_{2,3}}{B_K}, \quad B_{4,5}B_K \; \text{ and } B_{2,3}B_{4,5}
\end{equation}
have no leading order chiral logarithm.

For the matrix elements of the operators, we obtain the following expansions:
\begin{equation}
 \begin{aligned}
\la O_{\{2,3\}} \ra 
&= N_{\{2,3\}}  \B_i^\chi  \bar {B^{\chi}}^2 \left( 
1 + \frac{a+2b+c_{\{2,3\}}}{f^2} \chi_l - 2 \frac{\chi_l}{(4\pi f)^2} \log \frac{\chi_l}{\Lambda^2}  
\right),\\
&\equiv
\la O_{\{2,3\}} \ra^{\chi} \left( 
1 + \frac{A_{\{2,3\}}}{f^2} \chi_l - 2 \frac{\chi_l}{(4\pi f)^2} \log \frac{\chi_l}{\Lambda^2}  
\right), \\
\la O_{\{4,5\}} \ra 
&= N_{\{4,5\}}  \B_i^\chi  \bar {B^{\chi}}^2 \left( 
1 + \frac{a+2b+c_{\{4,5\}}}{f^2} \chi_l - \frac{\chi_l}{(4\pi f)^2} \log \frac{\chi_l}{\Lambda^2}  
\right), \\
&\equiv \la O_{\{4,5\}} \ra^{\chi} \left( 
1 + \frac{A_{\{4,5\}}}{f^2} \chi_l - \frac{\chi_l}{(4\pi f)^2} \log \frac{\chi_l}{\Lambda^2}  
\right).
 \end{aligned}
\end{equation}

Finally we consider the ratios $R_i$:
\begin{equation}
R_i = \frac{\la O_i \ra}{ \la O_1 \ra } \frac{m_K^2}{f_K^2}\;,
\end{equation}
this gives
\begin{equation}
 \begin{aligned}
  R_1 &=\frac{\bar B^\chi m_s}{ {f^{\chi}}^2 }
\left(
1 + \frac{C_{\{2,3\}}}{f^2}\chi_l + \frac{3}{2}\frac{\chi_l}{(4\pi f)^2} \log \frac{\chi_l}{\Lambda^2}  
\right), \\
R_{\{2,3\}} &= \frac{\la O_{\{2,3\}} \ra^{\chi}}{ \la O_1 \ra ^{\chi}} 
\left(
1 + \frac{C_{\{2,3\}}}{f^2}\chi_l + \frac{3}{2}\frac{\chi_l}{(4\pi f)^2} \log \frac{\chi_l}{\Lambda^2}  
\right), \\
R_{\{4,5\}} &= \frac{\la O_{\{4,5\}} \ra^{\chi}}{ \la O_1 \ra ^{\chi}} 
\left(
1 + \frac{C_{\{4,5\}}}{f^2}\chi_l + \frac{5}{2}\frac{\chi_l}{(4\pi f)^2} \log \frac{\chi_l}{\Lambda^2}  
\right).
 \end{aligned}
\end{equation}
We note that the chiral logarithms in $R_2$ and $R_3$ have the same coefficients as in $R_1$.
For completeness, we also give the following expressions:
\begin{equation}
 \begin{aligned}
  \frac{\la O_{\{2,3\}} \ra}{ \la O_1 \ra} 
&=
\frac{1}{m_s} \frac{\la O_{\{2,3\}} \ra^{\chi}}{ \la O_1 \ra ^{\chi}} 
\left(1 + \frac{D_{\{2,3\}}}{f^2} \right),
\\
\frac{\la O_{\{4,5\}} \ra}{ \la O_1 \ra}
&=
\frac{1}{m_s} \frac{\la O_{\{4,5\}} \ra^{\chi}}{ \la O_1 \ra ^{\chi}} 
\left(
1 + \frac{D_{\{4,5\}}}{f^2} + \frac{\chi_l}{(4\pi f)^2} \log \frac{\chi_l}{\Lambda^2} 
\right),
\\
\frac{ \la O_{\{2,3\}} \ra   \la O_{\{4,5\}} \ra }{f_K^4}
&=
\frac{ \la O_{\{2,3\}}^\chi \ra   \la O_{\{4,5\}}^\chi \ra }{{f_K^\chi}^4}
\left(1 + \frac{(C_{\{2,3\}} +  C_{\{4,5\}} - 2b )}{f^2} \chi_l 
\right).
 \end{aligned}
\end{equation}

%% file: Appendices/app_bare_results.tex
\subsection{Bare Results}
\label{appendix:bare_results}

\input Tables/table_bare24.tex
\input Tables/table_bare32.tex

Tables~\ref{table:barecoarse} and \ref{table:barefine}
show the fit results for the ratios of bare three-point function
as described in Section~\ref{sec:bare}. 
These quantities are obviously correlated, not only they
have been computed on the same gauge ensembles,
but they are normalised by the same quantity. 
Furthermore  $O_2$ and $O_3$ only differ by their colour structure
(and similarly for $O_4$ and $O_5$),
hence one expects them to have similar statistical fluctuations.
We find that the correlations depend very mildly on the quark masses,
so we only give the correlation matrices for the lightest unitary ensembles.
The numerical values can be found in Tables~\ref{table:Covbare24} and~\ref{table:Covbare32}.
%
\begin{table}[t]
  \be
  \begin{array}{c | c c c}
    \toprule
        &\;   \rat{3} \; &  \rat{4} \; & \rat{5} \; \\
  \hline
\rat{2} &-0.9947 &  -0.7008  &   -0.6906 \\
\rat{3} &        & \m0.6961  &  \m0.6861 \\
\rat{4} &        &           &  \m0.9948 \\
    \botrule
  \end{array}
  \nn
  \ee
  \caption{
Correlation matrix for the coarse lattice with $am_s = 0.04$ and $am_{ud}=0.005$.
}  
\label{table:Covbare24}
\end{table}
%
\begin{table}[t]
  \be
  \begin{array}{c | c c c}
    \toprule
        &\;   \rat{3} \; &  \rat{4} \; & \rat{5} \; \\
  \hline
\rat{2} &-0.9821 &  -0.5683  &  -0.5653 \\
\rat{3} &        & \m0.5684  & \m0.5645 1 \\
\rat{4} &        &           & \m0.9907 \\
    \botrule
  \end{array}
  \nn
\ee
\caption{
  Same as Table~\ref{table:Covbare24} for the fine lattice with $am_s = 0.03$ and $am_{ud}=0.004$.
}  
\label{table:Covbare32}
\end{table}
%
We also observe that the covariance matrices are very similar
for the two lattice spacings.
We find almost $100\%$ correlation between the colour partners  $(O_2,O_3)$
and $(O_4,O_5)$. The remaining correlation coefficients are of order $60-70\%$.


%% file: Tables/table_bare24.tex
\begin{table}[t]
\begin{tabular}{c | c c c | c c c}
  \toprule
  $am_{ud}$        &  0.005  &    0.010  &    0.020          &  0.005  &    0.010  &    0.020    \\
  \hline
  $am_{s}$ &  \multicolumn{3}{c|}{$ <O_2> / <O_1>$}      &  \multicolumn{3}{c}{$ <O_3> / <O_1>$} \\
 0.030 & -17.272(97) & -15.836(52) & -13.194(35)         &  4.1889(241)  & 3.8336(130)  &  3.1779(86) \\
 0.035 & -15.327(83) & -14.212(44) & -12.094(31)         &  3.7105(206)  & 3.4342(110)  &  2.9081(74) \\
 0.040 & -13.782(73) & -12.895(38) & -11.168(27)         &  3.3307(180)  & 3.1105(95)   &  2.6812(65)  \\
  \hline
  $am_{s}$ &  \multicolumn{3}{c|}{$ <O_4> / <O_1>$}     &  \multicolumn{3}{c}{$ <O_5> / <O_1>$}    \\
 0.030 & 32.418(124) & 29.079(87)  & 23.805(51)          &  10.703(40)   & 9.6505(279)  & 7.9879(166)   \\
 0.035 & 28.749(109) & 26.081(76)  & 21.798(45)          &  9.5504(355)  & 8.7068(244)  & 7.3565(145) \\
 0.040 & 25.826(98)  & 23.639(68)  & 20.101(39)          &  8.6312(319)  & 7.9374(218)  & 6.8219(128) \\  
\botrule
\end{tabular}
\caption{\raggedright{
  Fit results for the ratio of bare matrix elements on the coarse ensembles.
  The corresponding correlation matrix can be found in the text.
  }}\label{table:barecoarse}
\end{table}

%% file: Tables/table_bare32.tex
\begin{table}[t]
\begin{tabular}{c | c c c | c c c}
  \toprule
  $am_{ud}$        &  0.004  &    0.006  &    0.008          &  0.004  &    0.006  &    0.008    \\
  \hline
$am_{s}$ &  \multicolumn{3}{c|}{$ <O_2> / <O_1>$}      &  \multicolumn{3}{c}{$ <O_3> / <O_1>$} \\              
 0.025  & -18.947(92) & -17.548(62)  & -16.762(53)    &  4.6834(237) & 4.3303(159)  & 4.1325(135) \\   
 0.030  & -16.105(78) & -15.096(53)  & -14.526(46)    &  3.9713(200) & 3.7166(136)  & 3.5709(118) \\
  \hline
 $am_{s}$ &  \multicolumn{3}{c|}{$ <O_4> / <O_1>$}     &  \multicolumn{3}{c}{$ <O_5> / <O_1>$}    \\
 0.025  & 38.267(185) & 35.416(116)  & 33.398(87)     & 12.557(59)   & 11.651(36)  & 11.033(28)  \\
 0.030  & 32.541(171) & 30.371(102)  & 28.933(75)     & 10.760(55)   & 10.065(32)  &  9.625(24) \\
\botrule
\end{tabular}
\caption{ 
        Same as \ref{table:barefine} for the fine ensembles.
  }\label{table:barefine}
\end{table}

%% file: Appendices/app_MethodAB.tex
\subsection{Matrix elements from Methods A and B}
\label{appendix:MethodAB}

The SM matrix element is computed from $B_1=B_K$:
\be
\label{eq:SMME}
\la \Kb | O_1 |\K  \ra = \frac{8}{3}B_1m_K^2f_K^2 \quad .
\ee
For the BSM matrix elements $(i>1)$, we can either use the ratios $R_i$ 
\be
\label{eq:RtoME}
\la \Kb | O_i | \K \ra = R_i \, \la \Kb | O_1 |\K  \ra \;,
\ee
or the bag parameters $B_i$
\be
\label{eq:BtoME}
\la \Kb | O_i |\K  \ra = N_i \frac{m_K^4 f_K^2}{(m_s+m_d)^2} B_i\;.
\ee
In Eqs.~\ref{eq:SMME},\ref{eq:RtoME} and \ref{eq:BtoME}, we take $m_K = 495.6$ MeV, $f_K =156.2$ MeV.
For the value of $B_K$, we take the results obtained in this work,
but we checked that if we use the most recent value~\cite{Blum:2014tka},
the results are compatible within error and that the error remains the same.
For the quark masses, we take advantage of the precise values quoted
in~\cite{Blum:2014tka},
$m_d = 3.162(51)$ MeV, $m_s = 87.35(89)$ MeV for the SMOM-$\gmugmu$ scheme,
$m_d = 3.011(50)$ MeV, $m_s = 83.19(87)$ MeV in the SMOM-$\qq$ scheme.
and $m_s = 81.64(117)$ MeV and $m_d = 2.997(49)$ MeV in $\msbar$.
Our results are reported in Tables~\ref{table:ME1},\ref{table:ME2} and\ref {table:ME3}.\\
\begin{table}[t]
\centering
\begin{tabular}{c|l r|l r}
  \toprule
  & \multicolumn{2}{| c}{ from $R_i$ } & \multicolumn{2}{|c}{ from $B_i$ } \\
  \hline
 $\la \Kb | O_2 | \K \ra$ & $-0.1597(42)_{\rm stat}(34)_{\rm syst}$   & 3.4\% & $-0.1575(24)_{\rm stat}(63)_{\rm syst} $ & 4.3\%  \\
 $\la \Kb | O_3 | \K \ra$ & $\m0.0482(14)_{\rm stat}(15)_{\rm syst}$  & 4.2\% & $\m0.0464(9)_{\rm stat}(40)_{\rm syst}$ & 8.9\% \\
 $\la \Kb | O_4 | \K \ra$ & $\m0.3354(81)_{\rm stat}(163)_{\rm syst}$ & 5.4\% & $\m0.3377(42)_{\rm stat}(77)_{\rm syst}$ & 2.6\% \\
 $\la \Kb | O_5 | \K \ra$ & $\m0.0930(21)_{\rm stat}(70)_{\rm syst}$  & 7.9\% & $\m0.0941(11)_{\rm stat}(49)_{\rm syst}$ & 5.4\% \\
  \botrule
\end{tabular}
\caption{\raggedright{Four-quark operators Matrix elements in units of GeV$^4$ in the SMOM-$\gmugmu$ scheme at $\mu=3$ GeV.
    Results are obtained from the ratios $R_i$ and from the bag parameters $B_i$.
    The systematic errors combine the chiral and the discretisation errors,
    the percentage error is obtained by adding statistical and systematic errors in quadrature.
}}
\label{table:ME1}
\end{table}
\begin{table}[t]
\centering
\begin{tabular}{c|l r|l r}
  \toprule
  & \multicolumn{2}{| c}{ from $R_i$ } & \multicolumn{2}{|c}{ from $B_i$ } \\
  \hline
 $\la \Kb | O_2 | \K \ra$ & $ -0.1756(47)_{\rm stat}(33)_{\rm syst}$  & 3.2\% & $ -0.1726(26)_{\rm stat}(70)_{\rm syst}$ & 4.4\% \\
 $\la \Kb | O_3 | \K \ra$ & $\m0.0529(15)_{\rm stat}(18)_{\rm syst}$  & 4.4\% & $\m0.0509(10)_{\rm stat}(45)_{\rm syst}$ & 9.0\% \\
 $\la \Kb | O_4 | \K \ra$ & $\m0.3696(89)_{\rm stat}(173)_{\rm syst}$ & 5.3\% & $\m0.3715(46)_{\rm stat}(87)_{\rm syst}$ & 2.6\% \\
 $\la \Kb | O_5 | \K \ra$ & $\m0.0924(21)_{\rm stat}(72)_{\rm syst}$  & 8.1\% & $\m0.0935(11)_{\rm stat}(54)_{\rm syst}$ & 5.9\% \\
 \botrule
\end{tabular}
\caption{
    Same as the previous table but for the SMOM-$\qq$ scheme at $\mu=3$ GeV.}
\label{table:ME2}
  \end{table}
\begin{table}[t]
\centering
\begin{tabular}{c|l r|l r}
  \toprule
  & \multicolumn{2}{| c}{ from $R_i$ } & \multicolumn{2}{|c}{ from $B_i$ } \\
  \hline
 $\la \Kb | O_2 | \K \ra$ & $ -0.1636(43)_{\rm stat}(49)_{\rm syst}(36)_{\rm PT}$  & 4.5\% & $ -0.1671(25)_{\rm stat}(74)_{\rm syst}(8)_{\rm PT}$  & 4.8\% \\
 $\la \Kb | O_3 | \K \ra$ & $\m0.0510(14)_{\rm stat}(20)_{\rm syst}(12)_{\rm PT}$  & 5.3\% & $\m0.0509(10)_{\rm stat}(46)_{\rm syst}(4)_{\rm PT}$  & 9.3\% \\
 $\la \Kb | O_4 | \K \ra$ & $\m0.3619(87)_{\rm stat}(191)_{\rm syst}(94)_{\rm PT}$  & 6.4\% & $\m0.3781(47)_{\rm stat}(113)_{\rm syst}(48)_{\rm PT}$ & 3.5\% \\
 $\la \Kb | O_5 | \K \ra$ & $\m0.0923(21)_{\rm stat}(72)_{\rm syst}(27)_{\rm PT}$  & 8.7\% & $\m0.0969(12)_{\rm stat}(54)_{\rm syst}(27)_{\rm PT}$  & 6.9\% \\
 \botrule
\end{tabular}
\caption{\raggedright{Same as the previous tables but the results have been converted to $\msbar$ at $\mu=3$ GeV.
    The third error is the estimate of the error due the perturbative matching
    and is kept separate from the other systematic errors.
    For the percentage error, all the errors have been added in quadrature.}}
\label{table:ME3}
\end{table}
The two methods give very consistent results.
We also observe that for $O_{2,3}$ the ratios $R_i$ give more precise results,
whereas for $O_{4,5}$, the results obtained from the bag parameters $B_i$ have smaller error bars.
With this choice, we obtain the matrix element with a precision of $5\%$ or better.

%% file: Appendices/app_MethodC.tex
\subsection{Method C: Computing $B_i$ from $G_{ij}$}
\label{app:methodC}

The results for the quantities $G_{ij}$ given in~\ref{tab:syst3GeV}
have been  obtained by a linear extrapolation in $m_\pi^2$.
Combining these results with the numerical value of $B_1$,
we can can reconstruct the BSM bag parameters (see ~\ref{eq:G}).
We observe that effect of the chiral logarithm for $B_1$ is negligible
within our uncertainties. This has been confirmed by our recent
computation in which physical quark masses are included~\cite{Blum:2014tka}.
Here we find $B_1^{\gmugmu} = 0.523(11)$ at $\mu=3$ GeV, in complete agreement
with our new value $B_1^{\gmugmu} = 0.517(2)$.
Therefore the difference between the direct fit of the BSM bag parameters
and the bag parameters reconstructed from the quantities $G_{ij}$
is a direct indicator of the the chiral logarithms
potentially present in the BSM operators.
Using Eq.~(\ref{eq:G}) we find that
\begin{equation}
\begin{gathered}
B_2 = G_{21}B_K,\quad B_3 = \frac{G_{21}B_K}{G_{23}},\\
B_4 = \frac{G_{24}B_K}{G_{21}}, \quad B_5 = \frac{G_{24}B_K}{G_{45}G_{21}}.
\end{gathered}
\end{equation}
For three of the BSM bag parameters, we implement an alternative strategy, called {\bf Method $\bf C'$}.
We define other combinations of bag parameter 
(also free of leading chiral logarithm)
\begin{equation}
G_{31}=\frac{B_3}{B_K},\quad G_{41} = B_4 B_K ,\quad G_{51} = B_5 B_K.
\end{equation}
After extrapolation to the physical point, we extract the corresponding $B_{3,4,5}$
by inverting the previous system of equations.
\input Tables/table_MethodAC.tex

A couple of remarks are in order
\begin{itemize}
\item The difference between the various methods is smaller than
  our errors (actually smaller than the statistical error),
  showing that our chiral extrapolations are well under control
  within our precision.
\item The direct fit of the bag parameters give more precise results
  than the reconstruction from the combinations $G_{ij}$,
  with one notable exception: if we reconstruct $B_3$ from
  $G_{23}$, we obtain 
    $B_3^{\gmugmu} =     0.791(11)_{\rm stat} (45)_{\rm syst}  = 0.791(47)_{\rm combined}$\;.
  However the corresponding matrix element is better determined from the ratio $R_3$.
\item As mentioned in Section~\ref{sec:results},
  we have also computed $\tilde G_{23}$ (which is denoted by $G_{23}$ in \cite{Jang:2015sla}),
  the results are shown in Fig.~\ref{fig:tildeG23}, 
  We observe that $\tilde G_{23}$ exhibit large $a^2$ lattice artefacts,
  see the discussion in section~\ref{sec:results}. 
  Then $B_3$ can be computed from 
  \be
  B_3 = \left(5- \frac {3}{\tilde G_{23}}\right) G_{21} B_1 \;.
  \ee
  and we find
  \be 
  B_3 = 0.767(82) 
  \ee
\end{itemize}
Not surprisingly, the error quoted here is much larger than the obtained
from  $G_{23}$. Indeed, by changing the basis, the error varies by a factor two.

%% file: Tables/table_MethodAC.tex
\begin{table}[t]
\begin{tabular}{ c | c | c | c }
  \toprule
  Method & A & C & C' \\
  \hline
  $B_2$ & 0.526(20) &  0.526(26)  & - \\ 
  $B_3$ & 0.774(67) &  0.791(47)  & 0.774(76)  \\
  $B_4$ & 0.940(16) &  0.955(51)  & 0.954(29)  \\
  $B_5$ & 0.786(39) &  0.812(58)  & 0.801(52)  \\ 
  \botrule
\end{tabular}
\caption{\raggedright{
  Collection of results for the bag parameters using
  different methods. Results are given in $\gmugmu$ scheme at $3$ GeV
  and the errors combine statistical and systematic.}}
\end{table}

%% file: Appendices/app_correlations.tex
\subsection{Correlations}\label{app:correlations}

To provide the correlations between measurements we compute the correlation matrix from our bootstrapped data. We represent this data visually by a matrix plot for our various measurement techniques, orange illustrates positive correlation and blue indicates anti-correlation, the darker the colour the stronger the correlation. Black squares are by definition 1. The analysis is done with $500$ bootstrap samples. 

\begin{figure}[h]
\subfloat[$\gmugmu$ scheme $R_i$]
{
\includegraphics[type=pdf,ext=.pdf,read=.pdf,width=6cm]{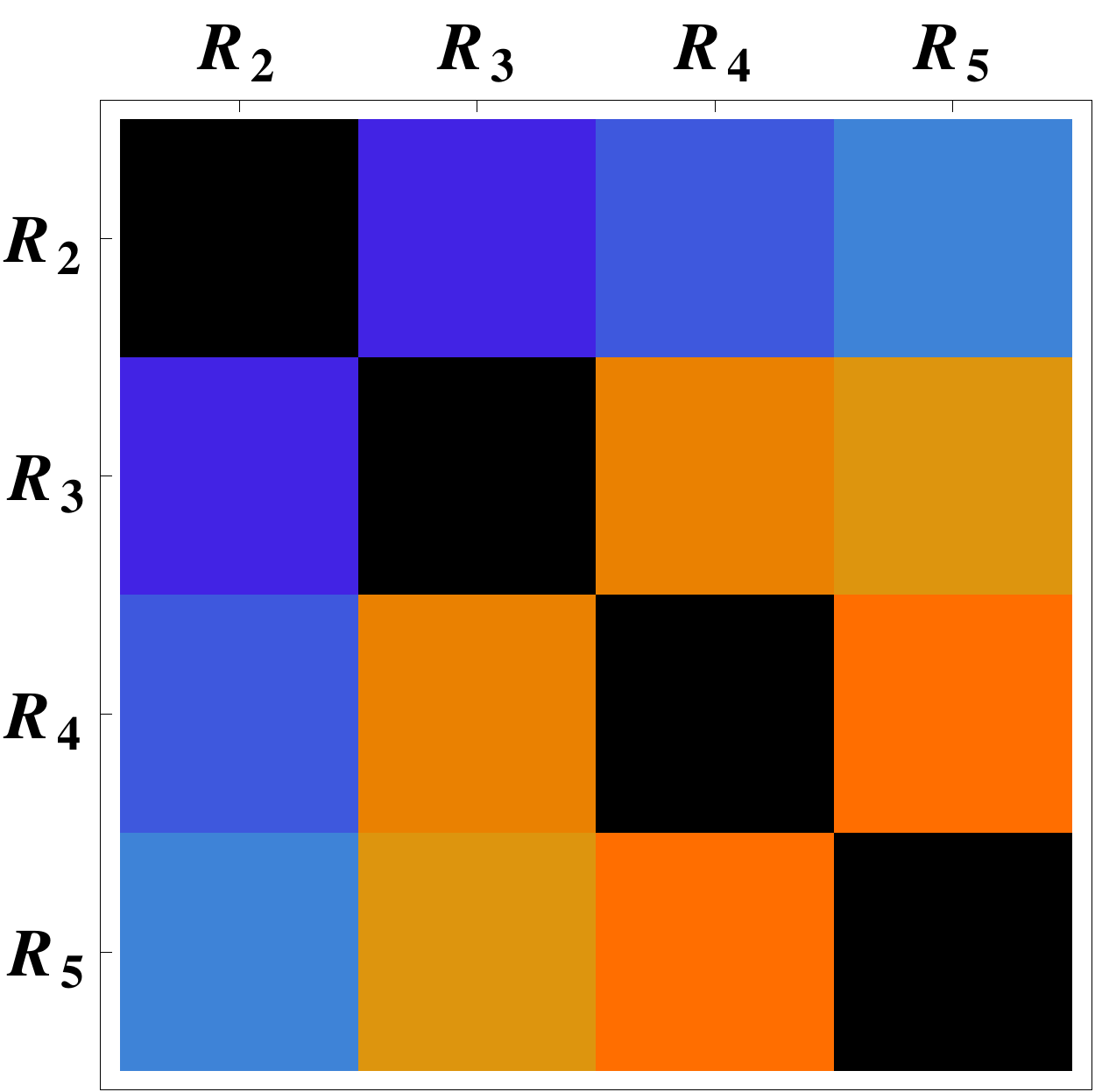}\label{app:fig:R_gg_correlations}
}
\subfloat[$\qq$ scheme $R_i$]
{
\includegraphics[type=pdf,ext=.pdf,read=.pdf,width=6cm]{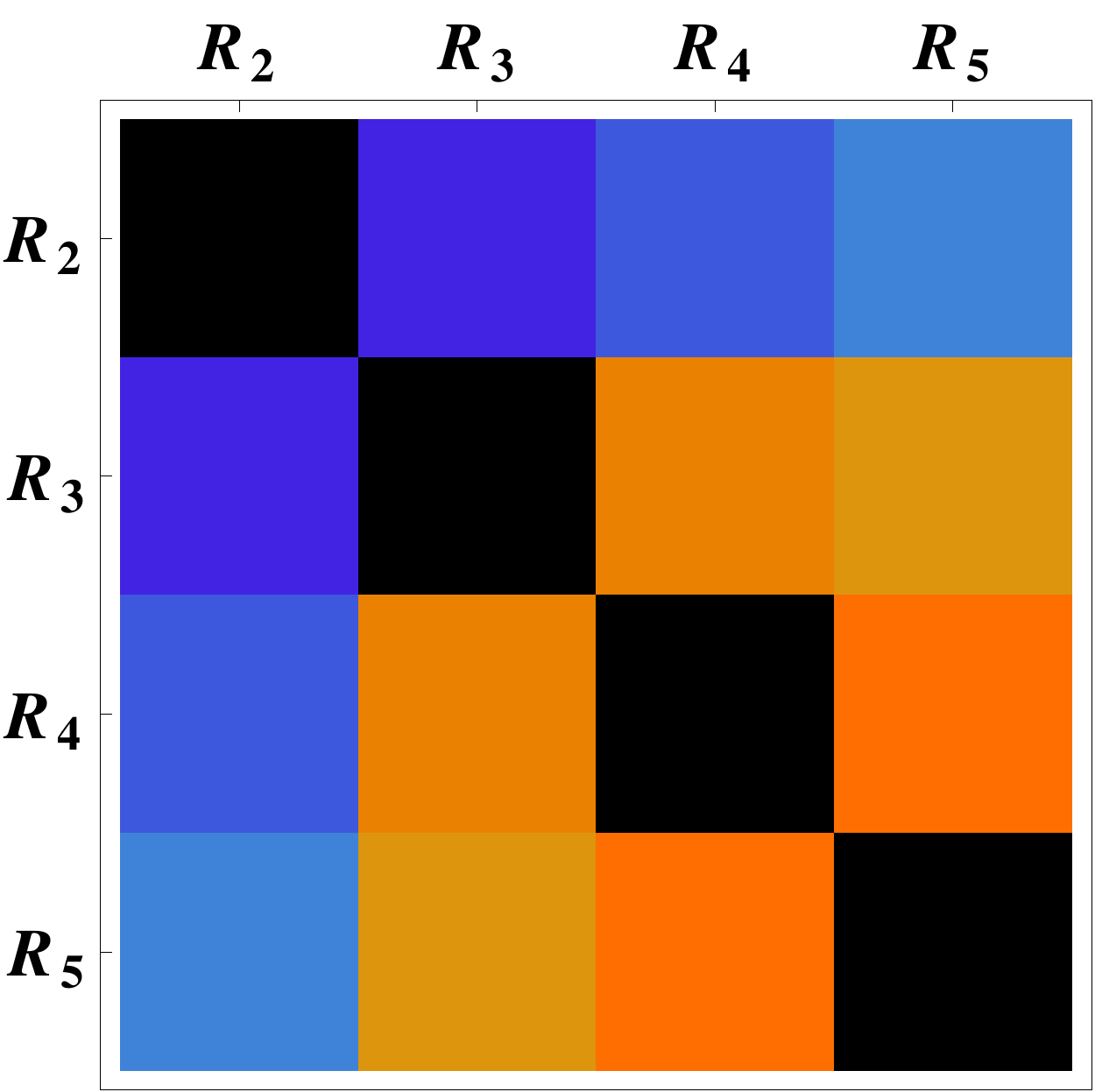}\label{app:fig:R_qq_correlations}
}
\caption{\raggedright{
Ratios $R_i$ for our two intermediate SMOM schemes renormalised at $\mu=3\text{ GeV}$, orange indicates positive correlation and blue anti-correlation, darker colours show a stronger correlation. These are a visualisation of the data from Table~\ref{app:table:CovR}.
}}\label{app:fig:R_correlations}
\end{figure}

\begin{table}[h!]
  \subfloat[$\gmugmu$ scheme $R_i$]
           {
             \begin{minipage}{0.4\textwidth}
               \be
               \begin{array}{c | c c c}
                 \toprule
                 &\;   R_3 \; &  R_4 \; & R_5 \; \\
                 \hline
                 R_2 \;& -0.9951   &  -0.9438   &  -0.9366 \\
                 R_3 \;&           & \m0.9342   & \m0.9216 \\
                 R_4 \;&           &            & \m0.9937 \\
                 \botrule
               \end{array}
               \nn
               \ee
             \end{minipage}
           }
  \subfloat[$\qq$ scheme $R_i$]
           {
             \begin{minipage}{0.4\textwidth}
               \be
               \begin{array}{c | c c c}
                 \toprule
                 &\;   R_3 \; &  R_4 \; & R_5 \; \\
                 \hline
                 R_2 \;&  -0.9882  &   -0.9398  &  -0.9329 \\
                 R_3 \;&           &  \m0.9338  & \m0.9208 \\
                 R_4 \;&           &            & \m0.9935 \\
                 \botrule
               \end{array}
               \nn
               \ee
             \end{minipage}
             }
           \caption{\raggedright{
               Correlation matrices for the ratios $R_i$ in our SMOM schemes at $\mu=3$ GeV.}}  
           \label{app:table:CovR}
\end{table}

In Fig.~\ref{app:fig:R_correlations} we compare the correlations between the ratios $R_i$ for our two SMOM schemes.
We observe that $R_2$ is strongly anti-correlated with all of the others due to the difference in sign with the others,
this has operator signature $SS-PP$ and most of the other ratios are strongly correlated with one-another.

We note that the correlations for the $\gmugmu$ and $\qq$ schemes are very similar, this is in fact a feature
for the other measurements so we will only show the $\gmugmu$ scheme evaluations for the $B$s and $G$s.

\begin{figure}[h]
\subfloat[$\gmugmu$ scheme $B_i$]
{
\includegraphics[type=pdf,ext=.pdf,read=.pdf,width=6cm]{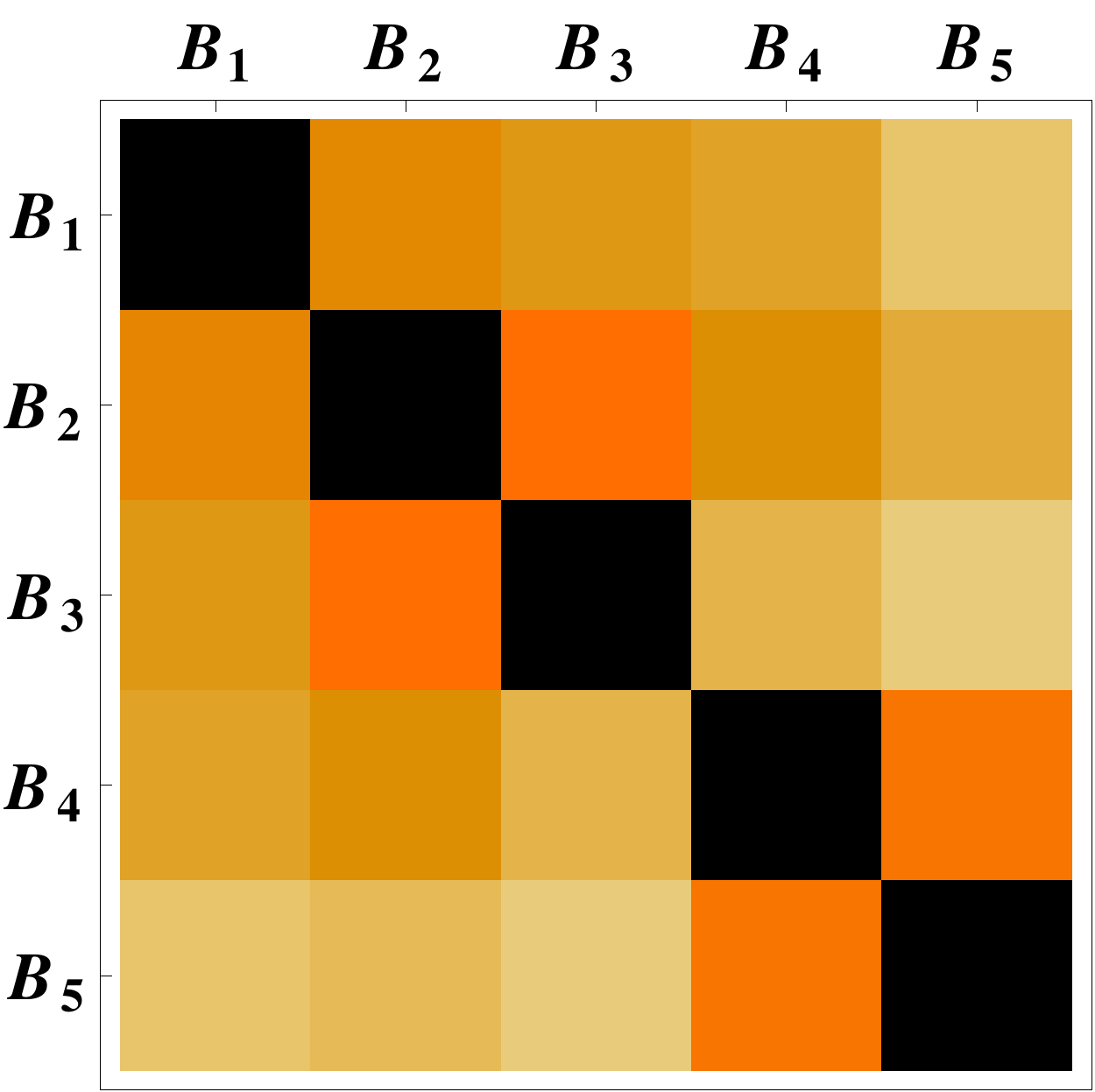}\label{app:fig:BGcorrelations_Bs}
}
\subfloat[$\gmugmu$ scheme $G_{ij}$]
{
\includegraphics[type=pdf,ext=.pdf,read=.pdf,width=6cm]{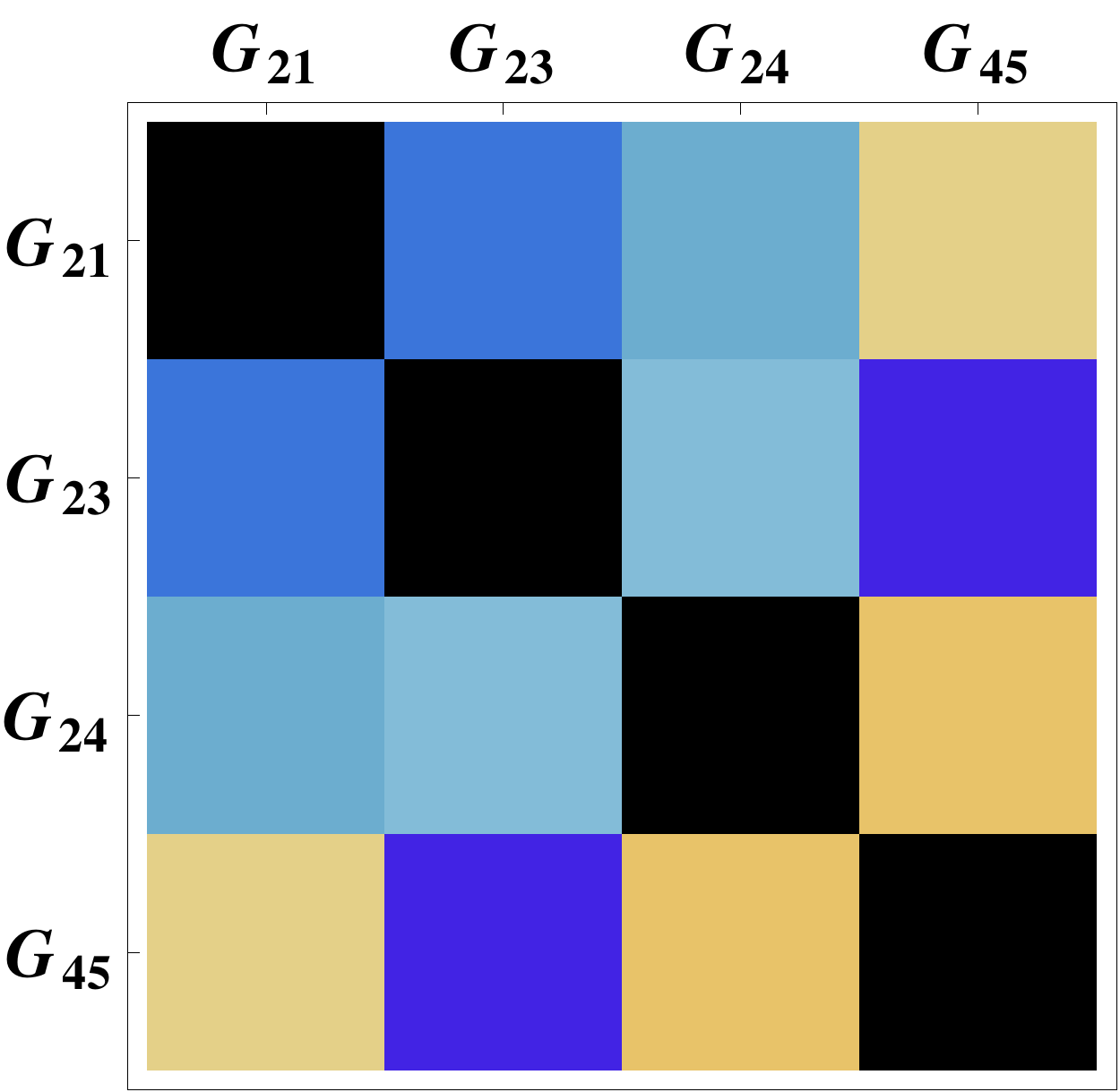}\label{app:fig:BGcorrelations_Gs}
}
\caption{\raggedright{
Correlation matrices in the $\gmugmu$ scheme renormalised at $\mu=3\text{ GeV}$ for the bag parameters $B_i$ and the combinations $G_{ij}$. This is a visualisation of the data in Table~\ref{app:table:CovBandG}.
}}\label{app:fig:BGcorrelations}
\end{figure}

\begin{table}[h!]
  \subfloat[$B_i$]
           {
             \begin{minipage}{0.4\textwidth}
               \be
               \begin{array}{c | c c c c}
                 \toprule
                 &\;   B_2 &\;   B_3 \; &  B_4 \; & B_5 \; \\
                 \hline
                 B_1 \;&   0.6762  &  0.6252 &   0.5961 &   0.5486 \\
                 B_2 \;&           &  0.9673 &   0.6356 &   0.5772 \\
                 B_3 \;&           &         &   0.5723 &   0.4974 \\
                 B_4 \;&           &         &          &   0.9016 \\
                 \botrule  
               \end{array}
               \nn
               \ee
             \end{minipage}
           }
           \subfloat[$G_{ij}$]
           {
             \begin{minipage}{0.4\textwidth}
               \be
               \begin{array}{c | c c c}
                 \toprule
                 &\;   G_{23} \; &  G_{24} \; & G_{45} \; \\
                 \hline
                 G_{21} \;&  -0.1883  &  -0.0333  &  \m0.0512 \\
                 G_{23} \;&           &  -0.0325  &   -0.2834 \\
                 G_{24} \;&           &            &  \m0.0739 \\
                 \botrule
               \end{array}
               \nn
               \ee
             \end{minipage}
           }
           \caption{\raggedright{
               Correlation matrices for the bag parameters $B_i$ and the combinations $G_{ij}$
               at $\mu=3$ GeV. We only show the SMOM-$\gmugmu$ results because the $\qq$ ones
               are almost identical.
           }}
           \label{app:table:CovBandG}
\end{table}

Similarly to what we found for the correlation between bare ratios,
we observe that the colour partners $(R_2,R_3)$ and $(R_4,R_5)$ are almost
$100\%$ correlated. (More precisely anti-correlated in the former case
because there is an relative sign in $R_2$ compared so the other ratios).
However the correlation between operators of different chirality
is enhanced compared to the bare rations ($\sim 90\%$).
As illustrated in Table~\ref{app:table:CovR}, the correlation matrix
does not depend on the renormalisation scheme. Although not shown here, the matching
to $\msbar$ also has almost no visible effects on the correlations.

For the bag parameters $B_i$, we observe a similar pattern,
however the correlations between $B_4$ and $B_5$ drops to $90\% $.
The correlations between operators of different chirality
are significantly lower than the ones for the ratios $R_i$,
namely around $50-60\%$. As expected, the quantities $G_{ij}$ do not exhibit significant correlation.